\documentclass[letterpaper,twocolumn,10pt]{article}
\usepackage{usenix-2020-09}
\usepackage[english]{babel}
\usepackage[utf8]{inputenc}
\usepackage[T1]{fontenc}

\usepackage{paralist}
\newcommand\repo{\url{https://github.com/proteus-core/prospect/tree/usenix_artifact}}

\usepackage[available,functional,reproduced]{usenixbadges}

\input{packages.sty}
\input{macros.sty}
\newcommand\stepRule[0]{
  \infer[step] { %
    \micro' \mydef \update(\leaked{\smemproj{\mem}}, \leaked{\sproj{\reg}}, \leaked{\sbufproj{\buf}}, \leaked{\micro}) \\
    \directive \mydef \nextdir(\micro') \\
    \smicro{\directive}{} {\conf{\mem, \reg, \buf, \micro'}} {\conf{\mem',
        \reg', \buf', \micro''}}
    {} \\
  }{
    \smicro{}{}
    {\conf{\mem, \reg, \buf, \micro}}
    {\conf{\mem', \reg', \buf', \micro''}}
    {}
  }
}

\newcommand\fetchPredictBranchJump[0]{
  \inferrule[fetch-predict-branch-jmp]
  {
    \slevel{\val{l}}{\_} \mydef \semexpr{\pc}{\apl(\buf, \reg)} \\
    \prog{\val{l}} \in \{\beqz{e}{\_}, \jmp{e}\} \\
    \val{l'} \mydef \predict(\micro) \\
  }
  {
    \smicro{\fetch}{}
    {\conf{\mem,\reg, \buf, \micro}}
    {\conf{\mem,\reg, \buf \cdot \pc \gets \leaked{\slevel{\val{l'}}{\botsec}} @ \leaked{\val{l}}, \micro}}
    {}
  }
}

\newcommand\fetchOthers[0]{
  \inferrule[fetch-others]
  {
    \reg' \mydef \apl(\buf, \reg) \\
    \slevel{l}{\_} \mydef \semexpr{\pc}{\reg'} \\
    \prog{\val{l}} \not\in \{\beqz{\_}{\_}, \jmp{\_}\} \\
    \slevel{l'}{\_} \mydef \semexpr{\pc+1}{\reg'} \\
  }
  {
    \smicro
    {\fetch}{}
    {\conf{\mem,\reg, \buf, \micro}} {\conf{\mem,\reg, \buf \cdot
        \prog{l}@\varepsilon \cdot \incrpc{\leaked{\slevel{l'}{\botsec}}}
          @\varepsilon, \micro}} {}
  }
}

\newcommand\branchCommit[0]{
  \infer[branch-commit]
  {
    |\buf| = i-1 \\
    \prog{\val{l_0}} = \beqz{e}{\val{l'}} \\
    \slevel{\val{c}}{\_} \mydef \semexpr{e}{\aplsan(\buf, \reg)} \\
    \val{l''} \mydef \ite{\val{c} = 0}{\val{l'}}{\val{l_0} + 1}\\
    \val{l} = \val{l''}
  }{
    \smicro
    {\execute{i}}{}
    {\conf{\mem,\reg, \buf \cdot  \pc \gets \slevel{l}{\_} @ \val{l_0} \cdot \buf', \micro}}
    {\conf{ \mem, \reg, \buf \cdot \pc \gets \leaked{\slevel{l}{\botsec}}@\varepsilon \cdot \buf', \micro}}
    {}
  }
}

\newcommand\branchRollback[0]{
  \infer[branch-rollback]
  {
    |\buf| = i-1 \\
    \prog{\val{l_0}} = \beqz{e}{\val{l'}} \\
    \slevel{\val{c}}{\_} \mydef \semexpr{e}{\aplsan(\buf, \reg)} \\
    \val{l''} \mydef \ite{\val{c} = 0}{\val{l'}}{\val{l_0} + 1}\\
    \val{l} \neq \val{l''}
    \\
  }{
    \smicro
    {\execute{i}}{}
    {\conf{\mem,\reg, \buf \cdot  \pc \gets \slevel{l}{\_} @ \val{l_0} \cdot \buf', \micro}}
    {\conf{ \mem, \reg, \buf \cdot \pc \gets \leaked{\slevel{l''}{\botsec}}@\varepsilon, \micro}}
    {}
  }
}

\newcommand\jmpCommit[0]{
  \infer[jmp-commit]
  {
    |\buf| = i-1 \\
    \prog{\val{l_0}} = \jmp{e} \\
     \slevel{\val{l'}}{\_} \mydef{} \semexpr{e}{\aplsan(\buf, \reg)} \\
     \val{l'} = \val{l} \\
  }
  {
    \smicro
    {\execute{i}}{}
    {\conf{\mem,\reg, \buf \cdot  \pc \gets \slevel{l}{\_} @ \val{l_0} \cdot \buf', \micro}}
    {\conf{\mem, \reg, \buf \cdot \pc \gets \leaked{\slevel{l}{\botsec}}@\varepsilon \cdot \buf', \micro}}
    {}
  }
}

\newcommand\jmpRollback[0]{
  \infer[jmp-rollback]
  {
    |\buf| = i-1 \\
    \prog{\val{l_0}} = \jmp{e} \\
     \slevel{\val{l'}}{\_} \mydef{} \semexpr{e}{\aplsan(\buf, \reg)} \\
     \val{l'} \neq \val{l} \\
  }
  {
    \smicro
    {\execute{i}}{}
    {\conf{\mem,\reg, \buf \cdot  \pc \gets \slevel{l}{\_} @ \val{l_0} \cdot \buf', \micro}}
    {\conf{\mem, \reg, \buf \cdot \pc \gets \leaked{\slevel{l'}{\botsec}}@\varepsilon, \micro}}
    {}
  }
}

\newcommand\executeAssign[0]{
  \infer[execute-assign]
  {
    |\buf| = i-1 \\
    e \not\in \SVals \\
    inst = \mov{r}{e} @ T\\
    \rulebox{\slevel{\val{v}}{s} \mydef \semexpr{e}{\apl(\buf, \reg)}}\\
    inst' \mydef \mov{r}{\slevel{\val{v}}{s}} @T
  }
  {
    \smicro
    {\execute{i}}{}
    {\conf{\mem,\reg, \buf \cdot inst \cdot \buf', \micro}}
    {\conf{ \mem, \reg, \buf \cdot inst' \cdot \buf', \micro}}
    {}
  }
}

\newcommand\executeLoadPredict[0]{
  \inferrule[execute-load-predict]
  {
    |\buf| = i-1 \\
    inst = \load{x}{e} @ T \\
    \slevel{\val{l}}{\_} \mydef \semexpr{\pc}{\apl(\buf, \reg)} \\
    \val{v} \mydef \predict(\micro) \\
    inst' \mydef \mov{x}{\leaked{\slevel{\val{v}}{\botsec}} @ \leaked{\val{l}}}
  }{
    \smicro
    {\execute{i}}{}
    {\conf{\mem,\reg, \buf \cdot inst \cdot \buf', \micro}}
    {\conf{ \mem, \reg, \buf \cdot inst' \cdot \buf', \micro}}
    {}
  }
}

\newcommand\executeLoadCommit[0]{
  \inferrule[execute-load-commit]
  {
    |\buf| = i-1 \\
    inst = \mov{x}{\slevel{\val{v}}{\_}} @ \val{l_0} \\
    \prog{\val{l_0}} = \load{x}{e} \\
    \store{\_}{\_} \not\in \buf\\
    \slevel{\val{a}}{\_} \mydef \semexpr{e}{\aplsan(\buf, \reg)} \\
    \mem(\val{a}) = \val{v} \\
    \memsec(\val{a}) = \botsec \\
    inst' = \mov{x}{\slevel{\val{v}}{\memsec(\val{a})}} @ \varepsilon \\
    \micro' \mydef \update(\micro, \leaked{\val{a}}) \\
  }
  {
    \smicro
    {\execute{i}}{}
    {\conf{\mem, \reg, \buf \cdot inst \cdot \buf', \micro}}
    {\conf{ \mem, \reg, \buf \cdot inst' \cdot \buf', \micro'}}
    {}
  }
}

\newcommand\executeLoadRollback[0]{
  \inferrule[execute-load-rollback]
  {
    |\buf| = i-1 \\
    inst = \mov{x}{\slevel{\val{v}}{\_}} @ \val{l_0} \\
    \prog{l_0} = \load{x}{e} \\
    \store{\_}{\_} \not\in \buf\\
    \slevel{\val{a}}{\_} \mydef \semexpr{e}{\aplsan(\buf, \reg)} \\
    \mem(\val{a}) \neq \val{v} \vee \memsec(\val{a}) = \topsec \\
    inst' = \mov{x}{\slevel{\mem(a)}{\memsec(\val{a})}} @ \varepsilon \\
    \micro' \mydef \update(\micro, \leaked{\val{a}}) \\
  }
  {
    \smicro
    {\execute{i}}{}
    {\conf{\mem, \reg, \buf \cdot inst \cdot \incrpc{\slevel{l}{s}}
          @\varepsilon \cdot \buf', \micro}}
    {\conf{\mem, \reg, \buf \cdot inst' \cdot \incrpc{\slevel{l}{s}}
          @\varepsilon, \micro'}}
    {}
  }
}

\newcommand\executeStore[0]{
  \inferrule[execute-store]
  {
    |\buf| = i-1 \\
    e_{a}, e_{v} \not\in \SVals \\
    \slevel{\val{a}}{\_} \mydef \semexpr{e_{a}}{\aplsan(\buf, \reg)} \\
    \rulebox{\slevel{\val{v}}{s} \mydef \semexpr{e_{v}}{\apl(\buf, \reg)}}\\
  }
  {
    \smicro
    {\execute{i}}{}
    {\conf{\mem, \reg, \buf \cdot \store{e_{a}}{e_{v}} @ T \cdot \buf', \micro}}
    {\conf{\mem, \reg, \buf \cdot \store{\leaked{\slevel{a}{\botsec}}}{\slevel{v}{s}} @ T \cdot \buf', \micro}}
    {}
  }
}

\newcommand\retireAssign[0]{
\infer[retire-assign]
  {
    \buf = \mov{r}{\slevel{v}{s}}  @ \varepsilon \cdot \buf' \\
  }
  {
    \smicro
    {\retire}
    {}
    {\conf{\mem,\reg, \buf, \micro}}
    {\conf{\mem, \reg\change{r}{\slevel{v}{s}}, \buf', \micro}}
    {}
  }
}

\newcommand\retireStoreLow[0]{
  \inferrule[retire-store-low] {
    \buf = \store{\slevel{a}{\_}}{\slevel{\val{v}}{\_}} @ \varepsilon \cdot \buf' \\
    \micro' = \update(\micro, \leaked{\val{a}}) \\
    \memsec(\val{a}) = \botsec \\
  }{ \smicro {\retire}{}
    {\conf{\mem, \reg, \buf, \micro}}
    {\conf{\mem\change{\val{a}}{\val{v}}, \reg, \buf', \micro'}}
    {\val{v}}
  }
}

\newcommand\retireStoreHigh[0]{
  \inferrule[retire-store-high] { %
    \buf = \store{\slevel{a}{\_}}{\slevel{\val{v}}{\_}} @ \varepsilon \cdot \buf' \\
    \micro' = \update(\micro, \leaked{\val{a}}) \\
    \memsec(\val{a}) = \topsec\\
  }{ \smicro {\retire}{}
    {\conf{\mem, \reg, \buf, \micro}}
    {\conf{\mem\change{\val{a}}{\val{v}}, \reg, \buf', \micro'}}
    {\varepsilon}
  }
}

\newcommand\retireStoreLowPatched[0]{
    \inferrule[retire-store-patched] {
    \buf = \store{\slevel{\val{a}}{\_}}{\slevel{\val{v}}{\_}} @ \varepsilon \cdot \buf' \\
    \micro' = \update(\micro, \leaked{\val{a}}) \\
    \declassified = \declassify{\val{v'}} \cdot \declassify{\declassified'}\\
    \memsec(\val{a}) = \botsec
  }{ \smicropatched {\retire}{}
    {\conf{\mem, \reg, \buf, \micro}}
    {\conf{\mem\change{\val{a}}{\declassify{\val{v'}}}, \reg, \buf', \micro'}}
    {\declassified}
    {\declassified'}
  }
}

\input{my_biblatex.sty}
\begin{document}
\title{\Large \bf \name{}: Provably Secure Speculation for the Constant-Time Policy \iftechreport\newline (Extended version)\fi}

\author[1]{Lesly-Ann Daniel}
\author[1]{Marton Bognar}
\author[1]{Job Noorman}
\author[2]{S{\'e}bastien Bardin}
\author[3]{Tamara Rezk}
\author[1]{Frank Piessens}
\affil[1]{imec-DistriNet, KU Leuven, 3001 Leuven, Belgium}
\affil[2]{CEA, List, Université Paris Saclay, France}
\affil[3]{INRIA, Université Côte d’Azur, Sophia Antipolis, France}

\renewcommand\Authands{ and }

\maketitle

\begin{abstract}
We propose \name{}, a generic formal processor model providing provably secure speculation for the constant-time policy.
For constant-time programs under a {\em non-speculative} semantics, \name{} guarantees that speculative and out-of-order execution cause no microarchitectural leaks.
This guarantee is achieved by tracking secrets in the processor pipeline and ensuring that they do not influence the microarchitectural state during speculative execution.
Our formalization covers a broad class of speculation mechanisms, generalizing prior work.
As a result, our security proof covers all known Spectre attacks, including load value injection (LVI) attacks.

In addition to the formal model, we provide a prototype hardware implementation of \name{} on a RISC-V processor and show evidence of its low impact on hardware cost, performance, and required software changes.
In particular, the experimental evaluation confirms our expectation that for a compliant constant-time binary, enabling ProSpeCT incurs no performance overhead.
\end{abstract}

\section{Introduction}
It is well-understood that microarchitectural optimization techniques commonly used in processors can lead to security vulnerabilities~\cite{Ge2018}.
One of the most recent and challenging problems in this space is the family of Spectre attacks~\cite{Kocher2019}, which abuse speculative execution to leak secrets to an attacker that can observe parts of the microarchitectural state of the platform on which the victim is executing.

In response to the discovery of Spectre, a wide range of countermeasures has already been proposed~\cite{Canella2019,DBLP:conf/usenix/AmitJW19,turnerRetpolineSoftwareConstruct,poimboeufPATCHV2Static2018,SpeculativeLoadHardening,indexmasking,DBLP:conf/asplos/TaramVT19,DBLP:conf/dac/KhasawnehKSEPA19,DBLP:conf/hpca/LiZH0M19,DBLP:conf/IEEEpact/BarberBZZT19,DBLP:conf/isca/Ainsworth020,DBLP:conf/isca/SakalisKRJS19,DBLP:conf/micro/SaileshwarQ19,DBLP:conf/micro/WeisseNLWK19,DBLP:conf/micro/YanCS0FT18,DBLP:conf/micro/YuYK0TF19,DBLP:conf/raid/ThomaFKGB21,DBLP:conf/secdev/KimM0MCMT020}.
It is an important and difficult challenge to understand the trade-offs offered by these mitigations in terms of security, performance, and applicability to legacy hardware or software.

On the one hand, software countermeasures targeting specific transient execution attacks
can still leave other attacks unmitigated~\cite{DBLP:conf/ndss/DanielBR21}, and they must be patched every time new
speculation mechanisms are introduced (e.g.,\ the predictive store forwarding
feature newly introduced in AMD Zen3
processors~\cite{SecurityAnalysisAMD2021}).
On the other hand, mainstream hardware
mitigations have been recently shown ineffective~\cite{barberis_branch_2022} against Spectre-v2 (BTB)
attacks~\cite{DBLP:conf/sp/KocherHFGGHHLM019}. %

\myparagraph{Hardware-based secure speculation} In a recent paper, Guarnieri et
al.~\cite{DBLP:conf/sp/GuarnieriKRV21} propose \emph{hardware-software
  contracts} to compare hardware-based mechanisms for secure speculation and
better understand how these defenses can enable software to provide end-to-end
security guarantees. For instance, they show that certain types of
hardware-level taint
tracking~\cite{DBLP:conf/micro/YuYK0TF19,DBLP:conf/micro/WeisseNLWK19,DBLP:conf/IEEEpact/BarberBZZT19}
provide secure speculation for the \emph{sandboxing} policy. %
On processors implementing one of these mechanisms, the software can simply enforce
the sandboxing policy under a non-speculative semantics and does not need to
consider the (error-prone and possibly expensive) placement of software
speculation barriers.

However, none of the hardware defenses studied under the
hardware-software contract framework enable secure speculation for the
constant-time policy, except for completely disabling speculative execution.
Hence, the classic cryptographic constant-time programming
model~\cite{DBLP:conf/uss/AlmeidaBBDE16} does not suffice to guarantee security
on processors with these countermeasures, and significantly more complex and
costly software programming models are required to recover
security~\cite{DBLP:journals/corr/abs-1902-05178,DBLP:conf/ccs/GuancialeBD20,
  DBLP:conf/pldi/CauligiDGTSRB20, DBLP:conf/sp/GuarnieriKMRS20,
  DBLP:conf/ndss/DanielBR21, DBLP:conf/sp/BartheCGKLOPRS21,
  DBLP:journals/pacmpl/VassenaDGCKJTS21}. %

\myparagraph{Problem statement} In this paper, we investigate how to
  provide efficient provably secure speculation for the
  constant-time policy under a wide range of speculation mechanisms.
Specifically, we apply the hardware-software contract framework to another
class of hardware taint-tracking mechanisms explicitly tracking {\em secrecy} of data in
the microarchitecture (e.g., systems like
ConTExT~\cite{DBLP:conf/ndss/0001LCSKG20},
SpectreGuard~\cite{DBLP:conf/dac/FustosFY19}, or
SPT~\cite{DBLP:conf/micro/ChoudharyYF021}). In such systems, a constant-time
program informs the processor about which memory cells contain secret data.
Using this additional information, hardware-based taint-tracking can provide
\emph{stronger} security guarantees than sandboxing
approaches~\cite{DBLP:conf/sp/GuarnieriKRV21}. Additionally, we consider a wide
variety of speculation mechanisms, whereas the model of Guarnieri et al.
considers only speculation on conditional branches.

\myparagraph{Our proposal} The main contribution of this paper is \name{}, a
generic processor model formalizing the essence of such secrecy-tracking
hardware mechanisms and a proof that it provides secure speculation for the
constant-time policy. %
Specifically, off-the-shelf constant-time cryptographic libraries can be run
securely on \name{} without additional protections for transient execution attacks.

\name{} is modular in the implementation of predictors and covers a broad class
of speculation mechanisms, including branch
prediction and store-to-load forwarding. As a novel aspect, \name{} additionally
covers new mechanisms like predictive store
forwarding~\cite{SecurityAnalysisAMD2021} and even mechanisms that are not (yet)
implemented in commercial processors, such as load value
prediction~\cite{DBLP:conf/asplos/LipastiWS96} or value
prediction~\cite{DBLP:conf/micro/LipastiS96}. %
In particular, we rigorously show that \name{} protects against Spectre-v2 (BTB)
attacks~\cite{DBLP:conf/sp/KocherHFGGHHLM019}, for which mainstream hardware
mitigations have recently been shown ineffective~\cite{barberis_branch_2022}. As
evidence for generality, we show that our mechanism even protects against Load
Value Injection (LVI) attacks~\cite{Bulck2020}, which are particularly
challenging to mitigate.

Another novel aspect of our formalization is the statement of our security condition,
which allows a program to declassify a ciphertext while still requiring the processor to make sure that the attacker does not learn anything about the plaintext or the key used to compute the ciphertext.

To demonstrate the viability of our proposed mechanism, we extend a RISC-V processor to
be \name{}-compliant and quantify the hardware costs. Results show that the
overhead of \name{} in area usage and critical path is reasonable. We also
demonstrate that the required software changes to cryptographic code are minimal and
that the performance impact is negligible if secrets are precisely annotated.
Our prototype is the first non-simulated hardware implementation of a
speculative and out-of-order processor that implements secure speculation for
the constant-time policy.

\myparagraph{Contribution} In summary, our contributions are:
\begin{itemize}
  \item We present \name, the first formal processor model %
        providing provably secure speculation for the constant-time policy
        (\cref{sec:formalization}).
        We propose a formal model of a processor that tracks secrets during
        execution and temporarily blocks speculative execution if secrets could
        leak. %
        The model is generic; it supports a wide range of
        speculation mechanisms and formalizes the guarantees provided by
        prior hardware-based secrecy tracking
        mechanisms~\cite{DBLP:conf/dac/FustosFY19, DBLP:conf/ndss/0001LCSKG20,
        DBLP:conf/micro/ChoudharyYF021}.

  \item We formally prove that \name{} provides secure speculation for the
        constant-time policy, i.e.,\ programs that comply with the
        classic cryptographic constant-time discipline will not leak secrets
        through microarchitectural channels (\cref{thm:security0}), including in the presence of declassification (\cref{thm:patched_declassification}). %
        The proof holds for a large variety of speculation mechanisms,
        encompassing all known Spectre and LVI attacks.

  \item We are the first to consider load value speculation. Interestingly, our
        formal analysis reveals that executions resulting from \emph{correct}
        load value speculation must sometimes be rolled-back to avoid attacks
        based on \emph{implicit resolution-based
        channels}~\cite{DBLP:conf/micro/YuYK0TF19}. We prove this formally (\cref{thm:security0}) and provide an example in \cref{sec:example}.

  \item We provide the first non-simulated hardware implementation of a
        processor offering secure speculation. We implement \name{} on a
        RISC-V processor supporting speculation (\cref{sec:implem}) and
        evaluate the costs of the proposed mechanism in terms of hardware,
        performance, and manual effort for precisely marking secret data
        (\cref{sec:eval}).
\end{itemize}
\myparagraph{Availability} Our implementation and the experimental evaluation
are open-sourced at \url{https://github.com/proteus-core/prospect}.
\iftechreport{}\else A technical report containing the full formalization and proofs is available at~\cite{techreport}. \fi %

\section{Problem statement}\label{sec:problem}
\subsection{Transient execution attacks}
\newcommand\secretvalue[0]{\textcolor{leak_color}{\lstinline{SecretVal}}}
Modern processors rely on heavy optimizations to improve performance. They can
execute instructions out-of-order to avoid stalling the pipeline when the
operands of an instruction are not available. %
Additionally, they employ \emph{speculation} mechanisms to predict the
instruction stream. %
The execution of instructions resulting from a misprediction, called
\emph{transient execution}, is reverted at the architectural level, but effects
on the microarchitectural state (e.g.,\ the cache) are persistent.

Spectre attacks~\cite{DBLP:conf/sp/KocherHFGGHHLM019} exploit these speculation
mechanisms to force a victim to leak secrets during transient
execution. An attacker can mistrain predictors to force a victim into
transiently executing a sequence of instructions, called a Spectre gadget,
chosen to encode secrets in the microarchitectural state. Finally, the attacker
can use microarchitectural attacks to extract the secret.
To this day, many variants of Spectre attacks have been discovered, exploiting a
wide variety of speculation mechanisms~\cite{DBLP:conf/ccs/MaisuradzeR18,
  DBLP:conf/woot/KoruyehKSA18,spectrev4,SecurityAnalysisAMD2021,
  DBLP:conf/sp/KocherHFGGHHLM019,
  DBLP:conf/uss/CanellaB0LBOPEG19,DBLP:conf/uss/RagabBBG21}.

Transient execution may also arise from incorrect data being forwarded by
faulting instructions. %
For instance, on some processors, the result of unauthorized loads is
transiently forwarded to subsequent instructions before the load is rolled back.
This mechanism has first been exploited in Meltdown-style
attacks~\cite{DBLP:conf/uss/Lipp0G0HFHMKGYH18,DBLP:conf/uss/BulckMWGKPSWYS18} to exfiltrate secret data from
another security domain. %
It is generally accepted that Meltdown-style attacks should be mitigated in
hardware by preventing such forwarding from faulting loads. We consider
Meltdown-style attacks out of scope for this paper.

However, these faulting loads have also been exploited to \textit{inject} incorrect data
into the victim's transient execution, and, similarly to Spectre attacks, lead the victim to leak their secrets into the microarchitectural state.
In particular, these so-called load value injection (LVI)
attacks~\cite{DBLP:conf/sp/BulckM0LMGYSGP20} are still possible in the presence
of Meltdown mitigations zeroing out the results of faulting loads at the silicon
level (i.e.,\ LVI-NULL). %
LVI attacks are related to Spectre attacks that would exploit {\em value
  speculation} during loads.

We illustrate variants of Spectre and LVI attacks in
\cref{lst:transient_execution_examples}, where programs in
\cref{lst:spectre-pht,lst:spectre-btb,lst:spectre-stl} abuse different sources
of transient execution ($\mathghost$) to encode \secretvalue{} in the cache
using the \lstinline{leak} function in \cref{lst:leak_function}. After encoding, the
attacker can extract the secret from the cache using cache attacks. Note that
while we illustrate these attacks using a cache side-channel, transient
execution vulnerabilities are independent of the microarchitectural
side-channel they exploit, such as branch predictor
state~\cite{chowdhuryy2021leaking}, SIMD
units~\cite{DBLP:conf/esorics/0001SLMG19}, port
contention~\cite{DBLP:conf/ccs/BhattacharyyaSN19,DBLP:conf/ccs/FustosBY20},
micro-op cache~\cite{renSeeDeadMu2021}, etc. %
Consequently, the \lstinline{leak(x)} function can be replaced with any other
function that reveals information on the value of \lstinline{x} via a timing or
microarchitectural side-channel. %

\begin{listing*}[ht]
\centering
\begin{minipage}{.28\linewidth}
  \begin{lstlisting}[style=nonumbers]
    0 - 15: A[16]
ptr_s (16): <@{\color{leak_color}{SecretVal}}@>
17 - 16400: B[256 * 64]
\end{lstlisting}
\vspace{-.9em}
\subcaption[listing]{Memory}\label{lst:memory}

  \begin{lstlisting}[firstnumber=1]
void leak(x):
  idx $\leftarrow$ x * 64
  y $\leftarrow$ load B + idx $\skull$ <@\label{line:leak}@>
\end{lstlisting}
\vspace{-.9em}
\subcaption[lstlisting]{Encode \lstinline{x} into the cache.}\label{lst:leak_function}

\end{minipage} \hfill
\begin{minipage}{.28\linewidth}
  \begin{lstlisting}[firstnumber=4]
if (idx < size_A) $\mathghost$ <@\label{line:pht:transient_cause}@>
  x $\leftarrow$ load A + idx <@\label{line:pht:load}@>
  leak(x) <@\label{line:pht:leak}@>
\end{lstlisting}
\vspace{-.9em}
\subcaption[listing]{Spectre-PHT (v1)}\label{lst:spectre-pht}

  \begin{lstlisting}[firstnumber=7]
f $\leftarrow$ trusted_func
x $\leftarrow$ load ptr_s <@\label{line:btb:load}@>
jmp f(x) $\mathghost$ <@\label{line:btb:transient_cause}@>
\end{lstlisting}
\vspace{-.9em}
\subcaption[listing]{Spectre-BTB (v2)}\label{lst:spectre-btb}

\end{minipage} \hfill
\begin{minipage}{.32\linewidth}
  \begin{lstlisting}[firstnumber=10]
store ptr_s 0 <@\label{line:stl:sanitize}@>
x $\leftarrow$ load ptr_s $\mathghost$ <@\label{line:stl:transient_cause}@>
leak(x) <@\label{line:stl:leak}@>
\end{lstlisting}
\vspace{-.9em}
\subcaption[listing]{Spectre-STL (v4)}\label{lst:spectre-stl}

  \begin{lstlisting}[firstnumber=13]
idx $\leftarrow$ load trusted_idx $\mathghost$ <@\label{line:lvi:transient_cause}@>
x $\leftarrow$ load A + idx <@\label{line:lvi:load}@>
leak(x) <@\label{line:lvi:leak}@>
\end{lstlisting}
\vspace{-.9em}
\subcaption[listing]{LVI}\label{lst:lvi}

\end{minipage}
\addtocounter{listing}{-1} %
\addtocounter{lstlisting}{1} %
\captionof{listing}{Examples of code snippets vulnerable to transient execution
  attacks. The memory layout given in \cref{lst:memory} where \secretvalue{} is
  the only secret input and \lstinline{ptr_s = 16} is common to
  \cref{lst:spectre-pht,lst:spectre-btb,lst:spectre-stl,lst:lvi}.
  \(\protect\mathghost\) indicates instructions triggering transient
  executions and $\protect\skull$ indicates a
  leakage.}\label{lst:transient_execution_examples}
\end{listing*}

The \textbf{Spectre-PHT} (Pattern History Table) or Spectre-v1 variant~\cite{DBLP:conf/sp/KocherHFGGHHLM019} exploits the conditional branch predictor
to transiently execute the wrong side of a conditional branch. %
For instance, in \cref{lst:spectre-pht}, an attacker can first mistrain the
conditional branch predictor to take the branch and then call the piece of code with
\lstinline{idx = 16} to make the victim transiently execute the branch,
accessing \secretvalue{} at line~\ref{line:pht:load} and encoding it to the
microarchitectural state at line~\ref{line:pht:leak}.

The \textbf{Spectre-BTB} (Branch
Target Buffer) or Spectre-v2 variant~\cite{DBLP:conf/sp/KocherHFGGHHLM019} exploits indirect branch prediction to
transiently redirect the control flow to an attacker-chosen location. For example,
the program in \cref{lst:spectre-btb} calls a trusted function, which
performs secure computations using \secretvalue{}. An attacker can mistrain the
branch predictor such that, after line~\ref{line:btb:transient_cause}, the victim
transiently jumps to the \lstinline{leak} function instead of the trusted
function and leaks \secretvalue{}.
The \textbf{Spectre-RSB} (Return Stack Buffer) variant~\cite{DBLP:conf/woot/KoruyehKSA18, DBLP:conf/ccs/MaisuradzeR18} is similar to Spectre-BTB
but exploits target predictions for \lstinline{ret} instructions.

The \textbf{Spectre-STL} (Store-To-Load-forwarding) or Spectre-v4 variant~\cite{spectrev4} exploits the fact that load
instructions can speculatively bypass preceding stores. In
\cref{lst:spectre-stl}, the secret located at \lstinline{ptr_s} is
overwritten at line~\ref{line:stl:sanitize}, followed by a \lstinline{load} to
the same address, which should return \lstinline{0}. With Spectre-STL, the
\lstinline{load} may bypass the \lstinline{store} at line~\ref{line:stl:sanitize} and
transiently load \secretvalue{}, which would then be leaked to the microarchitectural state at
line~\ref{line:stl:leak}.

Finally, \textbf{LVI} (Load Value
Injection) attacks~\cite{DBLP:conf/sp/BulckM0LMGYSGP20} exploit a faulting \lstinline{load} to directly
inject incorrect data into the victim's execution. %
For instance, in \cref{lst:lvi}, an attacker can prepare the microarchitectural
state so that the value \lstinline{16} is forwarded to \lstinline{idx} by the
\lstinline{load} instruction at line~\ref{line:lvi:transient_cause},
hence accessing \secretvalue{} at line~\ref{line:lvi:load} and leaking it at line~\ref{line:lvi:leak}.

\subsection{Secure speculation approaches}

Since transient execution attacks were discovered, several studies have focused
on adapting program semantics, security policies, and verification tools
to take into account the \emph{speculative semantics} of programs and place
extra software-level protections against Spectre
attacks, e.g.,~\cite{DBLP:conf/csfw/CheangRSS19, DBLP:conf/iccad/PescostaWZ21,
  DBLP:conf/pldi/CauligiDGTSRB20, DBLP:conf/pldi/Wu019,
  DBLP:conf/sp/BartheCGKLOPRS21, DBLP:conf/sp/GuarnieriKMRS20,
  DBLP:conf/uss/OleksenkoTSF20, DBLP:journals/pacmpl/VassenaDGCKJTS21,
  DBLP:journals/tosem/WangCBMR20, guoSpecuSymSpeculativeSymbolic2020,
  qiSpecTaintSpeculativeTaint2021, wangOo7LowoverheadDefense2020,
  DBLP:conf/ndss/DanielBR21}.
However, %
reasoning about transient execution attacks at the software level only can be
burdensome and fragile. Firstly, it necessitates knowledge of microarchitectural
details that are often not publicly available. Secondly, it requires changing
security policies and applying software patches every time new speculation
mechanisms are introduced (e.g.,\ the predictive store forwarding feature newly
introduced in AMD Zen3 processors~\cite{SecurityAnalysisAMD2021}). %
Finally, software countermeasures targeting specific transient execution attacks
can still leave the door open to other attacks~\cite{DBLP:conf/ndss/DanielBR21}.

Instead, we argue that, for a given policy \(P\), enforcement mechanisms at the
software level should only consider an architectural (non-speculative)
semantics, while the hardware should guarantee that transient execution does not
introduce additional vulnerabilities. We call this approach \emph{hardware-based
  secure speculation for \(P\)}.

\myparagraph{Hardware-based secure speculation for sandboxing} %
A sandboxing policy isolates a potentially malicious application by restricting
the memory range it can access. A program is said to be \emph{sandboxed}
if it never accesses memory outside its authorized address range.
Sandboxed programs are vulnerable to Spectre attacks, as out-of-bounds memory locations
may still be accessed transiently and have their contents leaked to the microarchitectural state.
As an example, the program in
\cref{lst:spectre-pht} is sandboxed but can still access and leak out-of-bounds
data when the condition is misspeculated.

Some hardware taint-tracking mechanisms %
~\cite{DBLP:conf/micro/YuYK0TF19,DBLP:conf/micro/WeisseNLWK19,
  DBLP:conf/IEEEpact/BarberBZZT19} have been shown to enable secure speculation
for sandboxing~\cite{DBLP:conf/sp/GuarnieriKRV21}. %
For instance, Speculative Taint Tracking (STT)~\cite{DBLP:conf/micro/YuYK0TF19}
taints speculatively accessed data and prevents tainted values from being
forwarded to instructions that may form a covert channel.
In \cref{lst:spectre-pht}, STT taints the variable \lstinline{x}
at line~\ref{line:pht:load} until the condition at
line~\ref{line:pht:transient_cause} is resolved. As \lstinline{x} is
tainted, its value is not forwarded to the insecure \lstinline{load} in the
\lstinline{leak} function. %

Unfortunately, hardware-based secure speculation for sandboxing {\em only protects
speculatively accessed data}, meaning that secret data loaded in
registers during sequential execution may still be transiently leaked.
For instance, STT does not protect the program in \cref{lst:spectre-btb} against
Spectre-BTB. At line~\ref{line:pht:load},
a secret is loaded during sequential execution.
As a result, \lstinline{x} is not
tainted by STT, and its value can still be forwarded to an insecure
instruction if the \lstinline{jmp} is
misspeculated.
Hardware-based secure speculation for sandboxing is therefore insufficient to
guarantee security for programs that compute on secrets, such as cryptographic
primitives. To protect these programs, we need to enable hardware-based secure
speculation for the constant-time policy.

\myparagraph{Hardware-based secure speculation for constant-time} %
A constant-time policy specifies that program secrets should not leak through
timing or microarchitectural side channels. %
Before the advent of transient execution attacks, the constant-time policy was
enforced with a coding discipline ensuring that the \emph{control-flow of the
  program}, \emph{addresses of memory accesses}, and \emph{operands of
  variable-time instructions} do not depend on secret data. %
This coding discipline is the de facto standard for writing cryptographic code;
it has been adopted in many cryptographic
libraries~\cite{DBLP:conf/latincrypt/BernsteinLS12,DBLP:conf/ccs/ZinzindohoueBPB17,BearSSLConstantTimeCrypto,
  DBLP:conf/ccs/AlmeidaBBBGLOPS17} and is supported by many
tools, e.g.,~\cite{DBLP:conf/acsac/WichelmannMES18,DBLP:conf/icse/BaoWLLW21,DBLP:conf/uss/AlmeidaBBDE16,
  DBLP:conf/pldi/DoychevK17,
  langleyImperialVioletCheckingThat2010,DBLP:conf/sp/DanielBR20,
  DBLP:conf/secdev/CauligiSBJHJS17,DBLP:conf/sp/BrotzmanLZTK19,DBLP:conf/uss/DoychevFKMR13,DBLP:conf/icst/HeEC20,DBLP:journals/corr/abs-2208-14942}.

A standard definition for constant-time programs (i.e.,\ programs
adhering to the constant-time policy), and the one we use in this paper, is the
following:
\begin{definition}[Constant-time program]\label{def:constant-time}
  A program is constant-time if the observation trace that it produces during
  \emph{sequential execution} is independent of secret data (where the
  observation trace records the control flow and memory accesses).\footnote{We give a
    formal definition in \cref{sec:theorems}. For simplicity, we do not
    include variable-time instructions in our security proofs but discuss how to handle them with \name{}.}
\end{definition}
\noindent%
Unfortunately, adhering to this definition
is insufficient to guarantee security on modern
processors vulnerable to transient execution attacks like Spectre or LVI. Indeed,
all programs in \cref{lst:transient_execution_examples} are
constant-time according to \cref{def:constant-time}, but they are vulnerable to transient execution attacks.

Hardware-based countermeasures guaranteeing secure speculation for sandboxing
do not guarantee secure speculation for the constant-time policy. Therefore, to
enforce the constant-time policy on speculative processors, it is still
necessary to insert specific protections (typically \lstinline{fence}
instructions or retpolines~\cite{turnerRetpolineSoftwareConstruct}) to protect
against transient execution attacks. Software developers still have to reason
about speculation when they want to enforce the constant-time policy. In this
paper, we address the problem of \emph{providing hardware-based provably secure
  speculation for the constant-time policy}.

\section{Informal overview of \name{}}\label{sec:overview}
In this section, we motivate our design choices, make explicit what guarantees have to be
enforced by software, and sketch the requirements
the hardware must enforce. Finally, we illustrate how \name{}
protects the programs in \cref{lst:transient_execution_examples}.

\subsection{Design choices}
\name{} relies on a hardware-software co-design where developers annotate their
secret data, and the hardware guarantees that no information about these secrets
can leak during transient execution. %
The design of \name{} is motivated by two main objectives.
The first objective is to support existing constant-time code with minimal software changes. To
  this end, we base our annotation and declassification mechanism on
  \context{}~\cite{DBLP:conf/ndss/0001LCSKG20} in which developers
  partition the memory into public and secret regions and can declassify secrets
  by writing them to public memory. %
  The second objective is to support secure code while maintaining full performance benefits of
  speculative and out-of-order execution. Specifically, \name{} delays
  speculative execution only when a secret is about to be leaked; hence \emph{in
    constant-time programs} (which do not leak secrets) \emph{\name{} only blocks mispredicted instructions}.

\myparagraph{Software contracts}
Software developers must comply with three contracts:
\begin{restatable}{contract}{hypannotations}\label{hyp:annotations}
  Secret memory locations are labeled.
\end{restatable}
\noindent%
For instance, in \cref{lst:transient_execution_examples},
address \lstinline{16} is labeled as \emph{secret} (or \emph{high}), denoted
\(\topsec\), whereas other addresses are labeled as \emph{public} (or
\emph{low}), denoted \(\botsec\).
\begin{restatable}{contract}{hypconstanttime}\label{hyp:ct-program}
  The program is constant-time.
\end{restatable}
\begin{restatable}{contract}{hypdeclassification}\label{hyp:declassification}
  Secret values written to public memory are \emph{intentionally declassified}
  by the program.
\end{restatable}
\noindent%
\Cref{hyp:declassification} allows, for instance, cryptographic code to declassify
ciphertexts. However, software developers must make sure to not
unintentionally declassify secrets by writing them to public memory.

We prove in \cref{sec:theorems} that if programs comply
with these three contracts, then execution on \name{} does not leak secrets
through timing and microarchitectural side channels.

\myparagraph{Hardware requirements}
On the hardware side, \name{} must realize the following:

\begin{requirement}\label{req:tainting}
  During the execution of a program, the processor tracks \emph{security levels}.
  Concretely, it labels values loaded from memory with their
  corresponding security level (\(\botsec\) or \(\topsec\)) and soundly propagates these
  security levels during computations.
\end{requirement}

\begin{requirement}\label{req:stall}
  The processor prevents values with security level \(\topsec\) to be
  leaked during speculative execution. Hardware developers identify
  \emph{insecure instructions} that may leak data through
  \begin{enumerate*}[label=(\arabic*)]
    \item changing the microarchitectural state,
    \item influencing the program counter, or
    \item exhibiting operand-dependent timing.
  \end{enumerate*}
  The processor prevents these instructions from being speculatively
  executed with secret operands.
\end{requirement}

\begin{requirement}\label{req:prediction}
  Predictions do not leak secret data, in particular:
\begin{enumerate*}[label=(\arabic*)]
    \item predictor states are only updated using public values, and
    \item speculations are rolled back (even the \emph{correct} ones) when their
    outcome depends on secrets (otherwise, it would leak whether the public
    prediction is equal to the secret value).
  \end{enumerate*}
\end{requirement}

\subsection{\name{} through illustrative examples}
Consider the program in \cref{lst:spectre-pht}, assuming that %
\lstinline{idx = 16} and the condition is misspeculated to \(true\). When
executing the \lstinline{load} instruction at line~\ref{line:pht:load}, \name{}
tags the register \lstinline{x} with the security level corresponding to address
\lstinline{16}, denoted
\(\text{\lstinline{x}} \mapsto \slevel{\text{\secretvalue{}}}{\topsec}\) (by
\cref{req:tainting}).\footnote{A more conservative design choice, adopted by
  \context{}~\cite{DBLP:conf/ndss/0001LCSKG20}, would be to prevent such
  speculative loads from accessing secret memory locations and to prevent the
  execution of line~\ref{line:pht:load}. However, we formally show that secure
  speculation is possilbe with this more liberal design choice.} %
Then, when the \lstinline{leak} function is executed, the \lstinline{load}
instruction (line~\ref{line:leak}, \cref{lst:leak_function}) is blocked because
it would leak a secret-labeled value during speculative execution (by
\cref{req:stall}). %
Conversely, if register \lstinline{x} contains a public-labeled value, i.e.,\
\(\text{\lstinline{x}} \mapsto \slevel{v}{\botsec}\), the \lstinline{load}
instruction is not blocked. \name{} only blocks speculative execution in a few
restricted cases, namely when secret data is about to be leaked.

Notice that, contrary to sandboxing-based approaches, \name{} also protects
secrets loaded in architectural registers from being
transiently leaked. For example, in \cref{lst:spectre-btb}, when the secret is
loaded at line~\ref{line:btb:load}, \lstinline{x} is labeled with \(\topsec\),
which prevents the secret from being transiently leaked later (by
\cref{req:stall}) if the \lstinline{jmp} instruction at
line~\ref{line:btb:transient_cause} transiently jumps to the \lstinline{leak}
function.

So far, we have seen examples of \name{} applied to Spectre-PHT and Spectre-BTB
(Spectre-RSB is similar to the latter). %
The protection generalizes to any other source of speculation, such as load value
prediction (which encompasses LVI and Spectre-STL). Take, for instance, the program
in \cref{lst:lvi}. Here, the source of speculation is the \lstinline{load}
instruction, which transiently forwards an incorrect value at
line~\ref{line:lvi:transient_cause}. Until the \lstinline{load} is resolved,
\name{} considers the following instructions speculative. Consequently,
(by \cref{req:stall}) it does not forward the secrets to the \lstinline{load} in
the \lstinline{leak} function (\cref{lst:leak_function}, line~\ref{line:leak})
and prevents the LVI attack.

Finally, \name{} also guarantees (by \cref{req:prediction}) that predicted
values do not depend on secrets. In particular, secret values
cannot be speculatively forwarded to other instructions. For example, in
\cref{lst:spectre-stl}, the \lstinline{load} instruction at
line~\ref{line:stl:transient_cause} cannot speculatively load
\secretvalue{}, %
because the corresponding address is labeled as secret.
Notice that \name{} still allows forwarding public values.

\section{Formalization and theorems}\label{sec:formalization}
This section presents one of the core contributions of this paper, the
formalization of \name{}. %
The \name{} semantics builds on prior semantics~\cite{DBLP:conf/sp/GuarnieriKRV21},
extended to consider a broader spectrum of prediction mechanisms. Moreover, it
 generalizes the standard constant-time leakage model; in addition
to disclosing control-flow and memory accesses, %
our semantics also discloses \emph{all public-labeled data}. %
Concretely, all public-labeled data can influence predictions and
is observable by an attacker.

\subsection{ISA language}\label{sec:isa_language}
The ISA is modeled using a small assembly language called
\muasm{}~\cite{DBLP:conf/sp/GuarnieriKMRS20}, described in \cref{fig:isa_instr}. %
\(\Vals\) is a set of values, including memory addresses and program
locations, and we let $\val{v}$ and $\val{l}$ range over \(\Vals\).  \pc{} denotes
the program counter register, and $\register{r}, \register{x}$ denote registers
with \(\register{x} \neq \pc\). A program \(\prog{}\) is a partial mapping from
locations to instructions. We use \(\prog{\val{l}}\) to denote the instruction at
location \(\val{l}\).
\begin{figure}[ht]
  \centering
\begin{equation*}
\begin{array}{r@{~}r@{~}c@{~}l}
  \text{(Expressions)}  & e & ::=& \val{v} \;|\; \register{r} \;|\; e_1 \otimes e_2 \\ %
  \text{(Instructions)} & inst & :: =  & \mov{x}{e}\;|\; \jmp{e} \;|\; \beqz{e}{\val{l}} \;|\;\\
                        &   &          & \load{x}{e} \;|\; \store{e_{a}}{e_{v}} \\
\end{array}
\end{equation*}
\caption{Syntax of \muasm{} programs where \(\otimes\) denotes a binary operation.}\label{fig:isa_instr}
\end{figure}

\myparagraph{Security levels} %
We assume a lattice $\Lattice$ with two security levels: public (low, \(\botsec\))
and secret (high, \(\topsec\)). We let $\level{s}, \level{s'}, \level{s_0}, \ldots$
range over security levels from $\Lattice$. $\sqcup$ denotes the least upper
bound operation on the lattice, with
\(\botsec \sqcup \topsec = \topsec\). Additionally, we let
\(\slevel{\val{v}}{s}\) denote a value \(\val{v}\) with security level
\(\level{s}\), ranging over the set \(\SVals = \Vals \times \Lattice\). For
simplicity, we restrict our description to this 2-level security lattice, but this work
generalizes to arbitrary security lattices.

\myparagraph{Location of secret data} As stated in
\cref{hyp:annotations}, programmers annotate secret data in the code. In our
formal semantics, we assume that they do so by specifying a \emph{security
  memory partition} $\memsec$,
which maps memory addresses %
to security levels in $\Lattice$. We assume this mapping to be fixed, it cannot
change over time. Hence, for the sake of readability, we do not explicitly
include it in the configurations of the semantics rules. %

\subsection{Hardware configurations}\label{sec:hardware_configurations}
Hardware configurations are of the form $\conf{\mem, \reg, \buf, \micro}$,
where %
\(\mem\) is the memory, which maps addresses to values in \(\Vals\); \(\reg\) is the
register map, which maps registers to pairs of a value and a security level in
\(\SVals\); $\buf$ is the reorder buffer, which is a sequence of (possibly
transient) instructions; and $\micro$ is the \emph{microarchitectural context}.

\myparagraph{Microarchitectural context} %
The microarchitectural context \(\micro\) can be thought of as the part of the
microarchitectural state that the attacker controls.
It is an abstract component that models both the \emph{observations} that the
attacker can make and the {\em influence} the attacker has on predictions and
scheduling.
Formally, it is a stateful deterministic component offering three functions: %
\begin{itemize}
  \item \(\update\), called by the semantics whenever microarchitectural state possibly leaks to the attacker;
  \item \(\predict\), giving the attacker control over predictions, for instance,
        to predict jump targets or load values;
  \item \(\nextdir\), giving the attacker control over scheduling decisions that determine the next
        processor step to execute.
\end{itemize}
Hence, our definition of the semantics must make sure that for each computation step,
\(\micro\) is updated with all information that
could leak from that computation. In particular, updates should include the
program counter and the addresses of memory accesses (which \emph{directly}
influence the instruction and data cache), but also operands of variable-time
instructions (which do not directly influence the microarchitectural state, but might
do so \emph{indirectly} via timing
variations~\cite{DBLP:conf/asplos/BehniaSPYZZUTR021}).

Importantly, to satisfy \cref{req:stall,req:prediction}, our evaluation rules
must satisfy the following invariant: if the program is constant-time (as stated
in \cref{hyp:ct-program}), secret data should never leak to \(\micro\). %
In particular, the \(\update\) function should never be given secret data as
input during speculative execution. It follows from our security theorems that this is
indeed the case.

\myparagraph{Reorder buffer} In out-of-order processors, program instructions
are fetched in order and placed in a \emph{reorder buffer} (ROB) where they can
be executed out-of-order. Contrary to ISA instructions, ROB instructions,
defined in \cref{fig:rob_instr}, keep track of the security levels of values. In
addition, \jmp{}{} and \beqz{}{} are directly translated to \pc{}
assignments when they are fetched and thus are not part of ROB instructions.
This also implies that, contrary to ISA instructions, assignments in the ROB can
target \(\pc\). Finally, instructions in the ROB with predicted values are
tagged with the address $\val{l}$ of the instruction that the prediction
resulted from; otherwise, they are tagged with \(\varepsilon\).

\begin{figure}[ht]
  \centering
\begin{equation*}
\begin{array}{r@{~}r@{~}c@{~}l}
\text{(Tags)}  & T & ::=& \val{l} \;|\; \varepsilon\\
\text{(ROB  exp)} & e & ::=& \slevel{\val{v}}{s} \;|\; \register{r} \;|\; e_1 \otimes e_2 \\ %
  \text{(ROB inst)} & i & :: =  & \mov{r}{e} @ T\;|\; \load{x}{e} @ T\;|\;\\
                    &   &       & \store{e_{a}}{e_{v}} @ T
\end{array}
\end{equation*}
  \caption{ROB instructions.}\label{fig:rob_instr}
\end{figure}

During the execution, changes that occur in the ROB are applied to the registers
using the function \(\apl\)~\cite{DBLP:conf/sp/GuarnieriKRV21}. When the value of
an assignment in the reorder buffer is not resolved yet, the corresponding
register is mapped to a special symbol \(\botval\), meaning that it is
undefined. Thus, the function \(\apl\) generates a new register map where the
value of some registers is undefined.

\begin{definition}[Apply function $\apl$]\label{def:apl}
  For all register maps \(\reg\) and reorder buffers \(\buf\):
\begin{equation*}
  \begin{array}{r@{\ }l@{\ }l@{}l@{}l}
    \apl(\varepsilon, \reg) &=& \reg \\
    \apl(\mov{r}{\slevel{v}{s}} @T\cdot \buf, \reg) &=& \apl(\buf, \reg\change{\register{r}}{\slevel{v}{s}}) \\
    \apl(\mov{r}{e} @T\cdot \buf, \reg) &=& \apl(\buf, \reg\change{\register{r}}{\botval}) {\sf ~if~} e \not\in \SVals \\
    \apl(\load{x}{e} @T \cdot \buf, \reg)  &=&
    \apl(\buf, \reg\change{\register{x}}{\botval}) \\
    \apl(\store{e_{a}}{e_{v}} @T \cdot \buf, \reg) &=& \apl(\buf, \reg)
  \end{array}
\end{equation*}
\end{definition}

\subsection{Sanitization of secret values}\label{sec:sanitize_secrets}
An important feature of \name{} is the ability to \emph{sanitize} secret values
during speculative execution and predictions.
To achieve sanitization, we define a low-projection for values denoted
\(\sproj{\slevel{\val{v}}{s}}\). It discloses public values but replaces
secret values with \(\botval\).
\begin{definition}[Low-projection]\label{def:sproj}
\begin{mathpar}
  \sproj{\slevel{\val{v}}{\botsec}} = \slevel{\val{v}}{\botsec}\and
  \sproj{\slevel{\val{v}}{\topsec}} = \botval\and
  \sproj{\botval} = \botval
\end{mathpar}
\end{definition}
We let \(\sproj{\reg}\) be the point-wise extension of
\(\sproj{\cdot}\) to register maps. Hence, a sanitized register map
\(\sproj{\reg}\) maps registers to either their value when the associated
security level is public or to \(\botval\) when the value is unresolved, or when it
is secret.

\begin{definition}[Low memory projection]\label{def:smemproj}
  We define the low-projection of a memory \(\smemproj{\mem}\) such that for all
  addresses \(\val{a}\): %
\begin{equation*}
  \smemproj{\mem}(\val{a}) =
  \begin{cases}
    \mem(\val{a}) & \textsf{\normalfont{if }} \memsec(\val{a}) = \botsec \\
    \botval & \textsf{\normalfont{otherwise}} \\
  \end{cases}
\end{equation*}
\end{definition}

The low-projection of a reorder buffer \(\buf\), denoted \(\sbufproj{\buf}\),
discloses all low values in \(\buf\). Values with security level \(\topsec\) are
replaced by \(\botval\). On the other hand, values with security level
\(\botsec\), unresolved expressions, and tags are not replaced. Details are
deferred to
\iftechreport \cref{app:full_rules}. \else the technical report~\cite{techreport}. \fi

\myparagraph{Sanitizing secrets in speculative execution} %
We define a function \(\aplsan\) that  selectively sanitizes
the register map returned by \(\apl\). In sequential execution (i.e.,\ when no instruction in \(buf\)
results from a prediction), it directly returns the result of \(\apl\). During
speculative execution (i.e.,\ where at least one instruction in \(\buf\) results from a
prediction), it returns the low-projection of \(\apl\), in which secrets are replaced with \(\botval\).
\begin{definition}[Apply function \(\aplsan(\cdot,\cdot)\)]\label{def:aplsan}
  \begin{equation*}
    \aplsan(\buf, \reg) = \left\{
      \begin{array}{ll}
        \apl(\buf, \reg) & {\sf if~} \forall inst @T \in \buf.\ T = \varepsilon \\
        \sproj{\apl(\buf, \reg)} & {\sf if~} \exists inst @T \in \buf.\ T \neq \varepsilon \\
      \end{array}\right.
  \end{equation*}
\end{definition}
\noindent
Concretely, the function \(\aplsan\) is a crucial part of \cref{req:stall}; it acts as a
filter that prevents forwarding secret data to insecure instructions (which
\(\update\) the microarchitectural context) during speculative execution.

\subsection{Evaluation rule}\label{sec:eval_expr}
\myparagraph{Expression evaluation} %
The evaluation of an expression \(e\) with a register map \(\reg\), denoted
\(\semexpr{e}{\reg}\), is a partial function from expressions to labeled values
in \(\SVals\). It is undefined if one of the sub-expressions is undefined.  Importantly, the evaluation of a binary
operation propagates the security level of its operands in a conservative way
(cf. \cref{req:tainting});   if at least one of the operands has security level
\(\topsec\), then the resulting security level is \(\topsec\). Details are available in \iftechreport \cref{app:full_rules}. \else the technical report~\cite{techreport}. \fi

\myparagraph{Instruction evaluation} %
The hardware semantics is given by a main relation
$\smicro{}{}{c_{1}}{c_{2}}{}$ and an auxiliary relation
\(\smicro{\directive}{}{c_{1}'}{c_{2}'}{}\) where $c_{1}$ and
$c_{2}$ are hardware configurations and \(\directive\) is a directive. %
Leaks are \leaked{highlighted} in the rules and \(\_\) is used to denote that
there exists an expression, but this expression is not important in the context.

The directive determines which processor step to execute: the \fetch{}
directive fetches an instruction and places it at the end of the ROB,
\execute{i} executes the \(i\)\textsuperscript{th} instruction in the ROB,
\retire{} removes the oldest instruction from the ROB and commits its changes to
the register map and memory.

The \textsc{step} rule selects the next directive to apply from the
microarchitectural context using the function $\nextdir(\micro)$ and updates the
hardware configuration accordingly:
\begin{mathpar}
\stepRule{}
\end{mathpar}
The rule updates the microarchitectural context using public-labeled values from
the memory \(\leaked{\smemproj{\mem}}\), the register map
\(\leaked{\sproj{\reg}}\), and the reorder buffer \(\leaked{\sbufproj{\buf}}\).
This means that \emph{any public-labeled value can influence
  subsequent predictions}. In \cref{sec:discussion}, we show how this
abstraction captures existing prediction mechanisms. %
It also means that any public-labeled value is leaked to the attacker,
which effectively leaks \emph{more} than the standard constant-time leakage
model corresponding to
\cref{def:constant-time}. %

We present some evaluation rules for each directive, the full
set of rules is available in \iftechreport \cref{app:full_rules}. \else our
technical report~\cite{techreport}. \fi %

\myparagraph{Fetch directive} %
The instructions are fetched in program order and placed in the ROB. We provide
here the rule \textsc{fetch-predict-branch-jmp}, which applies when the
instruction to fetch is a branch or jump.
\begin{mathpar}
  \centering
  \fetchPredictBranchJump{}
\end{mathpar}

The rule predicts the next program location \(\val{l'}\) and updates \pc{} in the ROB
accordingly.\footnote{Note that what we call prediction here also covers store-to-load forwarding, load value prediction, and load value injection, as
  discussed in \cref{sec:discussion}.} %
Notice that the next location \(\val{l'}\) is added to the ROB with
security level \(\botsec\), hence it will be leaked to the microarchitectural
context in the next \textsc{step} rule. In the same way, the current location
\(\val{l}\), added as a speculation tag in the ROB, is also leaked. %

\myparagraph{Execute assignments}
The rule \textsc{execute-assign} evaluates an expression \(e\) and updates the
value of a register \(\register{r}\) in the ROB accordingly.
The rule \textsc{execute-assign} assumes that the evaluation of the expression
\(e\) does not leak information about its operands---in particular, the execution time of the instruction is independent of the value of its operands.
In this case, the
instruction can securely execute using secret data. Therefore, the rule uses the
non-sanitized register map \(\apl(\buf, \reg)\) to evaluate \(e\) (cf.\ the
boxed hypothesis). %
For \emph{variable-time (insecure) instructions}~\cite{DBLP:conf/icisc/GrossschadlOPT09,
  DBLP:conf/sp/AndryscoKMJLS15, DBLP:conf/sp/CoppensVBS09}, %
 we can simply
replace the function \(\apl(\buf, \reg)\) in the box with the function
\(\aplsan(\buf, \reg)\) to prevent the instruction from executing using secret data
during speculative execution (cf.\ \cref{req:stall}). %

\begin{mathpar}
  \centering
  \executeAssign{}
\end{mathpar}

\myparagraph{Execute loads} %
The rule \textsc{execute-load-predict} predicts the value of a \load{}{}
instruction and updates the ROB with this predicted value. Note that the
predicted value \(\val{v}\) is added to the ROB with security level \(\botsec\)
and is hence observable. This is consistent with the fact that predictions can
only depend on public data (cf.\ \cref{req:prediction}).
\begin{mathpar}
  \centering
  \executeLoadPredict{}
\end{mathpar}

The rule \textsc{execute-load-commit} commits the result of a predicted
\(\load{}{}\) instruction to the ROB when the prediction is correct. It also
leaks the address of the load to the microarchitectural context. %
The rule uses the \emph{sanitized} value \(\val{a}\) of the address, which
effectively prevents the rule to be applied during speculative execution when
the address is secret (\cref{req:stall}).
Additionally, notice that the rule can only be applied if the address
\(\val{a}\) corresponds to public memory (i.e.,\ \(\memsec(\val{a}) = \botsec\)).
If the address \(\val{a}\) maps to secret memory (meaning that \(\mem(\val{a})\)
is secret), a rollback is performed to prevent leaking whether the
secret value \(\mem(\val{a})\) is equal to the predicted value \(\val{v}\), as
described in \cref{req:prediction} (a detailed example is provided in
\cref{sec:example}). %
\begin{mathpar}
  \centering
  \executeLoadCommit{}
\end{mathpar}

The complementary rule \textsc{execute-load-rollback} is applied when the
prediction is incorrect or when \(\mem(\val{a})\) is secret. It commits the
correct value to the ROB and drops younger instructions from \(\buf'\) (excluding the corresponding \(\pc\) update).

\myparagraph{Retire directives} %
Rules \textsc{retire-store-low} and \textsc{retire-store-high} retire a store
instruction on top of the ROB and update the memory accordingly. They also leak
the address of the store to the microarchitectural context, following the
constant-time leakage model. The rule \textsc{retire-store-low} is evaluated if the
address of the store corresponds to public memory (i.e.,\
\(\memsec(\val{a}) = \botsec\)). In this case, regardless of its original security level, the
value \(\val{v}\) becomes visible to the attacker: it is \emph{declassified}. As
stated in \cref{hyp:declassification}, it is the responsibility of the developer to make
sure that such declassification is intentional. In \cref{sec:example}, we
illustrate declassification in \name{} with an example. The declassified value
is recorded above the evaluation relation, denoted
\(\smicro{}{}{}{}{\declassify{\val{v}}}\). It does not affect the semantics but
will be used in the theorems of \cref{sec:theorems}. For secret store locations, the rule
\textsc{retire-store-high} is applied, which produces an empty declassification
trace.

\begin{mathpar}
  \retireStoreLow{}
  \and
  \retireStoreHigh{}
\end{mathpar}

\subsection{Theorems}\label{sec:theorems}
We first define a security
theorem for \name{} that can be applied to constant-time programs without declassification,
and then extend it to capture declassification. The full proofs are
available in \iftechreport \cref{app:proofs}. \else our technical report~\cite{techreport}. \fi %

An architectural configuration \(\aconf = \conf{\mem, \reg}\) is the subset of a hardware
configuration, consisting of
the memory and the register map. The (sequential) architectural semantics is given as
a relation \(\sarch{}{\declassified}{\aconf}{\aconf'}{\obs}\), which evaluates
an architectural configuration \(\aconf\) to another architectural configuration
\(\aconf'\). %
It produces a sequence of observations \(\leaked{\obs}\) that contains the
control flow (changes to the program counter) and the addresses of memory accesses. It also produces a declassification trace
\(\declassified\), which is the sequence of all values written to
public memory. Evaluation rules are otherwise standard and are provided
in \iftechreport \cref{app:arch}. \else the technical report~\cite{techreport}. \fi

Architectural configurations are said to be \emph{low-equivalent}, written
\(\sproj{\aconf} = \sproj{\aconf'}\), if they are identical in the low-projections of
their register maps and memories.

We let \(\smicro{}{n}{\mconf_{0}}{\mconf_{n}}{\declassified}\) denote an
\(n\)-step execution from a hardware configuration \(\mconf_{0}\) to a
configuration \(\mconf_{n}\), which produces a \emph{declassification trace}
\(\declassified\) \iftechreport{} (details in \cref{app:full_rules}). \else.\fi{} When \(\declassified\) is not needed in the
context, it is omitted. Similarly, we let
\(\sarch{n}{\declassified}{\aconf_{0}}{\aconf_{n}}{\leaked{\obs}}\) denote an
\(n\)-step execution in the architectural semantics.

\myparagraph{Security for constant-time programs}
\name{} guarantees that if a program is constant-time (\cref{hyp:ct-program}) and does not declassify
secret data (\cref{hyp:annotations,hyp:declassification}), then it does not
leak secret data when running on \name{}. %

The following definition formalizes the constant-time and no-declassification
policy that we expect to be enforced on the software side.
In that respect, it
establishes a \emph{hardware-software security
  contract}~\cite{DBLP:conf/sp/GuarnieriKRV21}.
\begin{definition}[Constant-time program]\label{def:ct}
  A program %
  is constant-time if for any initial architectural configurations
  \(\aconf_{0}\) and \(\aconf'_{0}\) such that \(\sproj{\aconf_{0}} = \sproj{\aconf_{0}'}\), and number of steps \(n\):
  \begin{equation*}
    \sarch{n}{\declassified}{\aconf_{0}}{\aconf_{n}}{\leaked{\obs}} \implies
    \sarch{n}{\declassify{\declassified'}}{\aconf_{0}'}{\aconf_{n}'}{\leaked{\obs'}} ~\wedge~ \leaked{\obs} = \leaked{\obs'}
  \end{equation*}
  Additionally, if \(\declassified = \declassify{\declassified'}\) for all possible \(\aconf_{0}\) and \(\aconf'_{0}\), we say that
  the program \emph{does not declassify secret data}.
\end{definition}

The goal of the attacker is to distinguish two low-equivalent initial configurations $\mconf$ and $\mconf'$, by providing a {\em strategy}, i.e., a concrete initial microarchitectural
context $\micro_{0}$ and implementations for \(\predict\), \(\update\), and \(\nextdir\)
(designed by the attacker to distinguish the configurations).
Under any such {\em deterministic} strategy, the attacker should have the same observations when running
in $\mconf$ and $\mconf'$.\footnote{Note that nondeterministic strategies would produce different observations due to nondeterminism,
not due to differences in secrets being leaked.}

\begin{restatable}{hypothesis}{hypdeterministic}\label{hyp:deterministic}
The functions \(\predict\), \(\update\), and \(\nextdir\) are deterministic.
\end{restatable}
\noindent%
Under this hypothesis, the hardware semantics is deterministic. %

The following theorem establishes end-to-end security for constant-time programs
without declassification, running on \name{}.
\begin{restatable}[Security for constant-time
  programs.]{theorem}{securityzero}\label{thm:security0}
  For any constant-time program that does not declassify secret
  data, microarchitectural state \(\micro_{0}\), initial configurations
  \(\mconf_{0} = \conf{\mem_{0}, \reg_{0}, \varepsilon, \micro_{0}}\) and
  \(\mconf_{0}' = \conf{\mem_{0}', \reg_{0}', \varepsilon, \micro_{0}}\) such
  that \(\sproj{\conf{\mem_{0}, \reg_{0}}} = \sproj{\conf{\mem_{0}', \reg_{0}'}}\) and number of steps \(n\),
  \begin{gather*}
    \smicro{}{n}{\mconf_{0}}{\mconf_{n}}{} \implies
    \smicro{}{n}{\mconf_{0}'}{\mconf_{n}'}{} ~\wedge~ %
    \micro_{n} = \micro_{n}'
  \end{gather*}
  where \(\micro_{n}\) and \(\micro_{n}'\) are the microarchitectural contexts in configurations \(\mconf_{n}\) and \(\mconf_{n}'\), respectively. %
\end{restatable}

\myparagraph{Security with declassification}
It is common for cryptographic programs to declassify ciphertexts after an
encryption primitive. As we show in \iftechreport \cref{app:declassification}\else the technical report~\cite{techreport}\fi, classic
definitions of
declassification~\cite{DBLP:conf/csfw/BartheDR04,DBLP:conf/isss2/SabelfeldM03,DBLP:conf/pldi/BanerjeeNR07,DBLP:conf/sp/AskarovS07}
allow a program to declassify more information than expected in the context of cryptographic
code.
Indeed, with
such definitions, declassifying \lstinline{f(m)} implicitly declassifies
\lstinline{m} when \lstinline{f} is an injective function (e.g.,\ a cryptographic
permutation). %
In contrast, we propose a novel definition of security \emph{up to} declassification
that captures the following intuition: %
if a program only declassifies ciphertexts, then plaintexts and keys remain
indistinguishable to an attacker (because they are {\em cryptographically}
indistinguishable) and should not be leaked.

As described above, the declassification trace of an execution
\(\smicro{}{n}{c}{c'}{\declassified}\) is the sequence of all values stored to
low (public) memory by the rule \textsc{retire-store-low}.
To express security up to declassification, we introduce a notion of
\emph{patched execution}, denoted
\(\smicropatched{}{}{c}{c'}{\declassified}{\declassify{\declassified'}}\), which
replaces values stored to low-memory by values from a declassification trace
\(\declassified\) (usually obtained by a low-equivalent run in the standard
semantics~\(\smicro{}{}{}{}{}\)). The patched execution ensures
that the low-memories of two low-equivalent executions remain equal, achieved by
patching the second execution with the declassification trace of the first execution.
For instance, a ciphertext that is declassified in the standard semantics
can be used to patch another (low-equivalent) execution, to obtain two
executions with the same declassified ciphertext (and to make sure that they leak the
same values).
Concretely, the patched execution only differs from the standard execution by
the rule \textsc{retire-store-low}, which is replaced by the following:
\begin{mathpar}
  \centering
  \retireStoreLowPatched{}
\end{mathpar}

We use
\(\smicropatched{}{n}{c}{c'}{\declassified}{\declassified'}\) to denote the
evaluation of \(n\) steps in the patched execution. Similarly, we define a
patched execution for the architectural semantics denoted
\(\sarchpatched{}{\aconf}{\aconf'}{\leaked{\obs}}{\declassified}{\declassified'}\) and provide the evaluation rules in \iftechreport\cref{app:arch}\else the technical report~\cite{techreport}\fi.

\begin{definition}[Constant-time up to declassification]\label{def:ct-decl}
  A program is constant-time up to declassification if for any pair of initial architectural configurations
  \(\aconf_{0}\) and \(\aconf'_{0}\) such that \(\sproj{\aconf_{0}} = \sproj{\aconf_{0}'}\), and number of steps \(n\):
  \begin{equation*}
  \sarch{n}{\declassified}{\aconf_{0}}{\aconf_{n}}{\leaked{\obs}} \implies
  \sarchpatched{n}{\aconf_{0}'}{\aconf_{n}'}{\leaked{\obs'}}{\declassified}{\varepsilon} ~\wedge~ \leaked{\obs} = \leaked{\obs'}
  \end{equation*}
\end{definition}

The following theorem establishes end-to-end security for constant-time programs
up to declassification running on \name{}.
\begin{restatable}{theorem}{thmpatcheddeclassification}\label{thm:patched_declassification}
  For any constant-time program up to declassification, microarchitectural state \(\micro_{0}\), initial configurations
  \(\mconf_{0} = \conf{\mem_{0}, \reg_{0}, \varepsilon, \micro_{0}}\) and
  \(\mconf_{0}' = \conf{\mem_{0}', \reg_{0}', \varepsilon, \micro_{0}}\) such
  that \(\sproj{\conf{\mem_{0}, \reg_{0}}} = \sproj{\conf{\mem_{0}', \reg_{0}'}}\) and number of steps \(n\),
  \begin{gather*}
    \smicro{}{n}{\mconf_{0}}{\mconf_{n}}{\declassified} \implies
    \smicropatched{}{n}{\mconf_{0}'}{\mconf_{n}'}{\declassified}{\varepsilon} ~\wedge~ %
    \micro_{n} = \micro_{n}'
  \end{gather*}
  where \(\micro_{n}\) and \(\micro_{n}'\) are the microarchitectural contexts in configurations \(\mconf_{n}\) and \(\mconf_{n}'\), respectively.
\end{restatable}
\noindent%
In the next section, we provide an example to demonstrate how
patched execution works.

\subsection{Examples}\label{sec:example}
This section showcases key aspects of \name{}'s semantics through small
examples. \Cref{ex:declassification} illustrates how declassification
works and how \name{} prevents forwarding secrets to potential side channels
during speculative execution. %
\Cref{{ex:rollback_correct_prediction}} demonstrates that when a predicted load
value is resolved and the actual value is secret, speculative execution must be
rolled back, even if the prediction was correct.

\newcommand\declassifiedValue[0]{\declassify{\texttt{f($\val{s_{1}}$)}}}
\begin{example}[Declassification]\label{ex:declassification}
  Consider an execution of the program in \cref{lst:patched_declassification}
  (in the standard hardware semantics~\(\smicro{}{}{}{}{}{}\)), where the
  register \(\register{s}\) evaluates to \(\slevel{\val{s_{1}}}{\topsec}\) in
  the initial configuration.
  Moreover, we assume that all conditions are
  predicted to be \(true\), but only \(\val{c_{1}}\) can architecturally evaluate to
  \(true\). Under this hypothesis, the program is constant-time up to
  declassification. %
  \begin{lstlisting}[caption=\protect{Illustration of declassification where
    \lstinline{s} is a secret input, \lstinline{f} is a one-way function, and
    \lstinline{a}$_{\botsec}$ is an address to a public memory location
    (\(\memsec({\text{\lstinline{a}}_{\botsec}}) = \botsec\)). For readability, \lstinline{beqz} instructions are replaced with \lstinline{if} constructs.},
  label={lst:patched_declassification}] %

store a$_{\botsec}$ <@\declassify{f(s)}@>       // Declassify f(s) <@\label{line:patched_declassification:declassify}@>
<@\declassify{d}@> $\leftarrow$ load a$_{\botsec}$ // Load declassified value<@\label{line:patched_declassification:load}@>
if $\val{c_{1}}$ { x $\leftarrow$ load <@\declassify{d}@> }         // Allowed <@\label{line:patched_declassification:leak}@>
if $\val{c_{2}}$ { x $\leftarrow$ load <@\leaked{s}@> }         // Blocked <@\label{line:patched_declassification:leak-insecure}@>
if $\val{c_{3}}$ { x $\leftarrow$ load <@\leaked{f(s)}@> }       // Blocked <@\label{line:patched_declassification:blocked}@>
\end{lstlisting}

  At line~\ref{line:patched_declassification:declassify}, the program computes
  \lstinline{f($\val{s_{1}}$)} and declassifies the result by storing it to
  public memory. By the rule \textsc{retire-store-low}, it also
  produces a declassification trace \declassifiedValue{}. %

  At line~\ref{line:patched_declassification:load}, the program loads the
  declassified value in register \(\declassify{\register{d}}\). By the rule \textsc{execute-load-commit}, \(\declassify{\register{d}}\) has
  security level \(\memsec({\text{\lstinline{a}}_{\botsec}}) = \botsec\), %
  hence \name{} can speculatively execute the \lstinline{load} on
  line~\ref{line:patched_declassification:leak} to speed up computations before
  $\val{c_{1}}$ is resolved. While it speculatively leaks
  \(\declassify{\register{d}}\) to the cache, it is not a security concern
  because the value has been intentionally declassified (cf.
  \cref{hyp:declassification}). %

  Notice that because \lstinline{f} is a one-way function, declassifying
  \declassifiedValue{} does not reveal information about \(\val{s_{1}}\). In
  particular, the \lstinline{load} on
  line~\ref{line:patched_declassification:leak-insecure} should certainly not be
  allowed to execute speculatively. %
  \name{} faithfully enforces this policy; the rule \textsc{execute-load-commit}
  uses \(\aplsan\) to compute the address of the load, and because this happens during
  speculative execution and \lstinline{s} is labeled secret, we get
  \(\semexpr{\register{s}}{\aplsan(\buf, \reg)} = \botval\) (cf.
  \cref{def:aplsan,def:sproj}).
  Hence, the \lstinline{load} cannot be
  executed. Similarly, the \lstinline{load} on
  line~\ref{line:patched_declassification:blocked} is also blocked because the
  value \lstinline{f(s)} is recomputed and inherits the secret label from
  \lstinline{s}.

  \smallskip
  Now, to illustrate \cref{thm:patched_declassification} (security up to declassification), consider a second execution
  of the program in the \emph{patched semantics} starting with a configuration
  low-equivalent to the previous one, but where \(\register{s}\)
  evaluates to \(\slevel{\val{s_{2}}}{\topsec}\) and where
  declassified values are patched with the declassification trace of the first
  execution, i.e.,\ \declassifiedValue{}. According to
  \cref{thm:patched_declassification}, the leakage of the first execution and
  this second patched execution should be the same.

  At line~\ref{line:patched_declassification:declassify}, the execution
  evaluates the rule \textsc{retire-store-low-patched}, which stores the value
  \declassifiedValue{} at address \lstinline{a$_{\botsec}$} (instead of storing
  \lstinline{f($\val{s_{2}}$)}).

  At line~\ref{line:patched_declassification:load}, the value
  \declassifiedValue{} is loaded into the register \(\declassify{\register{d}}\)
  and is assigned security level
  \(\memsec({\text{\lstinline{a}}_{\botsec}}) = \botsec\).

  At line~\ref{line:patched_declassification:leak} (cf.\ rule
  \textsc{execute-load-commit}), the value of \(\declassify{\register{d}}\) can
  be forwarded to the \load{}{} because its security level is \(\botsec\).
  Notice that because the second execution has been patched with the
  declassification trace of the first execution, the update to the
  microarchitectural context (i.e.,\ the leakage) is the same in both executions.

  At line~\ref{line:patched_declassification:leak-insecure}, similarly to the
  first execution, \name{} does not forward the value of \register{s} to the
  \load{}{} because its security level is \(\topsec\). %
  By contrast, if we would consider an insecure microarchitecture that forwards
  the value, the microarchitectural context would be updated with
  \leaked{$\val{s_{1}}$} in the first execution and \leaked{$\val{s_{2}}$} in
  the second execution, which would violate \cref{thm:patched_declassification}.

  Perhaps less intuitively, the leakage at
  line~\ref{line:patched_declassification:blocked} would also be considered
  insecure w.r.t.\ \cref{thm:patched_declassification}, even though it is
  semantically equivalent to the leakage at
  line~\ref{line:patched_declassification:leak}. Indeed, the leakage is not
  intentional w.r.t.\ \cref{hyp:declassification}, which our security definition
  takes into account.

\end{example}
In summary, using the notion of \emph{patched execution} allows us to express
that there are no microarchitectural leaks beyond public information and
explicitly declassified values. Even if some other secrets in the program are
information-theoretically derivable from declassified values (but, for instance,
are cryptographically protected against such derivation), they should still be
considered secret. %
In particular, our declassification condition guarantees that any attack exposing program secrets has to
do so based on public information and explicitly declassified values and hence
does not rely on any Spectre attack. %
We believe that for a
cryptographic primitive, our definition of declassification can be composed with
a standard notion of cryptographic indistinguishability to obtain a stronger
notion of cryptographic indistinguishability for the execution of the
  primitive on \name{}.

\smallskip
\begin{example}[Rolling back a correct prediction]\label{ex:rollback_correct_prediction}
{\hypersetup{hidelinks}
\hypersetup{colorlinks=false}
  Consider the following program, where \lstinline{a}$_{\topsec}$ is an address
  to a secret memory location
  (\(\memsec({\text{\lstinline{a}}_{\topsec}}) = \topsec\)):
  \begin{lstlisting}[%
  ] %

x $\leftarrow$ load a$_{\topsec}$ // Load secret value <@\label{line:rollback_correct_predictions:load}@>
y $\leftarrow$ x + 4 <@\label{line:rollback_correct_predictions:operation}@>
\end{lstlisting}

  We illustrate step-by-step a (possible) execution flow, focusing on the evolution
  of \(\buf\) and highlighting \leaked{changes} at each step.

  \begin{itemize}
    \item Consider that the scheduler first fetches all instructions: \\%
    \(\buf = %
    \load{x}{a_{\topsec}}@\varepsilon \cdot %
    \mov{\pc}{\slevel{\ref{line:rollback_correct_predictions:operation}}{\botsec}}@\varepsilon \cdot
    \mov{y}{\register{x} + 4}@\varepsilon\)
    \item The scheduler then applies the rule \textsc{execute-load-predict} and the predictor predicts the loaded value to be 0. Notice that the prediction is public, so we assume that the attacker knows (and can even influence) its value:\\
          \(\buf = %
          \leaked{\mov{x}{0}@\ref{line:rollback_correct_predictions:load}} \cdot %
          \mov{\pc}{\slevel{\ref{line:rollback_correct_predictions:operation}}{\botsec}}@\varepsilon \cdot \mov{y}{\register{x} + 4}@\varepsilon\)
    \item Next, the scheduler applies the rule \textsc{execute-assign}, which
          computes
          \(\mov{y}{\register{x} + 4}\): \\
          \(\buf = %
          \mov{x}{0}@\ref{line:rollback_correct_predictions:load} \cdot %
          \mov{\pc}{\slevel{\ref{line:rollback_correct_predictions:operation}}{\botsec}}@\varepsilon \cdot \leaked{\mov{y}{4}@\varepsilon}\)
    \item When resolving the prediction, because we have
          \(\memsec({\text{\lstinline{a}}_{\topsec}}) = \topsec\), the rule
          \textsc{execute-load-commit} cannot be applied, and the execution is
          rolled back, even if the predicted value (0) was correct:\\
          \(\buf_{rollback} = %
          \mov{x}{0}@\leaked{\varepsilon} \cdot %
          \mov{\pc}{\slevel{\ref{line:rollback_correct_predictions:operation}}{\botsec}}@\varepsilon\)\label{ex:rollback_correct_prediction:rollback}
  \end{itemize}
  Importantly, unconditionally rolling back the execution does not leak information about the
  secret value \(\mem(a_{\topsec})\), whereas allowing
  \textsc{execute-load-commit} to proceed would leak whether
  \(\mem(a_{\topsec}) = 0\). Indeed, if \(\mem(a_{\topsec}) \neq 0\), then a
  rollback would happen, and the final ROB would be \(\buf_{rollback}\).
  If \(\mem(a_{\topsec}) = 0\), the final ROB would be \(\buf_{commit} = %
  \mov{x}{0}@\varepsilon \cdot %
  \mov{\pc}{\slevel{\ref{line:rollback_correct_predictions:operation}}{\botsec}}@\varepsilon \cdot
  \leaked{\mov{y}{4}@\varepsilon}\). %
}

  Concretely, such a conditional rollback introduces a so-called \emph{implicit
    resolution-based channel}~\cite{DBLP:conf/micro/YuYK0TF19}: the rollback
  case and the commit case lead to distinct timing behaviors (indeed, contrary
  to the commit case, the rollback case has to recompute the instructions following the load,
  which takes extra cycles). %
  While prior solutions~\cite{DBLP:conf/micro/YuYK0TF19,
    DBLP:conf/micro/ChoudharyYF021} address implicit resolution-based channels
  (e.g.,\ from memory disambiguation) by delaying the squashing of the implicit
  branch %
  until prior speculations are resolved, this solution does not apply to load
  value prediction. Indeed, in our example, the implicit branch is already
  non-speculative (the processor knows for sure that the value can be
  committed).
\end{example}

\section{Discussion}\label{sec:discussion}
This section discusses prediction mechanisms supported by \name{},
limitations, and compatibility with legacy code.

\myparagraph{Prediction mechanisms} %
Prior work~\cite{DBLP:conf/micro/YuYK0TF19,DBLP:conf/isca/YuMT0F20} already stressed the importance of making predictions a function of public data. %
The novelty of \name{} is to allow predictions to depend on \emph{any public
  data}, hence generalizing standard models of speculative execution. Indeed,
public values are used in the \textsc{step} rule to update the
microarchitectural context \(\micro\), which is always given as an argument of
\(\predict\). We now discuss how this model encompasses known prediction
strategies.

Because the program counter is always public and therefore part of \(\micro\)
(cf.\ \iftechreport \cref{lemma:pc_low} in \cref{app:proofs} \else the technical report~\cite{techreport}\fi), \(\predict(\micro)\) always has
access to the current (and past) values of the program counter and corresponding
instructions. Hence, it can make control-flow predictions based on the full
control-flow history, %
which encompasses existing prediction strategies for conditional
branches~\cite{DBLP:conf/isca/Smith81}, indirect
branches~\cite{DBLP:journals/computer/LeeS84}, and return targets.

Speculation on memory disambiguation, related to memory-disambiguation machine
clears~\cite{DBLP:conf/uss/RagabBBG21} (i.e.,\
Speculative-store-bypass~\cite{spectrev4} and (predictive)
store-to-load-forwarding~\cite{SecurityAnalysisAMD2021}), is enabled by the
rule \textsc{execute-load-predict}. However, \name{} forwards only
\emph{public values} from the memory and microarchitectural buffers (the store buffer in particular). This restriction could be lifted by propagating
security levels in the store buffer and forwarding predicted values with their
security level, as done by SPT~\cite{DBLP:conf/micro/ChoudharyYF021}. We leave
this optimization for future work.

Via the rule \textsc{execute-load-predict}, our semantics also encompasses the
more futuristic load value prediction~\cite{DBLP:conf/asplos/LipastiWS96}, which
can be implemented as forwarding a simple constant or forwarding a value based
on the \emph{public} history of load operations.
Value prediction~\cite{DBLP:conf/micro/LipastiS96} (also related to
floating-point machine clears~\cite{DBLP:conf/uss/RagabBBG21}), which we do not
formalize here for simplicity, is similar to load value
prediction. In particular, the prediction should not depend on secrets, and the
execution must always be rolled back if the actual value is secret.

Finally, \(\predict\) might also return any arbitrary value, which accounts for
predictor states that have been poisoned by an attacker or that forward dummy
values, hence encompassing Spectre, as well as LVI (and LVI-NULL) attacks.

\myparagraph{Limitations of \name{}} %
Memory-ordering machine clears~\cite{DBLP:conf/uss/RagabBBG21} are another
source of transient execution, sometimes requiring rolling back memory
operations to preserve memory consistency for concurrent programs. Even
though our semantics does not support concurrency,
this kind of mechanism could be supported by tracking whether memory instructions might
be
rolled back w.r.t.\ to some memory consistency model, similar to speculations.
For instance, to support the total-store-ordering model (TSO), the rule
\textsc{execute-load-commit} should additionally make sure that all prior
\load{}{} operations in the ROB are resolved (e.g.,\ with an additional
hypothesis \(\load{\_}{\_} \not\in \buf\)).

Self-modifying code machine clears~\cite{DBLP:conf/uss/RagabBBG21} can also cause
transient execution in self-modifying code when the
instruction cache (queried in the fetch stage) and the data cache (modified by
prior \(\store{}{}\) instructions) are desynchronized. Because our semantics
assumes instruction memory to be fixed, it does not apply to self-modifying
code. 

\myparagraph{Legacy software compatibility} %
\name{} is fully compatible with legacy software. Code without secret annotations
works on \name{} as is, but without additional
security over the base processor. Security and performance can also be
traded off: the entire memory (or the stack) can be marked as secret, but it
will likely result in additional performance overhead. Finally, to achieve security and optimal
performance, only secret-handling code needs to be annotated. 
For instance, an
annotated cryptosystem could be linked securely with (memory-safe) legacy
code if the legacy code architecturally accesses only public or declassified information.

\section{Implementation and evaluation}\label{sec:implem_eval}
\subsection{Implementation}\label{sec:implem}

To better understand the costs and benefits of \name{}, we built a prototype hardware implementation using the Proteus RISC-V processor framework.\footnote{\url{https://github.com/proteus-core/proteus}}
Proteus is implemented in SpinalHDL~\cite{spinalhdl}, a Scala-based hardware
description language (HDL). SpinalHDL generates Verilog or VHDL code that can be
run in a simulator or synthesized for an FPGA.

From the many existing open-source RISC-V CPU implementations,\footnote{\url{https://github.com/riscv/riscv-isa-manual/blob/master/marchid.md}} we selected Proteus
because it is designed to be extended with new hardware mechanisms via
a plugin system (inspired by VexRiscv~\cite{vexriscv}). It is easily
configurable in the number of ROB entries and execution units, and it supports
branch target prediction and speculative execution, making it
vulnerable to Spectre-PHT, -BTB, and -RSB attacks.

\name{} is implemented as a Proteus plugin, with some additional modifications
in the base processor. In future work, we are looking into combining \name{} with other security extensions on Proteus.
Our implementation is open-sourced at \url{https://github.com/proteus-core/prospect}.

\myparagraph{Simplifications} %
Our prototype adopts memory partitioning: secrets are co-located in one or
more secret memory regions, and we manually inform the hardware of the
region boundaries.
While it is possible to hardcode the boundaries of arbitrary secret regions
in hardware, we implemented a more flexible approach that enables the
configuration of secret region boundaries via control and status registers (CSRs).
The number of CSRs can affect the hardware costs of the implementation; we
report on setups allowing one and two secret regions.
Note that in our benchmarks, we could co-locate all secrets in a single region
when the stack is public and in two regions when the stack is secret. In
future work, we plan to develop compiler support for co-locating all secrets
in a single region and automatically setting up the secret region boundaries through CSRs.

The prototype takes a conservative approach and
stalls every speculative instruction operating on secret data,
not only \emph{insecure instructions} (cf. \cref{req:stall}).
Finally, interrupts are not supported.

\myparagraph{Code size} %
The full implementation of a \name{}-enabled processor consists of 5275 lines of SpinalHDL code, which generates approximately 104,000 lines of Verilog code.
The \name{} plugin is written in 90 lines of SpinalHDL, and approximately 270 additional lines of the base Proteus code were modified.
For our evaluation, we use 5 execution units and a reorder buffer of size 16.
To encourage further experimentation with different configurations, we open-source our evaluation setup.

\subsection{Evaluation}\label{sec:eval}

The security benefit of \name{} comes with a tradeoff in three different aspects:
\begin{enumerate}
  \item \emph{Hardware cost}: \name{} uses additional hardware to track secret data
        and restrict its propagation.
        This can have an impact on hardware cost metrics like area used
        and critical path.
  \item \emph{Runtime overhead}: \name{} can delay the forwarding of secret data,
        which might impact the execution time of applications.
  \item \emph{Labeling of secrets}: Software needs to declare what data is
        secret, possibly requiring additional developer effort to avoid
        unintentional declassification (cf.\ \cref{hyp:declassification}).
\end{enumerate}

To validate the security claims of the implementation~\cite{bognar2022gap}, we executed code samples vulnerable against Spectre on
both the base Proteus and the \name{}-extended implementation and did not observe
leakage in the latter case.

\myparagraph{Hardware cost}
To assess the overhead of the area used and the critical path, we synthesized
Proteus without and with the \name{} modifications for an \texttt{Artix-7 XC7A35T} FPGA
with a speed grade of -1 using Xilinx Vivado. Our experiments showed a reasonable
overhead for \name{}.
The version supporting a single secret region increases
the number of slice LUTs from 16,847 to 19,728~($+17\%$) and slice registers from
11,913 to 12,600~($+6\%$). The critical path increases from 30.1~ns to 30.7~ns~($+2\%$).
We did not observe any additional increase in these numbers when adding a second set of
CSRs to support a second secret region.

\myparagraph{Runtime overhead}\label{sec:runtime_overhead}
The runtime overhead of \name{} depends on two main factors.
First, the amount of data marked as secret: if a program only accesses public
data, \name{} incurs no overhead, while if the whole memory is marked as secret,
the overhead is maximal. Second, the performance benefit of speculative
execution on the program: if the performance of a program heavily benefits from
speculative execution, the overhead of \name{} will be higher than if the program
does not benefit from speculative execution.

Making a binary \name{}-compliant requires co-locating secrets in memory, which could
impact performance, e.g. via caching effects. We leave an evaluation of these secondary
effects for future work.

Configuring the secret regions from software using CSRs only requires a few additional instructions (loading the boundary addresses into the CSRs before starting the program), resulting in negligible overhead.

We evaluate the first two main factors using the synthetic benchmarks from
SpectreGuard~\cite{DBLP:conf/dac/FustosFY19}. These benchmarks simulate a mix of
computations on public data (whose performance heavily benefits from speculative
execution) and an encryption routine (whose performance benefits less from
speculative execution) with different fractions of
speculation/crypto (i.e., \textbf{S}/\textbf{C}). We modify the benchmarks in two
ways:
\begin{enumerate*}[label=(\arabic*)]
  \item because we specifically target constant-time code, we replace the
  non-constant-time \texttt{AES} primitive with the constant-time
  \texttt{chacha20} primitive from HACL*~\cite{DBLP:conf/ccs/ZinzindohoueBPB17},
  \item we annotate not only the key and plaintext as secret but also all
  variables that may contain secrets to avoid unintentional declassification
  (cf.\ \cref{hyp:declassification}).
\end{enumerate*}

We ran the benchmarks in three different configurations: the base processor with
no \name{} extension (our baseline), precisely defining the secret region and
defining a secret region to cover the entire address space.
More precisely, for the second configuration, \texttt{P(key)}, we co-locate all secrets
in a single region and load the boundaries of this secret region into the CSRs.
In the third configuration, \texttt{P(all)}, we load the first and last address of
the memory
into the CSRs, protecting the entire address space.
Results are given in \cref{table:chacha},
where the percentages denote the relative execution time compared to the
baseline for each configuration.

\begin{table}
  \centering
  \caption{Relative SpectreGuard benchmark performance on \name{}.}
  \label{table:chacha}
  \begin{tabular}{@{}l  c  c  c  c@{}}
    \toprule
    Setting & 25\textbf{S}/75\textbf{C} & 50\textbf{S}/50\textbf{C} & 75\textbf{S}/25\textbf{C} & 90\textbf{S}/10\textbf{C} \\
    \midrule
    baseline & 100\% & 100\% & 100\% & 100\% \\
    \(\prog{}\)(\texttt{key}) & 100\% & 100\% & 100\% & 100\% \\
    \(\prog{}\)(\texttt{all}) & 110\% & 125\% & 136\% & 145\% \\
    \bottomrule
  \end{tabular}
\end{table}

In line with our expectations, our results show that for a \name{}-compliant binary,
enabling the defense incurs no runtime
overhead when secret values are only accessed in constant-time code.
When marking the whole memory secret, the overhead ranges from 10\% to 45\%,
which is comparable (but lower)
to the overhead of SpectreGuard~\cite{DBLP:conf/dac/FustosFY19} when enabled for the entire address space.\footnote{Of course, no direct comparison can be made as the compiled programs, architectures, and microarchitectures are different.}
Overall, we conclude that \name{} incurs a low overhead when secret data is
precisely annotated, especially for programs where only a restricted part of the
code computes on secrets, which is a common scenario~\cite{DBLP:conf/dac/FustosFY19,DBLP:conf/ndss/0001LCSKG20} (in SSH clients, web
servers, etc.).

\myparagraph{Labeling of secrets}
To benefit from the security guarantees provided by our security theorems, code
must be verified to be constant-time in the sequential execution model.
Fortunately, verified constant-time implementations of cryptographic primitives
are readily available~\cite{DBLP:conf/ccs/ZinzindohoueBPB17}, and such verified
code also makes explicit what program data should be marked as secret.
However, %
it is still not trivial to identify which memory addresses should be marked as
secret. %
For instance, if the compiler spills secret arguments on the stack, that
stack memory must also be marked as secret to avoid unintentional
declassification and comply with \cref{hyp:declassification}. While it is secure
to conservatively over-approximate the memory areas marked as secret (and, for
instance, always mark the stack as secret), this has a performance cost.

Hence, we evaluate how difficult it is to obtain precise information about which
memory addresses should be labeled secret for a set of representative
constant-time cryptographic primitives given in \cref{tab:bench}. To do so, we
manually annotate (in the C code) all local variables that may contain secret
data, to place them in a dedicated memory section. Additionally, we patch the generated assembly code to clear secret values from
registers after declassification. The number of required annotations and
assembly lines %
is reported in \cref{tab:bench}. As a sanity check, we validate that secret data
is not written by the compiler to public memory locations outside of the
dedicated declassification memory.\footnote{We track whether secret data is
  written to non-secret locations using a hardware plugin that we built on top
  of \name{}. } %
In the worst case, the time required to annotate secret variables and to validate
that no secrets are written on the stack was less than 1 hour. %

\begin{table}[ht]
  \centering
  \setlength\tabcolsep{3pt}
  \caption{Cryptographic primitives used for the experimental evaluation,
    reporting the lines of C code (LoC), whether the stack (S) is labeled public (\(\botsec\)) or secret
    (\(\topsec\)), the number of annotations manually (\(A_{m}\)), and automatically (\(A_{a}\)) inserted for
    marking variables, and the number of assembly instructions (\(I\)) manually inserted.}\label{tab:bench}
  \begin{tabular}{@{} r r c r@{ }r r l@{}}
    \toprule
    & LoC & S & \(A_{m}\) & \(A_{a}\) & \(I\) & Description \\
    \midrule
    \texttt{djbsort}~\cite{djbsort}
                              &246& \(\botsec\) & 3 & 0  & 6 & Constant-time sort \\
    \texttt{sha256}~\cite{DBLP:conf/ccs/ZinzindohoueBPB17}
                              &1795& \(\botsec\) & 34 & 0  & 6 & Hash function \\
    \texttt{chacha20}~\cite{DBLP:conf/ccs/ZinzindohoueBPB17}
                              &1864& \(\botsec\) & 51 & 0 & 6 & Encryption \\
    \texttt{curve25519}~\cite{DBLP:conf/ccs/ZinzindohoueBPB17}
                              &3026& \(\topsec\) & 9 & 67 & 0 & Elliptic curve \\
    \bottomrule
  \end{tabular}
\end{table}

Interestingly, in 3 of the 4 cryptographic primitives, secret
registers are not spilled on the stack by the primitive itself but by the
surrounding code. Therefore, for these examples, it is possible to isolate
  secrets from public data and \emph{keep the stack public} to minimize performance
impact. %

Finally, for the more complex primitive \texttt{curve25519}, secret register
spilling cannot easily be avoided manually. Hence, we label the stack as secret,
and instead of annotating secret variables, we annotate \emph{public} variables
to place them out of the stack and limit the performance impact. Notice that because the
program is constant-time, pointers are public and we can automate their
annotation in 67 cases.

In summary, with reasonable manual effort, we were able to keep the stack public
for 3 out of 4 cryptographic primitives and isolate public variables from the (secret)
stack for the remaining primitive. %
We expect that this manual effort can easily be automated with compiler support
along the lines of existing work for
\texttt{x86}~\cite{DBLP:conf/ndss/0001LCSKG20,DBLP:conf/eurosp/SimonCA18}.

\section{Related work and conclusion}
We discussed transient execution attacks throughout the paper,
more details can be found in existing
surveys~\cite{DBLP:conf/uss/CanellaB0LBOPEG19,DBLP:conf/uss/RagabBBG21}.

\myparagraph{Formal microarchitectural semantics} Many studies have proposed
operational semantics for speculative execution to formally reason about Spectre
attacks (see~\cite{surveysunjay} for a detailed comparison up to 2021). Most
previous semantics only capture Spectre-PHT, with a few capturing other
variants such as
Spectre-STL~\cite{DBLP:conf/ndss/DanielBR21,jasmin,DBLP:conf/pldi/CauligiDGTSRB20,DBLP:conf/ccs/GuancialeBD20,DBLP:conf/ccs/FabianGP22},
Spectre-BTB, and Spectre-RSB~\cite{DBLP:conf/pldi/CauligiDGTSRB20,
  DBLP:conf/ccs/GuancialeBD20}. Contrary to previous operational
semantics, \name{} handles \emph{arbitrary} load value prediction~\cite{DBLP:conf/asplos/LipastiWS96} and can additionally capture LVI~\cite{DBLP:conf/sp/BulckM0LMGYSGP20}.
Using \emph{axiomatic semantics}, Ponce-de-Le\'on and Kinder~\cite{catspectre} can accommodate new leakage models for different prediction mechanisms and thus cover such cases.
In contrast to our work, they can also model attacks based on memory-ordering machine clears~\cite{DBLP:conf/uss/RagabBBG21}.
Yet, it is an open question how to model non-interference and declassification, as in our work, using axiomatic semantics.

\myparagraph{Declassification definitions}
Existing leakage models for secure speculation do not permit any kind of  leakage~\cite{surveysunjay} that depends on secrets.
Yet, in practice, it is common to treat encrypted secrets as observable.
In the literature before transient execution attacks, security properties with
intentional leakage (i.e., declassification) have been widely
studied~\cite{DBLP:journals/jcs/SabelfeldS09}. When declassifying ciphertexts is
a goal, declassification definitions that are not built for this, such as
delimited release~\cite{DBLP:conf/isss2/SabelfeldM03}, may consider programs
that unintentionally leak secrets (including the cryptographic keys), as secure,
as we show in \iftechreport \cref{app:declassification}\else our technical report~\cite{techreport}\fi. Our definition does not suffer from
this limitation and is closer to cryptographically masked
flows~\cite{maskedflows} since it considers a symbolic model for declassified
values. Laud~\cite{laud01} pioneered work in the area of security conditions
composable with indistinguishability properties of encryption, which was later
generalized to other cryptographic primitives~\cite{DBLP:conf/popl/FournetR08,
  DBLP:conf/ccs/FournetPR11}. Later, Laud~\cite{Laud08} devised necessary
conditions to compose cryptographically masked flows with standard cryptographic
indistinguishability properties. We stipulate that our security property will
require similar conditions to compose. None of these previous works consider
transient execution attacks.

\myparagraph{Hardware defenses against Spectre} %
Many defenses specifically target the cache
hierarchy~\cite{DBLP:conf/micro/SaileshwarQ19,DBLP:conf/secdev/KimM0MCMT020,DBLP:conf/dac/KhasawnehKSEPA19,DBLP:conf/isca/Ainsworth020,DBLP:conf/micro/YanCS0FT18,
  DBLP:conf/asplos/LiuHMHTS15,DBLP:conf/micro/KirianskyLADE18,DBLP:conf/isca/SakalisKRJS19,DBLP:conf/asplos/TaramVT19,DBLP:conf/micro/Ainsworth21},
yet, these defenses are still vulnerable to attacks exploiting other
side channels~\cite{chowdhuryy2021leaking,DBLP:conf/ccs/BhattacharyyaSN19,DBLP:conf/ccs/FustosBY20,DBLP:conf/esorics/0001SLMG19,renSeeDeadMu2021}.

Speculative taint-tracking
approaches~\cite{DBLP:conf/micro/YuYK0TF19,DBLP:conf/isca/YuMT0F20,DBLP:conf/micro/WeisseNLWK19,DBLP:conf/IEEEpact/BarberBZZT19}
delay instructions that depend on speculatively loaded data. As shown
in~\cite{DBLP:conf/sp/GuarnieriKRV21}, these approaches enforce hardware-based
secure speculation for sandboxing, but offer no protection for non-speculatively
accessed data. Hence, they do not provide secure speculation for the
constant-time policy.

DOLMA~\cite{DBLP:conf/uss/LoughlinNMTWNK21} additionally protects non-speculatively accessed data during
speculation, but its performance relies on optimizations allowing (under certain
conditions) speculative execution of variable-time instructions and memory
operations, which might still be exploited via resource contention. %

Data oblivious ISA extensions (OISA)~\cite{DBLP:conf/ndss/YuHHF19} are a
hardware-based secrecy-tracking mechanism that prevents secret data from leaking,
\emph{including during non-speculative execution}. Software must be updated to use the ISA extensions
to make sure that secret data is
not used as an unsafe operand. %
In contrast, \name{} can be retrofitted into existing ISAs and supports
existing (constant-time) cryptographic code with minimal (software and hardware)
changes.

\myparagraph{Secure speculation for the constant-time policy} The idea of
propagating security levels from software to hardware and using this
information to delay speculative instructions originates from
\context{}~\cite{DBLP:conf/ndss/0001LCSKG20} and
SpectreGuard~\cite{DBLP:conf/dac/FustosFY19}. %
Our work contributes the formalization, security proof,
and hardware implementation. %
While our implementation tracks secret memory regions via
CSRs, \context{} and SpectreGuard track secrets at a
page-level granularity through, for instance, page table entry bits.
\context{} also tracks public security labels in the cache to reduce
over-tainting e.g., when public values are written to the secret
stack.
As a minor difference, \context{} and SpectreGuard completely block the forwarding
of secret data during speculative execution, whereas \name{} allows
executing instructions that do not leak information on their operands. For
instance, in the program \lstinline|if(c){h $\leftarrow$ h + 1}|, if
\lstinline|h| is secret, \context{} and SpectreGuard would stall the
instruction \lstinline|h $\leftarrow$ h + 1| until speculations are resolved,
whereas \name{} would allow \lstinline|h $\leftarrow$ h + 1| to speculatively
execute.
Given a whitelist of secure instructions
(e.g.,~\cite{DataOperandIndependentTimingInstructionsIntel}), \context{} and SpectreGuard could adopt
this less conservative approach while still being secure.
Finally, contrary to \context{} and SpectreGuard, our work
addresses load value speculation and shows that correct predictions must
sometimes be rolled back for security.

Speculative Privacy Tracking (SPT)~\cite{DBLP:conf/micro/ChoudharyYF021} is
another taint-tracking mechanism providing secure speculation for the
constant-time policy, but without requiring support from applications. %
SPT initially considers all data as secret, and whenever a register is
architecturally leaked, SPT declassifies (i.e., untaints) the register and propagates the information through the
microarchitecture with (forward and backward) untainting. %
SPT also tracks security labels dynamically in the L1D cache.
As an example, consider the program in \cref{lst:spt} such that only public
values are accessed. When $\register{x}_{1}$ and $\register{x}_{2}$ are
loaded from memory for the first time, SPT marks them as secret. At
line~\ref{line:spt:declassify}, $\register{x}_{1}$ is architecturally leaked to
the microarchitectural state, meaning that it gets untainted. Hence, the
\lstinline{load} at line~\ref{line:spt:continue} can be executed speculatively.
However, because $\register{x}_{2}$ is tainted, the \lstinline{load} at
line~\ref{line:spt:stall} cannot be executed speculatively. %
In contrast, on \name{}, if the \lstinline{load} at
line~\ref{line:spt:declassify} corresponds to a public location, then $\register{x}_{2}$ is set to public and the
\lstinline{load} at line~\ref{line:spt:stall} can be executed speculatively.
Contrary to SPT, \name{} requires
annotations but can label data more precisely.

\begin{lstlisting}[caption={Taint tracking in SPT where
    $\leaked{\register{x}_{i}:\topsec}$ indicates that the register
    $\register{x}_{i}$ gets a secret label and
    $\declassify{\register{x}_{i}:\botsec}$ indicates that the register
    $\register{x}_{i}$ gets untainted.},label={lst:spt},float=ht] %

x$_{1}$ $\leftarrow$ load a      // $\leaked{\register{x}_{1}:\topsec}$
x$_{2}$ $\leftarrow$ load x$_{1}$      // $\leaked{\register{x}_{2}:\topsec}$, $\declassify{\register{x}_{1}:\botsec}$  <@\label{line:spt:declassify}@>
if (c) $\mathghost$ <@\label{line:spt:transient_cause}@>
  z$_{1}$ $\leftarrow$ load $\declassify{\register{x}_{1}}$   // Continue <@\label{line:spt:continue}@>
  z$_{2}$ $\leftarrow$ load $\leaked{\register{x}_{2}}$   // Stall <@\label{line:spt:stall}@>
\end{lstlisting}

The above defenses
have been implemented in simulators and target the \texttt{x86} platform. In
contrast, we provide a hardware implementation for RISC-V, which 
allows us to evaluate hardware costs. %
\name{} generalizes these prior efforts with a formalization
supporting a wide range of microarchitectural optimizations capturing
recent attacks, and a proof that this enables secure speculation for the
constant-time policy.

\marton{make sure to fix references to the tech report}

\section*{Acknowledgments}
We are grateful for the valuable feedback of our shepherd and the other reviewers, which helped us improve our paper.
This research was partially funded by the ORSHIN project (Horizon Europe grant agreement No. 101070008) and the Flemish Research Programme Cybersecurity.
It has also received funding from ANR TAVA, Carnot Flexsecurity, and PEPR Cyber/Secureval.

\printbibliography{}

\iftechreport{}
\vfill
\onecolumn
\fi{}
\appendix

%

\appendix
\section{Artifact Appendix}


\subsection{Abstract}

The artifact contains the source code of the base Proteus processor extended with \name, alongside the benchmarks and security tests from our paper.
All materials (except for the tool required for hardware cost measurements) are bundled into a Docker container and distributed on GitHub.

\subsection{Description \& Requirements}


\subsubsection{Security, privacy, and ethical concerns}
None, our artifact is contained in a Docker container, it does not perform any attacks against the host system and it does not use user data.

\subsubsection{How to access}

The artifact is available on GitHub at the following URL: \repo.

\subsubsection{Hardware dependencies}
None.

\subsubsection{Software dependencies}

Our artifact uses the following two tools, which are available for both Windows and Linux.
\begin{itemize}
    \item Docker and 7 GB of disk space for the container (\url{https://docs.docker.com/engine/install/}).
    \item Xilinx Vivado 2022.2 Standard Edition, requiring approximately 55 GB of disk space (\url{https://www.xilinx.com/products/design-tools/vivado/vivado-ml.html}).
\end{itemize}

\subsubsection{Benchmarks}
Our evaluation uses modified benchmarks from the SpectreGuard paper, which are included in our artifact.

\subsection{Set-up}


\subsubsection{Installation}
\begin{enumerate}
    \item Install the two dependencies (Docker and Vivado). Our repository contains detailed instructions on setting up Vivado to minimize the required disk space.
    \item Clone our GitHub repository or download the Dockerfile from the root directory (\repo).
    \item Build the Docker container by following the instructions in the \texttt{README.md} of the repository (building takes approximately 2 hours on a mid-range desktop).
\end{enumerate}

\subsubsection{Basic Test}
\label{basic-test}

The security evaluation can be run from the Docker container using the following commands:
\iftechreport
\begin{verbatim}
// first, launch the container
$ docker run -i -t prospect

// inside the container, run the tests
# cd /prospect/tests/spectre-tests/
# ./eval.py /proteus-base/sim/build/base /prospect/sim/build/prospect
TEST secret-before-branch
SECURE VARIANT:  	Secret did not leak!
INSECURE VARIANT:	Secret leaked!
[...]
\end{verbatim}
\else
\begin{verbatim}
// first, launch the container
$ docker run -i -t prospect

// inside the container, run the tests
# cd /prospect/tests/spectre-tests/
# ./eval.py /proteus-base/sim/build/base \
    /prospect/sim/build/prospect
TEST secret-before-branch
SECURE VARIANT:  	Secret did not leak!
INSECURE VARIANT:	Secret leaked!
[...]
\end{verbatim}
\fi

\subsection{Evaluation workflow}

\subsubsection{Major Claims}

\begin{compactdesc}

    \item[(C1):] \name{} prevents the leakage of secrets from well-annotated programs via Spectre attacks. This is shown by experiment (E1) described in Section~6.2, which executes programs vulnerable to Spectre on the baseline and the extended secure implementation.

    \item[(C2):] \name{} incurs no overhead on precisely annotated constant-time code. This is shown by experiment (E2), described in Section~6.2 (Runtime overhead) and Table~1.

    \item[(C3):] \name{} only incurs a small overhead in terms of hardware cost. This is shown by experiment (E3), described in Section~6.2 (Hardware cost).

\end{compactdesc}

\subsubsection{Experiments}

\begin{compactdesc}

    \item[(E1):] [Security tests, 5 human-minutes]:
    \begin{asparadesc}
        \item[How to:] The experiment is performed in the container by launching a script (identical to the basic test \ref{basic-test}).

        \item[Preparation:] Launch the container with \texttt{docker run -i -t prospect} and navigate to the experiment with \texttt{cd /prospect/tests/spectre-tests}.

        \item[Execution:]
        Run the following command:
        \iftechreport
        \begin{verbatim}
./eval.py /proteus-base/sim/build/base /prospect/sim/build/prospect
\end{verbatim}
        \else
        \begin{verbatim}
./eval.py /proteus-base/sim/build/base \
/prospect/sim/build/prospect
\end{verbatim}
\fi
This will run and evaluate the experiments with both the baseline implementation (first argument) and the \name{}-extended version (second argument).

        \item[Results:] The results are displayed as text. The security evaluation should fail with the baseline implementation and succeed with the extension, validating claim (C1).
    \end{asparadesc}

    \item[(E2):] [Runtime overhead, 5 human-minutes + 9 compute-hours]:
    \begin{asparadesc}
        \item[How to:] The experiment is performed in the container by launching a script.

        \item[Preparation:] Launch the container with \texttt{docker run -i -t prospect} and navigate to the experiment with \texttt{cd /prospect/tests/synthetic-benchmark}.

        \item[Execution:]
        Run the following command:
        \iftechreport
        \begin{verbatim}
./eval.py /proteus-base/sim/build/base_nodump /prospect/sim/build/prospect_nodump
\end{verbatim}
        \else
        \begin{verbatim}
./eval.py \
/proteus-base/sim/build/base_nodump \
/prospect/sim/build/prospect_nodump
\end{verbatim}
\fi
        This will run and evaluate the experiments with both the baseline implementation (first argument) and the \name{}-extended version (second argument), using the variants compiled with no waveform dumping to save disk space.

        \item[Results:] The results are displayed as text. The generated table should reflect Table 1 from the paper, validating claim (C2).
    \end{asparadesc}

    \item[(E3):] [Hardware cost, 1 human-hour + 2 compute-hours]:
    \begin{asparadesc}
        \item[How to:] The experiment is performed in Vivado, using generated Verilog files from the Docker container.

        \item[Preparation:] Follow the instructions under the heading \emph{Hardware overhead} in \texttt{README.md} to obtain the Verilog files used for the synthesis and to set up the Vivado project (\emph{Creating the Vivado project}).

        \item[Execution:]
        Follow the instructions under the heading \emph{Running the Vivado evaluation} in \texttt{README.md} to (iteratively) obtain the hardware costs of both the baseline and the \name{}-extended hardware design.

        \item[Results:] The results of the synthesis should be interpreted according to the description under the heading \emph{Interpreting the results} in the \texttt{README.md} and compared to the reported numbers in the paper under the heading \emph{Hardware cost} (Section 6.2).
    \end{asparadesc}

\end{compactdesc}


\subsection{Notes on Reusability}
\label{sec:reuse}


Using the \texttt{newlib} board support package included in this repository and building on the scripts used for our benchmarks, it is possible to run other benchmarks on Proteus and \name{}, making additional benchmarking and security tests possible.
The source code of \name{} can also be modified to investigate tradeoffs or to extend the offered security guarantees.


\subsection{Version}
Based on the LaTeX template for Artifact Evaluation V20220926. Submission,
reviewing and badging methodology followed for the evaluation of this artifact
can be found at \url{https://secartifacts.github.io/usenixsec2023/}.

\iftechreport %
The appendix includes full rules for the hardware semantics
(\cref{app:full_rules}), the architectural semantics (\cref{app:arch}), an
illustration of why a classical definition of declassification is not strong
enough to capture security properties of some cryptographic primitives
(\cref{app:declassification}), and the full proofs of our security theorems
(\cref{app:proofs}). \fi

\iftechreport
\section{Hardware Semantics}\label{app:full_rules}
This section provides some details on the hardware semantics that are not
crucial for understanding of the main content of the paper, including the
definition of initial configurations (\cref{sec:initial_configurations}),
definition of the low-projection of reorder buffers (\cref{sec:sbufproj}), the
full evaluation rules of the hardware semantics (\cref{sec:full_rules}), the
definition of n-step execution (\cref{sec:n-step}).

\subsection{Initial configurations}\label{sec:initial_configurations}
\begin{definition}[Initial configuration]\label{def:initial_configuration}
  An initial configuration is of the form
  \(\conf{\mem_{0}, \reg_{0}, \buf_{0}, \micro_{0}}\) where \(\mem_{0}\) and
  \(\micro_{0}\) are arbitrary memory and microarchitectural contexts,
  \(\buf_{0}\) is empty (\(\buf_{0} = \varepsilon\)), and \(\reg_{0}\) is a
  register map such that for all register \register{x},
  \(\reg(\register{x}) \in \SVals\) (i.e.,\ no register maps to \(\botval\)), and
  \(\reg_{0}(\pc) = \slevel{ep}{\botsec}\) where \(\val{ep}\) is the entrypoint
  of the program.
\end{definition}

\subsection{Low buffer projection}\label{sec:sbufproj}
The low projection of a reorder buffer \(\buf\), discloses all low values in
\(\buf\). Values with security level \(\botsec\) are replaced by \(\botval\). On
the contrary, values with security level \(\botsec\), unresolved expressions,
and tags are not replaced.

\begin{definition}[Low buffer projection \(\sbufproj{\buf}\)]\label{def:sbufproj}
For a reorder buffer \(\buf\) its low-projection \(\sbufproj{\buf}\) is given by:
  \begin{equation*}
    \begin{array}{rl}
      \sbufproj{e} = &
                      \begin{cases}
                        \botval & {\sf if~} e = \slevel{\val{v}}{\topsec} \\
                        e & {\sf otherwise} \\
                      \end{cases}
      \\
      \sbufproj{\varepsilon} = & \varepsilon \\
      \sbufproj{\mov{r}{e}} = & \mov{r}{\sbufproj{e}} \\
      \sbufproj{\load{x}{e}} = & \load{x}{\sbufproj{e}} \\
      \sbufproj{\store{e_{a}}{e_{v}}} = & \store{\sbufproj{e_{a}}}{\sbufproj{e_{v}}} \\
      \sbufproj{inst @ T \cdot \buf} = & \sbufproj{inst} @ T \cdot \sbufproj{\buf}
  \end{array}
\end{equation*}
\end{definition}

\subsection{Full hardware semantics}\label{sec:full_rules}
This section details the evaluation of expressions and the evaluation rules for each directive, \(\fetch\), \(\execute\), and \(\retire\).
In this section and in the proof, we mark ROB assignments that increment the
program counter as \(\incrpc{\slevel{\val{v}}{s}}\), to differentiate them from
\(\pc\) assignments resulting from a control-flow instruction (denoted
\(\mov{\pc}{\slevel{\val{v}}{s}}\)). This change is purely syntactic and is made
to facilitate the proofs; in particular, it does not influence the semantics.

\paragraph{Expression evaluation}
The evaluation of an expression \(e\) with a register map \(\reg\), denoted
\(\semexpr{e}{\reg}\), is given in \cref{fig:eval_expr}. It is a partial
function from expressions to labeled-values in \(\SVals\): it is undefined if
one of the sub-expressions is undefined (\(\reg(\register{r}) = \botval\)).

\begin{figure}[h]
  \centering
\begin{mathpar}
  \semexpr{\slevel{\val{v}}{s}}{\reg} = \slevel{\val{v}}{s} \and
  \infer[]{
    \reg(\register{r}) \in \SVals\\
  }{
  \semexpr{\register{r}}{\reg} = \reg(\register{r})
} %
\and
\infer[]{
    \semexpr{e_1}{\reg} = \slevel{\val{v}_1}{s_1}\\
    \semexpr{e_2}{\reg} = \slevel{\val{v}_2}{s_2}
  }{
    \semexpr{e_1 \otimes e_2}{\reg} =
    \slevel{\val{v_1} \otimes \val{v_2}}{s_1 \sqcup s_2}
  }
\end{mathpar}
\caption{Evaluation of expressions.}\label{fig:eval_expr}
\end{figure}

Importantly, the evaluation of a binary operation propagates the security level
of its operands in a conservative way (cf. \cref{req:tainting}): if at least one
of the operands has security level \(\topsec\), then the resulting
security level is \(\topsec\).

Notice that operations using \(\botval\) operands are undefined which
constitutes a side channel that differentiates \(\botval\) from values in
\(\SVals\). %
However this is not a security concern because it only reveals public
information: the security level of operands and whether they are resolved or
not. Indeed, \(\botval\) is only introduced in two cases: in the function
\(\aplsan\) when the value is marked as secret and is sanitized; and in the
function \(\apl\) when an operand is not fully resolved.

\paragraph{Fetch directive}
\begin{itemize}
  \item \textsc{fetch-predict-branch-jmp} %
        is the evaluation of a \fetch{} directive when the instruction to fetch
        is a branch. The rule predicts the next location \(\val{l'}\) and updates
        \pc{} in the ROB accordingly. Notice that the next location \(\val{l'}\) is
        added to the ROB with security level \(\botsec\), hence it will be
        leaked to the microarchitectural context in the next \textsc{step} rule.
        In the same way, the current location \(\val{l}\), which is added as a tag in
        the ROB is also leaked.
  \item \textsc{fetch-others} is the evaluation of a \fetch{} directive when the
        instruction to fetch is not a branch. The rule adds the current
        instruction to the ROB, increments \pc{},\footnote{For simplicity, we use
        \(\pc + 1\) to denote the increment of the program counter, but the
        actual expression depends on the architecture. Additionally, it may also
        depend on the current program location \(\prog{l}\) without hindering
        the security guarantees given in \cref{sec:theorems}.} and updates the
        ROB accordingly. We assume that if the fetched instruction contains an
        immediate value \(\val{v}\), it is transformed to the ROB expression
        \(\slevel{v}{\botsec}\). Notice that the next location (\(\val{l'}\)) is added
        to the ROB with security level \(\botsec\), hence it will be leaked to
        the microarchitectural context in the next \textsc{step} rule.
\end{itemize}

\begin{figure}[ht]
  \begin{mathpar}
    \centering
    \fetchPredictBranchJump{}
    \and
    \fetchOthers{}
  \end{mathpar}
\caption{Evaluation rules for the \textsc{fetch} directive.}\label{fig:fetch-rules}
\end{figure}

\myparagraph{Execute directive for control-flow instructions}
\begin{itemize}
  \item \textsc{branch-commit} (resp.\ \textsc{jmp-commit}) is the execution of
        a conditional branch (resp.\ indirect jump) in the ROB. It evaluates the
        condition (resp.\ jump target), ensures that the predicted target
        location \(\val{l}\) corresponds to the actual target, and replaces the
        tag \(\val{l_{0}}\) with \(\varepsilon\) to mark the instruction as resolved.
  \item \textsc{branch-rollback} (resp.\ \textsc{jmp-rollback}) is the execution
        of a conditional branch (resp.\ indirect jump) in the ROB when the
        target is incorrectly predicted. The rule discards the instruction that
        has been mispredicted from the ROB, together with more recent ROB
        instructions.
\end{itemize}

\begin{figure}[ht]
  \centering
  \begin{mathpar}
    \branchCommit{}
    \and
    \branchRollback{}
    \and
    \jmpCommit{}
    \and
    \jmpRollback{}
  \end{mathpar}
\caption{Evaluation rules for the \textsc{execute} directive (for control-flow
  instructions).}\label{fig:execute-jmp-rules}
\end{figure}

\myparagraph{Execute directive for memory and assignments}
\begin{itemize}
  \item \textsc{execute-assign} executes an assignment of an expression \(e\) to
        a register \register{r}. The rule evaluates \(e\) and updates the value
        of \(\register{r}\) in the ROB accordingly.
  \item \textsc{execute-load-predict} predicts the value of a \load{}{}
        instruction and updates the ROB accordingly. The prediction is based on
        the microarchitectural context. Notice that the predicted value
        \(\val{v}\) is added to the ROB with security level \(\botsec\) and will
        therefore be leaked to the microarchitectural context in the \textsc{step}
        rule. This is consistent with the fact that predictions only depend on
        public data.
  \item \textsc{execute-load-commit} commits the result of a predicted \load{}{}
        instruction to the ROB. The rule ensures that the ROB does not contain
        \store{}{} instructions preceding the \load{}{}, evaluates the address
        \(\val{a}\), retrieves the corresponding value \(\mem(\val{a})\) from
        the memory, determines the corresponding security level
        \(\memsec(\val{a})\), and updates the ROB. The rule also leaks the value
        of the index to the microarchitectural context. Notice that the rule can
        only be applied if the address \(\val{a}\) corresponds to public memory
        (\(\memsec(\val{a}) = \botsec\)). If the address \(\val{a}\) maps to
        secret memory (meaning that \(\mem(\val{a})\) is secret), the
        \textsc{execute-load-rollback} is executed instead, in order to prevent
        leaking whether the secret value \(\mem(\val{a})\) is equal to the
        predicted value \(\val{v}\).
  \item \textsc{execute-load-rollback} discards the result of a predicted
        \load{}{} instruction from the ROB. Similarly to
        \textsc{execute-load-commit}, the rule computes the address and value of
        the \load{}{}. Contrary to \textsc{execute-load-commit}, the predicted
        value \(\val{v}\) is replaced by the actual value \(\mem(\val{a})\) and
        younger instructions in the ROB are discarded (excluding the corresponding \(\pc\) update). Notice that the rule is
        applied if the \load{}{} value has been mispredicted or if it is secret.
  \item \textsc{execute-store} evaluates both operands of a store instruction
        and adds the result to the reorder buffer. %
        Notice that the value of the store is not yet committed to the memory
        (it will be committed in the rule \textsc{retire-store}). However, if
        its security level is public, it can be used to the update
        microarchitectural context in the \textsc{step} rule. In particular, it
        means that a store value can be forwarded to a load instruction, only if
        it is public.
\end{itemize}

\begin{figure}[ht]
  \centering
  \begin{mathpar}
    \executeAssign{}
    \and
    \executeLoadPredict{}
    \and
    \executeLoadCommit{}
    \and
    \executeLoadRollback{}
    \and
    \executeStore{}
  \end{mathpar}
\caption{Evaluation rules for the \textsc{execute} directive (assignment and
  memory instructions).}\label{fig:execute-mem-rules}
\end{figure}

\myparagraph{Retire directive}
\begin{itemize}
  \item \textsc{retire-assign} retires an assignment on top of the ROB and
        updates the register map accordingly;
  \item \textsc{retire-store} and \textsc{retire-store-decl} retire a store
        instruction on top of the ROB and updates the memory accordingly. These
        rules also updates the microarchitectural context with the index of the
        store. The rule \textsc{retire-store-decl} is applied when a secret is
        declassified, meaning that the value is marked as secret while the
        address of the store corresponds to a part of the memory that is
        observable by the attacker (i.e.,\ \(\memsec(\val{a}) = \botval\)). In
        this case, the value \(\val{v}\) becomes visible to the attacker (i.e.,\
        it is declassified). The rule also produces a declassification trace
        with the declassified value \(\val{v}\);
  \item \textsc{retire-patched} replaces the rule \textsc{retire-store-decl} in
        the patched semantics. Similarly to the rule \textsc{retire-store-decl},
        it is applied when a secret is declassified. It replaces the value that
        is declassified with a value from a declassification trace \(\declassified\).
\end{itemize}

\begin{figure}[ht]
\begin{mathpar}
  \centering
  \retireAssign{}
  \and
  \retireStoreLow{}
  \and
  \retireStoreHigh{}
  \and
  \retireStoreLowPatched{}
\end{mathpar}
\caption{Evaluation rules for the \textsc{retire} directive.}\label{fig:retire-rules}
\end{figure}

\subsection{n-step execution}\label{sec:n-step} %
A \(n\)-step execution in the hardware semantics, denoted
\(\smicro{}{n}{\mconf{}}{\mconf{}'}{\declassified}\), makes \(n\) evaluation
steps from a configuration \(\mconf{}\) to a configuration \(\mconf{}'\).
Additionally, it produces a \emph{declassification trace} \(\declassified\) that
is the sequence of declassified values:
\begin{definition}[\(n\)-step execution]\label{def:n-step}
  For any microarchitectural configuration \(\mconf{}\):
  \begin{mathpar}
    \inferrule[]{}{ %
      \smicro{}{0}{\mconf{}}{\mconf{}}{\varepsilon} %
    }
    \and
    \inferrule[] { %
      \smicro{}{n-1}{\mconf{}}{\mconf{}''}{\declassified}\\
      \smicro{}{}{\mconf{}''}{\mconf{}'}{\val{v}}
    }{ %
      \smicro{}{n}{\mconf{}}{\mconf{}'}{\declassified \cdot \val{v}} %
    }
  \end{mathpar}
\end{definition}

\noindent%
A \(n\)-step patched execution, denoted
\(\smicropatched{}{n}{\mconf{}}{\mconf{}'}{\declassified}{\declassified'}\) is defined as
follows:
\begin{definition}[\(n\)-step patched execution]\label{def:n-step-patched}
  For any microarchitectural configuration \(\mconf{}\):
  \begin{mathpar}
    \inferrule[]{}{ %
      \smicropatched{}{0}{\mconf{}}{\mconf{}}{\declassified}{\declassified} %
    }
    \and
    \inferrule[] { %
      \smicropatched{}{n-1}{\mconf{}}{\mconf{}''}{\declassified}{\declassified''}  \\    \smicropatched{}{}{\mconf{}''}{\mconf{}'}{\declassified''}{\declassified'}
    }{ %
      \smicropatched{}{n}{\mconf{}}{\mconf{}'}{\declassified}{\declassified'}
    }
  \end{mathpar}
\end{definition}

\newpage
\fi{}

\iftechreport
\section{Architectural Semantics}\label{app:arch}
This section defines the architectural semantics, including the definition of
architectural configurations (\cref{sec:initial_configurations:arch}), the full
evaluation rules (\cref{sec:full_rules:arch}), and the definition of n-step
executions (\cref{sec:n-step:arch}).

\subsection{Architectural configurations}\label{sec:initial_configurations:arch}
The sequential execution operates on \emph{architectural configurations}
\(\conf{\mem, \reg}\) where \(\mem\) and \(\reg\) are a memory and a register
map, similar to hardware configurations. Notice that even though register maps
keep track of the taint of registers, the sequential semantics simply ignores
the taint. Therefore, we write \(\reg(\register{r}) = \val{v}\) instead of
\(\reg(\register{r}) = \slevel{\val{v}}{s}\) when \(\level{s}\) is not needed.

\begin{definition}[Initial configuration]\label{def:initial_configuration:arch}
  An initial configuration is of the form \(\conf{\mem_{0}, \reg_{0}}\) where
  \(\mem_{0}\) is an arbitrary memory, and \(\reg_{0}\) is a register map such
  that for all register \register{x}, \(\reg(\register{x}) \in \SVals\) (i.e.,\ no
  register maps to \(\botval\)), and \(\reg_{0}(\pc) = \slevel{ep}{\botsec}\)
  where \(\val{ep}\) is the entrypoint of the program.
\end{definition}
Notice that for any initial architectural configuration
\(\conf{\mem_{0}, \reg_{0}}\), and microarchitectural context \(\micro_{0}\),
\(\conf{\mem_{0}, \reg_{0}, \varepsilon, \micro_{0}}\) is also an initial hardware
configuration as defined in \cref{def:initial_configuration}.

\subsection{Full evaluation rules of the architectural semantics}\label{sec:full_rules:arch}

The architectural semantics is given by a relation
\(\sarch{}{\declassified}{\conf{\mem, \reg}}{\conf{\mem', \reg'}}{\obs}\) which
evaluates an instruction in a configuration \(\conf{\mem, \reg}\) to a
configuration \(\conf{\mem', \reg'}\) and produces an observation \(\obs\) and a
declassification trace \(\declassified\). When the declassification trace is
empty, it is omitted. The rules are given in \cref{fig:sequential-semantics}.

The leakage model that we consider in sequential semantics corresponds to the
standard constant-time leakage model~\cite{DBLP:conf/latincrypt/BernsteinLS12,
  DBLP:conf/uss/AlmeidaBBDE16,DBLP:conf/ccs/BartheBCLP14}. In particular, the
rule \textsc{branch} leaks the outcome of the condition, the rule \textsc{jmp}
leaks the jump target, the rule \textsc{load} and \textsc{store} leak the memory
address that is accessed. The \textsc{store} rule additionally produces a
declassification trace when writing to the low part of the memory (i.e.,\
\(\memsec(\val{a}) = \botsec\)).

\begin{figure}[ht]
  \centering
  \input{rules/sequential-semantics.tex}
  \caption{Evaluation rules of the architectural semantics.}\label{fig:sequential-semantics}
\end{figure}

Additionally, to express security with declassification in a similar way as for
the hardware semantics, we define a patched architectural semantics, denoted
\(\sarchpatched{}{\conf{\mem, \reg}}{\conf{\mem', \reg'}}{\obs}{\declassified}{\declassified'}\).
The patched architectural semantics is similar to the sequential semantics given
in \cref{fig:sequential-semantics} but replaces the \textsc{store} rule which
the rules \textsc{store-high} and \textsc{store-patched} given in
\cref{fig:sequential-patched-semantics}. The rule \textsc{store-high} applies
when the store address corresponds to the secret part of the memory and behaves
like a standard store. The rule \textsc{store-patched} applies when the store
address corresponds to the public part of memory and replaces the value with a
value from the declassification trace \(\declassified\).

\begin{figure}[ht]
  \centering
  \input{rules/sequential-semantics-patched.tex}
  \caption{Evaluation rules of the patched architectural
    semantics.}\label{fig:sequential-patched-semantics}
\end{figure}

\subsection{n-step execution}\label{sec:n-step:arch} %
A \(n\)-step execution in the architectural semantics, denoted
\(\sarch{n}{\leaked{\obs}}{\aconf{}}{\aconf{}'}{\declassified}\), makes \(n\) evaluation
steps from a configuration \(\aconf{}\) to a configuration \(\aconf{}'\). It
produces a \emph{observation trace} \(\leaked{\obs}\) that is the sequence of
individual observations produced by each rule, and a \emph{declassification
  trace} \(\declassified\) that is the sequence of declassified values:
\begin{definition}[\(n\)-step execution]\label{def:n-step:arch}
  For any architectural configuration \(\aconf{}\):
  \begin{mathpar}
    \inferrule[]{}{ %
      \sarch{0}{\declassify{\varepsilon}}{\aconf{}}{\aconf{}}{\leaked{\varepsilon}} %
    }
    \and
    \inferrule[] { %
      \sarch{n-1}{\declassified}{\aconf{}}{\aconf{}''}{\obs}\\
      \sarch{}{\declassify{\val{v}}}{\aconf{}''}{\aconf{}'}{\obs'}
    }{ %
      \sarch{n}{\declassified \cdot \declassify{\val{v}}}{\aconf{}}{\aconf{}'}{\obs \cdot \obs'} %
    }
  \end{mathpar}
\end{definition}

\noindent%
A \(n\)-step patched execution, denoted
\(\sarchpatched{n}{\aconf{}}{\aconf{}'}{\leaked{\obs}}{\declassified}{\declassified'}\) is defined as
follows:
\begin{definition}[\(n\)-step patched execution]\label{def:n-step-patched:arch}
  For any architectural configuration \(\aconf{}\):
  \begin{mathpar}
    \inferrule[]{}{ %
      \sarchpatched{0}{\aconf{}}{\aconf{}}{\leaked{\varepsilon}}{\declassified}{\declassified} %
    }
    \and
    \inferrule[] { %
      \sarchpatched{n-1}{\aconf{}}{\aconf{}''}{\leaked{\obs}}{\declassified}{\declassified''}  \\    \sarchpatched{}{\aconf{}''}{\aconf{}'}{\leaked{\obs'}}{\declassified''}{\declassified'}
    }{ %
      \sarchpatched{n}{\aconf{}}{\aconf{}'}{\leaked{\obs} \cdot \leaked{\obs'}}{\declassified}{\declassified'}
    }
  \end{mathpar}
\end{definition}

\newpage
\fi{}

\iftechreport
\section{Problem with Classical Declassification}\label{app:declassification}
Classic definitions of declassification require the absence of leakage for pairs
of executions agreeing on their declassification
traces~\cite{DBLP:conf/csfw/BartheDR04,DBLP:conf/isss2/SabelfeldM03,DBLP:conf/pldi/BanerjeeNR07,DBLP:conf/sp/AskarovS07}:
\iftechreport
\begin{theorem}\label{thm:security_declassification}
  For all constant-time program \(\prog{}\), number of steps \(n\), memories
  \(\mem_{0}, \mem_{0}'\), register maps \(\reg_{0}, \reg_{0}'\), and
  microarchitectural state \(\micro_{0}\),
  \begin{gather*}
    \smemproj{\mem_{0}} = \smemproj{\mem_{0}'} \wedge \sproj{\reg_{0}} = \sproj{\reg_{0}'} ~\wedge~
    \smicro{}{n}{\conf{\mem_{0}, \reg_{0}, \varepsilon, \micro_{0}}}{\conf{\mem_{n}, \reg_{n}, \buf_{n},  \micro_{n}}}{\declassified} \implies\\
    \smicro{}{n}{\conf{\mem_{0}', \reg_{0}', \varepsilon, \micro_{0}}}{\conf{\mem_{n}', \reg_{n}', \buf_{n}', \micro_{n}'}}{\declassify{\declassified'}} \wedge (\declassified = \declassify{\declassified'}
    \implies %
    \mu_{n} = \mu_{n}')
  \end{gather*}
\end{theorem}
\else
\begin{theorem}\label{thm:security_declassification}
  For all constant-time program \(\prog{}\), number of steps \(n\), memories
  \(\mem_{0}, \mem_{0}'\), register maps \(\reg_{0}, \reg_{0}'\), and
  microarchitectural state \(\micro_{0}\),
  \begin{gather*}
    \smemproj{\mem_{0}} = \smemproj{\mem_{0}'} ~\wedge~ %
    \sproj{\reg_{0}} = \sproj{\reg_{0}'} ~\wedge\\
    \smicro{}{n}{\conf{\mem_{0}, \reg_{0}, \varepsilon, \micro_{0}}}{\conf{\mem_{n}, \reg_{n}, \buf_{n},  \micro_{n}}}{\declassified} \implies\\
    \smicro{}{n}{\conf{\mem_{0}', \reg_{0}', \varepsilon, \micro_{0}}}{\conf{\mem_{n}', \reg_{n}', \buf_{n}', \micro_{n}'}}{\declassified} ~\wedge\\ %
    (\declassified = \declassify{\declassified'} \implies \mu_{n} = \mu_{n}')
  \end{gather*}
\end{theorem}
\fi

\paragraph{Problem} The problem with \cref{thm:security_declassification} is
that it allows for declassifying more information than intended. %
For instance, declassifying the output of a one-way function
\lstinline{f(m)}, does not reveal information on the secret \lstinline{m}.
Consequently, it should not be allowed to leak the value of \lstinline{m}
because \lstinline{f(m)} has been declassified. In
\cref{ex:need_better_declassification}, we show that the notion of
declassification given in \cref{thm:security_declassification} fails to
capture this intuition and that declassifying \lstinline{f(m)} can also
implicitly declassify \lstinline{m}.

\newcommand\declassifiedValueA[0]{\declassify{\texttt{f($\val{m_{1}}$)}}}
\newcommand\declassifiedValueB[0]{\declassify{\texttt{f($\val{m_{2}}$)}}}

\begin{example}\label{ex:need_better_declassification}
  Consider the execution of the program in \cref{lst:better_declassification} from two
  low-equivalent initial configuration with respective secret input
  \(\reg(\val{m}) = \slevel{\val{m_{1}}}{\topsec}\) and
  \(\reg'(\val{m}) = \slevel{\val{m_{2}}}{\topsec}\) on an insecure speculative
  out-of-order processor.

  At line~\ref{line:better_declassification:declassify}, the first execution
  declassifies \declassifiedValueA{} and the second execution declassifies
  \declassifiedValueB{}. Consider that the code at
  line~\ref{line:better_declassification:leak} is speculatively executed. In
  this case \(\leaked{\val{m_{1}}}\) and \(\leaked{\val{m_{2}}}\) are leaked
  during speculative execution in the first and second executions
  respectively. %
  Notice that because \lstinline{f} is a one-way function, declassifying the
  value of \lstinline{f(m)} does not give away information on the value of
  \lstinline{m}. Therefore, the program should be considered insecure.

  However, because \cref{lst:better_declassification} restricts to pairs of
  traces with the same declassification trace, it only check the absence of
  leakage under the condition \(\declassifiedValueA{} = \declassifiedValueB{}\).
  Under this condition, we have \(\val{m_1} = \val{m_{2}}\) because
  \lstinline{f} is injective. Consequently, these executions satisfy
  \cref{thm:security_declassification}. In other words, declassifying
  \lstinline{f(m)} also implicitly declassifies \lstinline{m}.
\end{example}
\begin{lstlisting}[caption={Illustration of declassification where \lstinline{m} is a secret
    input, \lstinline{f} is a injective one-way function, and
    \lstinline{a}$_{\botsec}$ is an address to a public memory location.},label={lst:better_declassification},float=ht] %

store a$_{\botsec}$ <@\declassify{f(m)}@>     // Declassify f(m) <@\label{line:better_declassification:declassify}@>
if v1 { load <@\leaked{m}@> } // Leak secret m <@\label{line:better_declassification:leak}@>
\end{lstlisting}

Notice that \name{}, provides stronger guarantees than what is captured by
\cref{thm:security_declassification}. Because \lstinline{m} is labeled as
secret, it is not forwarded to the speculative \load{}{} at
line~\ref{line:better_declassification:leak} and cannot be speculatively leaked.
As a consequence, we need an alternative definition of declassification that
captures more complex notions of security relevant to cryptographic primitives.

\fi

\iftechreport
\newpage
\section{Proofs}\label{app:proofs}
\newcommand\newpar[0]{\bigskip

\noindent}

The architectural semantics, together with our definition of constant-time
programs can be thought of as a hardware-software security contract, similar to
contracts proposed in prior work~\cite{DBLP:conf/sp/GuarnieriKRV21}. %
Hence, our proofs naturally builds on the proofs given in prior
work~\cite{DBLP:conf/sp/GuarnieriKRV21} that relate security contracts to
hardware semantics, in particular \cref{sec:proof-corr} and
\cref{sec:proof-corr}. The former establishes a correspondence
relation between the architectural semantics and the hardware semantics, and the
latter establishes a correspondence relation between the patched architectural
semantics and the patched hardware semantics. %
The main adaptations are related to the load value speculation and the patched
semantics.
\Cref{sec:proof-low-eq-lemmas} establishes some useful lemmas to reason about low-equivalent pairs of executions.
\Cref{sec:proof-ct} contains the proof for our security theorem applicable to
constant-time programs \emph{without} declassification (cf.
\cref{thm:security0}), and \cref{sec:proof-decl} contains the proof for our
security theorem applicable to constant-time programs \emph{with}
declassification (cf. \cref{thm:patched_declassification}).

\subsection{Correspondence between architectural and hardware semantics}\label{sec:proof-corr}
\paragraph{Notations} %
We let \(\buf\bufindex{i}\) with \(0 \leq i \leq |\buf|\) denote the \(i^{th}\)
instruction in the reorder buffer \(\buf\) and let
\(\buf\bufindex{0} = \varepsilon\). %
Additionally, we let \(\buf\bufrange{i}{j}\) with \(0 \leq i <= j \leq |\buf|\)
be the restriction of the buffer \(\buf\) to instructions between \(i\) and
\(j\) (included).
We let \(\arun\) (resp.\ \(\mrun\)) denote a sequence of architectural
configuration or (resp.\ hardware configuration) resulting from a execution in
the architectural (resp. hardware) semantics. %
Given a sequence of architectural configurations \(\arun\), we let
\(\aconf_{j}\) with \(0 \leq j < |\arun|\) denote the \(j^{th}\)
configuration in \(\arun\) (the same holds for \(\mrun\)).

The following lemma directly follows from the syntax of the ISA
language (cf.\ \cref{fig:isa_instr}).
\begin{lemma}\label{prop:no_pc_assign}
  A program does not modify its control-flow with direct assignments to \(\pc\).
\end{lemma}

The deep update of an architectural configuration \(\conf{\mem,\reg}\) with a
reorder buffer \(\buf\), denoted \(\deepapl{\conf{\mem,\reg}}{\varepsilon}\),
applies all pending instructions in \(\buf\) to the configuration
\(\conf{\mem,\reg}\). Notice that predicted values in \(\buf\) are first
resolved to their correct value before being applied.
\begin{definition}[Deep update]\label{def:deepapl} For any reorder buffer \(\buf\) and architectural configuration \(\conf{\mem,\reg}\):
  \begin{align*}
    \deepapl{\conf{\mem,\reg}}{\varepsilon} & \mydef \conf{\mem,\reg} \\
    \deepapl{\conf{\mem,\reg}}{\register{x} \gets e @ T} & \mydef
    \begin{cases}
     \conf{\mem,\reg\change{\register{x}}{\semexpr{e}{\reg}}} &
     \text{if } T = \varepsilon\\
     \conf{\mem,\reg\change{\register{x}}{\slevelnormal{\mem(\semexpr{e_{a}}{\reg})}{\memsec{\semexpr{e_{a}}{\reg}}}}} &
     \text{if } T = \val{l} \wedge \prog{\val{l}} = \load{\register{x}}{e_{a}}\\
    \end{cases} \\
    \deepapl{\conf{\mem,\reg}}{\pc \gets e @ T} & \mydef
    \begin{cases}
      \conf{\mem,\reg\change{\pc}{\semexpr{e}{\reg}}} & %
      \text{if } T = \varepsilon\\
      \conf{\mem,\reg\change{\pc}{\semexpr{e_{l}}{\reg}}} & %
      \text{if } T = \val{l} \wedge \prog{\val{l}} = \jmp{e_{l}}\\
      \conf{\mem,\reg\change{\pc}{\val{l'}}} & %
      \text{if } T = \val{l} \wedge \prog{\val{l}} = \beqz{e_{c}}{\val{l'}} %
      \wedge \semexpr{e_{c}}{\reg} = 0\\
      \conf{\mem,\reg\change{\pc}{\val{l + 1}}} & %
      \text{if } T = \val{l} \wedge \prog{\val{l}} = \beqz{e_{c}}{\val{l'}} %
      \wedge \semexpr{e_{c}}{\reg} \neq 0\\
    \end{cases} \\
    \deepapl{\conf{\mem,\reg}}{\load{x}{e_{a}} @ T} & \mydef \conf{\mem,\reg\change{\register{x}}{\slevelnormal{\mem(\semexpr{e_{a}}{\reg})}{\memsec{\semexpr{e_{a}}{\reg}}}}}\\
    \deepapl{\conf{\mem,\reg}}{\store{e_{a}}{e_{v}} @ T} & \mydef \conf{\mem\change{\semexpr{e_{a}}{\reg}}{\semexpr{e_{v}}{\reg}},\reg}\\
    \deepapl{\conf{\mem,\reg}}{(inst @ T \cdot \buf)} & \mydef
    \deepapl{(\deepapl{\conf{\mem,\reg}}{inst @ T})}{\buf}
  \end{align*}
\end{definition}

The predicate \(\wf(\buf, \conf{\mem, \reg})\) defines what it means for a
reorder buffer \(buf\) to be well-formed with respect to a configuration
\(\conf{\mem,\reg}\).
\begin{definition}[Well-formed reorder-buffers]\label{def:wfbuf}
  A reorder buffer \(\buf\) is well-formed for an architectural configuration
  \(\conf{\mem, \reg}\) the following conditions hold:
  \begin{align*}
    \wf(\varepsilon, \conf{\mem, \reg}) &  \\
    \wf(inst \cdot \buf, \conf{\mem, \reg}) & %
    ~\text{ if } \wf(inst, \conf{\mem, \reg}) %
    \wedge \wf(\buf, \deepapl{\conf{\mem, \reg}}{inst}) \\
    \wf(\pc \gets \slevel{\val{l}}{\botsec} @ \varepsilon, \conf{\mem, \reg})  & %
    ~\text{ if } \prog{\semexpr{\pc}{\reg}} \in \{\beqz{x}{l'}, \jmp{e}\}\\
    \wf(\pc \gets \slevel{\val{l}}{\botsec} @ \val{l_{0}}, \conf{\mem, \reg}) & %
    ~\text{ if } \semexpr{\pc}{\reg} = \val{l_{0}} %
    \wedge \prog{l_{0}} \in \{\beqz{x}{l'}, \jmp{e}\}\\
    \wf(\incrpc{\slevel{l}{\botsec}} @ \varepsilon, \conf{\mem, \reg}) & %
    \\
    \wf(\mov{x}{e} @ \varepsilon \cdot \incrpc{\slevel{l}{\botsec}} @ \varepsilon, \conf{\mem, \reg})  & %
    ~\text{ if } \prog{\semexpr{\pc}{\reg}} \in \{\mov{x}{e}, \load{x}{e}\} %
    \wedge \val{l} = \prog{\semexpr{\pc+1}{\reg}}\\
    \wf(\mov{x}{\slevel{v}{\botsec}} @ \val{l_{0}} \cdot \incrpc{\slevel{l}{\botsec}} @ \varepsilon, \conf{\mem, \reg})  & %
    ~\text{ if } \semexpr{\pc}{\reg} = \val{l_{0}}
    \wedge \prog{l_{0}} = \load{x}{e} %
    \wedge \val{l} = \prog{\semexpr{\pc+1}{\reg}}\\
    \wf(\load{x}{e} @ \varepsilon \cdot \incrpc{\slevel{l}{s}} @ \varepsilon, \conf{\mem, \reg})  & %
    ~\text{ if } \prog{\semexpr{\pc}{\reg}} = \load{x}{e} %
    \wedge \val{l} = \prog{\semexpr{\pc+1}{\reg}}\\
    \wf(\store{e_{a}}{e_{v}} @ \varepsilon \cdot \incrpc{\slevel{l}{s}} @ \varepsilon, \conf{\mem, \reg})  & %
    ~\text{ if } \prog{\semexpr{\pc}{\reg}} = \store{e_{a}}{e_{v}} %
    \wedge \val{l} = \prog{\semexpr{\pc+1}{\reg}}
    \wedge e_{a}, e_{v} \not\in \SVals \\
    \wf(\store{\slevel{a}{\botsec}}{\slevel{v}{s}} @ \varepsilon \cdot \incrpc{\slevel{l}{s}} @ \varepsilon, \conf{\mem, \reg})  & %
    ~\text{ if } \prog{\semexpr{\pc}{\reg}} = \store{e_{a}}{e_{v}} %
    \wedge \val{l} = \prog{\semexpr{\pc+1}{\reg}} \wedge \val{a} = \semexpr{e_{a}}{\reg} \wedge \val{v} = \semexpr{e_{v}}{\reg}\\
  \end{align*}
\end{definition}

The following lemma states that hardware executions produce well-formed reorder
buffers.
\begin{lemma}[Reorder-buffers are well-formed]\label{lemma:wfbuf}
  For any initial configuration
  \(\conf{\mem_{0}, \reg_{0}, \varepsilon, \micro_{0}}\) and number of steps
  \(n\) such that
  \(\smicro{}{n}{\conf{\mem_{0}, \reg_{0}, \varepsilon, \micro_{0}}}{\conf{\mem_{n}, \reg_{n}, \buf_{n}, \micro_{n}}}{}\), then
  \(\wf(\buf_{n}, \conf{\mem_{n}, \reg_{n}})\).
\end{lemma}
\begin{proof}
  The proof goes by induction on the number of steps and case analysis on the
  evaluation rules of the hardware semantics.
\end{proof}

It follows, that at any point of the execution \(\pc\) is always defined and its
security level is always \(\botsec\). %
\begin{restatable}[Security level of \(\pc\) is \(\botsec\)]{corollary}{pclow}\label{lemma:pc_low}
  For all initial configuration
  \(\conf{\mem_{0}, \reg_{0}, \varepsilon, \micro_{0}}\) and number of steps
  \(n\), such that
  \(\smicro{}{n}{\conf{\mem_{0}, \reg_{0}, \varepsilon, \micro_{0}}}{\conf{\mem_{n}, \reg_{n}, \buf_{n}, \micro_{n}}}{}\),
  then:
  \begin{gather}
    \reg_{n}(\pc) = \slevel{\_}{\botsec} %
    \text{ and }
    \mov{\pc}{e} @ T \in \buf_{n} \implies e = \slevel{\_}{\botsec} %
  \end{gather}
\end{restatable}
\begin{proof}
  The proof follows from the fact that reorder-buffers are well-formed
  (cf.\ \cref{lemma:wfbuf}), and well-formed buffers only contain resolved assignments to \(\pc\) with security level \(\botsec\) (cf.\ \cref{def:wfbuf}).

  Additionally, \(\pc\) is initially set to security level \(\botsec\) in the
  register map and is only updated with well-formed ROB assignments.
\end{proof}

The following lemma expresses that undefined values are never committed to the
register map.
\begin{lemma}[Well defined architectural register maps]\label{lemma:reg_sval}
  For all initial configuration
  \(\conf{\mem_{0}, \reg_{0}, \varepsilon, \micro_{0}}\) and number of steps
  \(n\) such that
  \(\smicro{}{n}{\conf{\mem_{0}, \reg_{0}, \varepsilon, \micro_{0}}}{\conf{\mem_{n}, \reg_{n}, \buf_{n}, \micro_{n}}}{}\),
  then for all register \register{r}, \(\reg_{n}(\register{r}) \in \SVals\).
\end{lemma}
\begin{proof}
  The proof goes by induction on the number of steps. The base case follows from
  \cref{def:initial_configuration}. The inductive case follows from the
  instruction evaluation rules. In particular the rule \textsc{retire-assign} is
  the only rule that modifies the register map and only updates the mapping of a
  register with a value in \(\SVals\).
\end{proof}

The following predicates define what it means for a hardware configuration
\(\mconf\) to be transient, denoted \(\transient(\mconf)\), and what it means
for a ROB instruction \(inst @ \varepsilon\), to be correctly predicted with
respect to a configuration \(\conf{\mem, \reg}\), denoted
\(\goodpred(inst @ \varepsilon, \conf{\mem, \reg})\).
\begin{definition}[Transient ROB]\label{def:trans}
  For all hardware configuration \(\conf{\mem, \reg, \buf, \_}\), and architectural configuration \(\conf{\mem, \reg}\):
  \begin{align*}
    &&\transient(\conf{\mem, \reg, \buf, \_}) & \mydef %
      \begin{cases}
      false & \text{if } \forall 1 \leq i \leq |\buf|.\  %
      \goodpred(\buf\bufindex{i}, %
      \deepapl{\conf{\mem, \reg}}{\buf\bufrange{0}{i-1}}) \\
      true & \text{otherwise}
      \end{cases} \\
    &&\goodpred(inst @ \varepsilon, \conf{\mem, \reg}) & \\%
    &&\goodpred(\pc \gets \slevel{\val{l}}{\_} @ \val{l_{0}}, \conf{\mem, \reg})
     &~\text{ if }~ \prog{\val{l_{0}}} = \beqz{e}{\val{l'}} %
      \wedge \semexpr{e}{\reg} = 0 \wedge \val{l} = \val{l'}\\ %
    &&\goodpred(\pc \gets \slevel{\val{l}}{\_} @ \val{l_{0}}, \conf{\mem, \reg}) %
     &~\text{ if }~ \prog{\val{l_{0}}} = \beqz{e}{\val{l'}} %
      \wedge \semexpr{e}{\reg} \neq 0 \wedge \val{l} = \val{l_{0}} + 1 \\
    &&\goodpred(\pc \gets \slevel{\val{l}}{\_} @ \val{l_{0}}, \conf{\mem, \reg}) %
     &~\text{ if }~ \prog{\val{l_{0}}} = \jmp{e} %
      \wedge \semexpr{e}{\reg} = \val{l}\\
    &&\goodpred(\register{x} \gets \slevel{\val{v}}{\_} @ \val{l_{0}}, \conf{\mem, \reg}) %
     &~\text{ if }~ \prog{\val{l_{0}}} = \load{x}{e} %
       \wedge \semexpr{e}{\reg} = \val{a} %
       \wedge \memsec(a) = \botsec %
       \wedge \mem(\val{a}) = \val{v}\\
  \end{align*}
\end{definition}

The prefixes of a reorder buffer \(\buf\) with respect to a configuration
\(\conf{\mem, \reg}\), denoted \(\prefix(\buf, \conf{\mem, \reg})\). Notice that
the prefixes of a ROB only contain correctly predicted sequences of instructions
with respect to \(\conf{\mem, \reg}\). If \(\buf\) contains a transient
instructions it will be included as the last instruction of the longest prefix
but subsequent instructions will not be included.
\begin{definition}[Prefix of reorder buffer]\label{def:prefixes}
  For a configuration \(\conf{\mem, \reg}\) and well-formed reorder buffer
  \(\buf\), \(\prefix(\buf, \conf{\mem, \reg})\) is defined as:
  \begin{align*}
    \prefix(\varepsilon, \conf{\mem, \reg}) \mydef~& \{ \varepsilon \} \\
   \prefix(\pc \gets \slevel{\val{l}}{s} @ \varepsilon \cdot \buf, \conf{\mem, \reg}) \mydef~& %
    \{\varepsilon \} %
    \cup \{\pc \gets \slevel{\val{l}}{s} @ \varepsilon \cdot \buf' ~|~ %
      \buf' \in \prefix(\buf, \deepapl{\conf{\mem, \reg}}{\pc \gets \slevel{\val{l}}{s} @ \varepsilon}) \\
    \prefix(\pc \gets \slevel{\val{l}}{s} @ \val{l_{0}} \cdot \buf, \conf{\mem, \reg}) \mydef~& %
    \{\varepsilon,  \pc \gets \slevel{\val{l}}{s} @ \val{l_{0}} \} %
    \cup~ \{\mov{\pc}{\slevel{\val{l}}{s}} @ \val{l_{0}} \cdot \buf' ~|\\ %
    & ~\buf' \in \prefix(\buf, \deepapl{\conf{\mem, \reg}}{\pc \gets \slevel{\val{l}}{s} @ \val{l_{0}}}) \wedge \goodpred(\pc \gets \slevel{\val{l}}{s}, \val{l_{0}}, \conf{\mem, \reg})\} \\
    \prefix(inst @ \varepsilon \cdot \incrpc{\slevel{l}{s}} @ \varepsilon \cdot \buf, \conf{\mem, \reg}) \mydef~& %
    \{\varepsilon \} \cup %
    \{inst @ \varepsilon \cdot \incrpc{\slevel{\val{l}}{s}} @ \varepsilon \cdot \buf' ~|~ %
      \buf' \in \prefix(\buf, \deepapl{\conf{\mem, \reg}}{inst @ \varepsilon \cdot \incrpc{\slevel{l}{s}} @ \varepsilon})\} \\
    \prefix(inst @ \val{l_{0}} \cdot \incrpc{\slevel{l}{s}} @ \varepsilon \cdot \buf, \conf{\mem, \reg}) \mydef~& %
    \{\varepsilon,  inst @ \val{l_{0}} \cdot \incrpc{\slevel{\val{l}}{s}} @ \varepsilon\} \cup
    \{inst @ \val{l_{0}} \cdot \incrpc{\slevel{\val{l}}{s}} @ \varepsilon \cdot \buf' ~|~\\ %
      &~\buf' \in \prefix(\buf, \deepapl{\conf{\mem, \reg}}{inst @ \val{l_{0}} \cdot \incrpc{\slevel{l}{s}} @ \varepsilon}) \wedge \goodpred(inst @ \val{l_{0}}, \conf{\mem, \reg})\} \\
      \prefix(\incrpc{\slevel{l}{s}} @ \varepsilon \cdot \buf, \conf{\mem, \reg}) \mydef~& %
    \{\incrpc{\slevel{\val{l}}{s}} @ \varepsilon\}
    \cup~ \{\incrpc{\slevel{\val{l}}{s}} @ \varepsilon \cdot \buf' ~|~ %
      \buf' \in \prefix(\buf, \deepapl{\conf{\mem, \reg}}{\incrpc{\slevel{l}{s}} @ \varepsilon})\}
\end{align*}
where \(inst \not\in \{\mov{\pc}{e}, \incrpc{e}\}\).
\end{definition}

Given a contract run \(\arun\) and a hardware run \(\mrun\), the correspondence
relation \(\corr{\arun}{\mrun}\) maps prefixes of reorder buffers of hardware
states in \(\mrun\) to their corresponding architectural states in \(\arun\). In
particular, \(\corr{\arun}{\mrun}(n)(i) = j\) indicates that the prefix of
length \(i\) of \(\buf_{n}\) corresponds
to the architectural state \(\aconf_{j}\) (where \(\buf_{n}\) denotes the ROB in the \(n^{th}\) hardware state in \(\mrun\) and \(\aconf_{j}\) denoted the \(j^{th}\) architectural state in \(\arun\)).
\begin{definition}[Correspondence relation]\label{def:corr}
  Let \(\aconf{}_{0} = \conf{\mem, \reg}\) be an initial architectural
  configuration and \(\micro\) be a microarchitectural context. Additionally, let
  \(\arun\) be the longest contract run, starting from \(\aconf_{0}\), and
  \(\mrun\) be the longest \name{} run starting from
  \(\mconf_{0} = \conf{\mem, \reg, \varepsilon, \micro}\). Additionally, for all
  \(0 \leq n < |\mrun|\), we let \(\mconf_{n}\) denote the \(n^{th}\)
  configuration in \(\mrun\).

  The correspondence relation \(\corr{\arun}{\mrun}\) is defined as follows:
  \begin{align*}
    &&\corr{\arun}{\mrun}(0) & \mydef \{0 \mapsto 0\} \\
    &&\corr{\arun}{\mrun}(n) & \mydef
      \begin{cases}
        \corrfetch{\arun}{\mrun}(n) &
              \text{if } \nextdirconf(c_{n-1}) = \fetch \wedge \neg \transient(c_{n-1})\\
        \corr{\arun}{\mrun}(n-1) &
              \text{if } \nextdirconf(c_{n-1}) = \fetch \wedge \transient(c_{n-1})\\
        \corr{\arun}{\mrun}(n-1) &
              \text{if } \nextdirconf(c_{n-1}) = \execute{i} \wedge \neg \transient(c_{n-1}) \\
        \shift(\corr{\arun}{\mrun}(n-1)) &
              \text{if } \nextdirconf(c_{n-1}) = \retire \wedge \neg \transient(c_{n-1}) \\
      \end{cases} \\
\text{where}& \\
    &&\corrfetch{\arun}{\mrun}(n) & \mydef
      \begin{cases}
        \corr{\arun}{\mrun}(n-1)\change{|buf_{n-1}| + 1}{\corr{\arun}{\mrun}(n-1)(|buf_{n-1}|) + 1} &
              \text{if } \prognott{\lstpc(\mconf_{n-1})} \in \{\beqz{\_}{\_}, \jmp{\_}\} \\
        \corr{\arun}{\mrun}(n-1)\change{|buf_{n-1}| + 2}{\corr{\arun}{\mrun}(n-1)(|buf_{n-1}|) + 1} &
              \text{if } \prognott{\lstpc(\mconf_{n-1})} \not\in \{\beqz{\_}{\_}, \jmp{\_}\} \\
      \end{cases} \\
    &&\shift(map) & \mydef \lambda i \in \mathbb{N}.\ map(i+1) \\
    &&\nextdirconf(\conf{\mem, \reg, \buf, \micro}) & \mydef
       \nextdir(\update(\smemproj{\mem}, \sproj{\reg}, \sbufproj{\buf}, \micro)) \\
    &&\lstpc(\conf{\mem, \reg, \buf, \micro}) & \mydef
      \reg'(\pc) \text{ where } \conf{\_, \reg'} = \deepapl{\conf{\mem,\reg}}{\buf}
  \end{align*}
\end{definition}

The following lemma states that for all non-transient buffer \(buf\) with
respect to a configuration \(\conf{\mem, \reg}\), the register map obtained by
the function \(\apl\) is equivalent (when defined) to the register map obtained
by the deep update of \(\buf\) with \(\conf{\mem, \reg}\).
\begin{lemma}\label{lemma:deepapl_apl}
  Let \(\conf{\mem,\reg}\) be an architectural configuration and \(\buf\) a
  reorder buffer such that \(\neg\transient(\conf{\mem,\reg,\buf,\_})\). Let
  \(\conf{\_,\reg'} = \deepapl{\conf{\mem,\reg}}{\buf}\). We have that for all
  \(\register{r}\) if \(\apl(\buf, \reg)(\register{r}) = \slevel{\val{v}}{s}\)
  then \(\reg'(\register{r}) = \slevel{\val{v}}{s}\).
\end{lemma}
\begin{proof}
  The proof goes by induction on the size of the ROB, \cref{def:apl} and
  \cref{def:deepapl}. In particular, the following cases require particular
  attention:

  For the case \(inst = \pc \gets \val{l} @ T\), because we have
  \(\neg\transient(\conf{\mem,\reg,\buf,\_})\), then
  \(\deepapl{\conf{\mem,\reg}}{\pc \gets \val{l} @ T} =
  \conf{\mem,\reg\change{\pc}{\val{l}}}\). %
  Additionally, we also have
  \(\apl(\pc \gets \val{l} @ T) = \reg\change{\pc}{\val{l}}\) Hence applying
  \(inst\) with \(\apl\) or \(\deepapl{}{}\) gives the same register map.

  For the case \(inst = \load{x}{e_{a}} @ T\), we get with \(\apl\) that
  \(\register{x}\) is undefined, which immediately concludes our goal.

  For the case \(inst = \store{e_{a}}{e_{v}} @ T\), the instruction is ignored
  by \(\apl\) but the memory is modified in \(\deepapl{}{}\). Note that it only
  impacts the result of subsequent \(inst = \load{x}{e} @ T\) instructions but
  these instructions result in undefined \(\register{x}\) with \(apl\), which
  immediately concludes our goal.
\end{proof}

\begin{corollary}\label{lemma:deepapl_apl_expr}
  Let \(\conf{\mem,\reg}\) be an architectural configuration and \(\buf\) a
  reorder buffer such that \(\neg\transient(\conf{\mem,\reg,\buf,\_})\). Let
  \(\conf{\_,\reg'} = \deepapl{\conf{\mem,\reg}}{\buf}\). We have that for all
  expression \(e\), \(\semexpr{e}{\apl(\buf, \reg)} = \slevel{\val{v}}{\_}\)
  then \(\semexpr{e}{\reg'} = \slevel{\val{v}}{\_}\).
\end{corollary}
\begin{proof}
  This follows directly from \cref{lemma:deepapl_apl} and by structural
  induction on the expression evaluation (cf. \cref{fig:eval_expr}).
\end{proof}

\begin{corollary}\label{lemma:deepapl_aplsan_expr}
  Let \(\conf{\mem,\reg}\) be an architectural configuration and \(\buf\) a
  reorder buffer such that \(\neg\transient(\conf{\mem,\reg,\buf,\_})\). Let
  \(\conf{\_,\reg'} = \deepapl{\conf{\mem,\reg}}{\buf}\). We have that for all
  expression \(e\), \(\semexpr{e}{\aplsan(\buf, \reg)} = \slevel{\val{v}}{\_}\)
  then \(\semexpr{e}{\reg'} = \slevel{\val{v}}{\_}\).
\end{corollary}
\begin{proof}
  Observe by definition of \(\aplsan\) (cf. \cref{def:aplsan}), that
  \(\semexpr{e}{\aplsan(\buf, \reg)} = \slevel{\val{v}}{\_} \implies \semexpr{e}{\apl(\buf, \reg)} = \slevel{\val{v}}{\_}\).
  Form there, \(\semexpr{e}{\reg'} = \slevel{\val{v}}{\_}\) follows directly by
  application of \cref{lemma:deepapl_apl_expr}.
\end{proof}

The following lemma states that for each hardware state in an hardware run, for
all the prefixes of its reorder buffer, there exists a corresponding
architectural state in the corresponding architectural run.
\begin{lemma}[Correctness of the \(\corr{\arun}{\mrun}\) relation]\label{lemma:corr}
  Let \(\aconf{}_{0} = \conf{\mem, \reg}\) be an initial architectural
  configuration and \(\micro\) be a microarchitectural context. Additionally, let
  \(\arun\) be the longest contract run, starting from \(\aconf_{0}\), and
  \(\mrun\) be the longest \name{} run starting from
  \(\mconf_{0} = \conf{\mem, \reg, \varepsilon, \micro}\). Additionally, for all
  \(0 \leq n < |\mrun|\), we let \(\mconf_{n}\) denote the \(n^{th}\)
  configuration in \(\mrun\).

  For all \(0 \leq n < |\mrun|\), we have:
  \begin{equation}
    \text{For all } \buf \in \prefix(\buf_{n}, \conf{\mem_{n},\reg_{n}}),
    \deepapl{\conf{\mem_{n},\reg_{n}}}{\buf} = \aconf_{j} \text{ where } j = \corr{\arun}{\mrun}(n)(|\buf|). \tag{G}\label{lemma:corr:goal}
  \end{equation}
\end{lemma}
\begin{proof}
  The proof goes by induction on \(n\).

  \proofcase{Base case \((n = 0)\).} We have
  \(\mconf_{0} = \conf{\mem, \reg, \varepsilon, \micro}\), hence
  \(\buf_{0} = \varepsilon\). %
  From \cref{def:prefixes}, we have that
  \(\prefix(\varepsilon) = \{\varepsilon\}\). Additionally, from
  \cref{def:corr}, we have that %
  \(j = \corr{\arun}{\mrun}(0)(|\varepsilon|) = 0\).
  From \cref{def:deepapl}, we have that
  \(\deepapl{\conf{\mem,\reg}}{\varepsilon} = \conf{\mem,\reg} = \aconf_{0}\),
  which concludes\cref{lemma:corr:goal}.

  \proofcase{Inductive case.} %
  Assume that the hypothesis holds for hardware runs of length \(n-1\), namely:
  \begin{equation}
    \text{For all } \buf \in \prefix(\buf_{n-1}, \conf{\mem_{n-1},\reg_{n-1}}).\
    \deepapl{\conf{\mem_{n-1},\reg_{n-1}}}{\buf} = \aconf_{j} \text{ where } j = \corr{\arun}{\mrun}(n-1)(|\buf|). \tag{IH}\label{lemma:corr:IH}
  \end{equation}
  We show that \cref{lemma:corr:goal} holds at step \(n\), more precisely after
  taking a step \(\smicro{}{}{c_{n-1}}{c_{n}}{}\), \cref{lemma:corr:goal} holds
  for the configuration \(c_{n}\).

  First, notice that \(c_{n}\) is obtained by application of the \textsc{step}
  rule, which applies the directive
  \(\directive \mydef \nextdirconf(c_{n-1})\). %
  Hence the proof follows by case analysis on the directive \(\directive\).

  \proofcase{Fetch directive.} When \(\directive = \fetch\), two rules may
  apply: \textsc{fetch-predict-branch-jmp} and \textsc{fetch-predict-others}. We
  consider both cases separately.

  \newrule{fetch-predict-branch-jmp}: from the hardware evaluation rules (cf.\
  \cref{fig:fetch-rules}), we have
  \(\buf_{n} = \buf_{n-1} \cdot \pc \gets \slevel{\val{l'}}{\botsec} @ \val{l}\)
  where
  \(\slevel{\val{l}}{\_} \mydef \semexpr{\pc}{\apl(\buf_{n-1}, \reg_{n-1})}\),
  and \(\prog{\val{l}} \in \{\beqz{\_}{\_}, \jmp{\_}\}\). Additionally, we have
  \(\mem_{n} = \mem_{n-1}\) and \(\reg_{n} = \reg_{n-1}\). We have to consider
  two cases:
\begin{itemize}
  \item \(\transient(\mconf_{n-1})\). From \cref{def:corr}, we have that
  \(\corr{\arun}{\mrun}(n) = \corr{\arun}{\mrun}(n-1)\). Additionally,
  from \cref{def:prefixes}, because there is a transient instruction in
  \(\buf_{n-1}\) and because
  \(\conf{\mem_{n-1},\reg_{n-1}} = \conf{\mem_{n},\reg_{n}}\), we have
  \(\prefix(\buf_{n}, \conf{\mem_{n},\reg_{n}}) = \prefix(\buf_{n-1}, \conf{\mem_{n-1},\reg_{n-1}})\).
  From there, \cref{lemma:corr:goal} directly follows from
  \cref{lemma:corr:IH}.
  \item \(\neg\transient(\mconf_{n-1})\). Because
        \(\conf{\mem_{n-1},\reg_{n-1}} = \conf{\mem_{n},\reg_{n}}\), we have
        \(\prefix(\buf_{n}, \conf{\mem_{n}, \reg_{n}}) = \prefix(\buf_{n}, \conf{\mem_{n-1}, \reg_{n-1}})\)
        From this, \cref{def:prefixes}, the definition of \(\buf_{n}\) and the
        fact that \(\buf_{n}\) is well formed (from \cref{lemma:wfbuf}),
        we get  %
        \[\prefix(\buf_{n}, \conf{\mem_{n}, \reg_{n}}) = \prefix(\buf_{n-1}, \conf{\mem_{n-1}, \reg_{n-1}}) \cup \{\buf_{n}\}\]
        Let \(\buf\) be an arbitrary prefix in
        \(\prefix(\buf_{n}, \conf{\mem_{n}, \reg_{n}})\). We can consider two
        cases:
        \begin{enumerate}
          \item
          \(\buf \in \prefix(\buf_{n-1}, \conf{\mem_{n-1}, \reg_{n-1}})\).
          From \cref{lemma:corr:IH} and
          \(\conf{\mem_{n-1},\reg_{n-1}} = \conf{\mem_{n},\reg_{n}}\),
          we have
          \(\deepapl{\conf{\mem_{n},\reg_{n}}}{\buf} = \aconf_{\corr{\arun}{\mrun}(n-1)(|\buf|)}\).
          Moreover, from \cref{def:corr} and because
          \(|\buf| \leq |\buf_{i+1}|\), we have
          \(\corr{\arun}{\mrun}(n)(|\buf|) = \corr{\arun}{\mrun}(n-1)(|buf|)\),
          which entails
          \(\deepapl{\conf{\mem_{n},\reg_{n}}}{\buf} = \aconf_{\corr{\arun}{\mrun}(n)(|\buf|)}\) and concludes \cref{lemma:corr:goal} for this case.
          \item
          \(\buf \not\in \prefix(\buf_{n-1}, \conf{\mem_{n-1}, \reg_{n-1}})\),
          in which case
          \(\buf = \buf_{n} = \buf_{n-1} \cdot \pc \gets \slevel{\val{l'}}{\botsec} @ \val{l}\).
                Because \(\neg\transient(\mconf_{n-1})\) and because reorder
                buffers are well-formed (\cref{lemma:wfbuf}), we have from
                \cref{def:prefixes} that
                \(\buf_{n-1} \in \prefix(\buf_{n-1}, \conf{\mem_{n-1},
                \reg_{n-1}})\). %
                Hence, from \cref{lemma:corr:IH}, we have
                \[\deepapl{\conf{\mem_{n-1},\reg_{n-1}}}{\buf_{n-1}} = \aconf_{j}
                \text{ where } j = \corr{\arun}{\mrun}(n-1)(|buf_{n-1}|)\] %
                In the following, we let
                \(\aconf_{j} = \conf{\reg_{j}, \mem_{j}}\). From
                \cref{lemma:deepapl_apl_expr} and
                \(\slevel{\val{l}}{\_} = \semexpr{\pc}{\apl(\buf_{n-1}, \reg_{n-1})}\),
                we have \(\reg_{j}(\pc) = \slevel{\val{l}}{\_}\) and hence
                \(\lstpc(\mconf_{n-1}) = \val{l}\), which means
                \(\prognott{\lstpc(\mconf_{n-1})} \in \{\beqz{\_}{\_}, \jmp{\_}\}\).
                From this, \cref{def:corr}, and \(|\buf| = |\buf_{n-1}| + 1\),
                we get \(\corr{\arun}{\mrun}(n)(|\buf|) = \corr{\arun}{\mrun}(n-1)(|buf_{n-1}|) + 1 = j + 1\). %

          Additionally, from \cref{def:deepapl}, we have that
          \(\deepapl{\conf{\mem_{n},\reg_{n}}}{\buf} = \deepapl{(\deepapl{\conf{\mem_{n},\reg_{n}}}{\buf_{n-1}})}{\pc \gets \slevel{\val{l'}}{\botsec} @ \val{l}}\). From \(\conf{\mem_{n-1},\reg_{n-1}} = \conf{\mem_{n},\reg_{n}}\),
          and the definition of \(\aconf_{j}\), this
          gives us
          \(\deepapl{\conf{\mem_{n},\reg_{n}}}{\buf} = \deepapl{\aconf_{j}}{\pc \gets \slevel{\val{l'}}{\botsec} @ \val{l}}\).
          Hence, in order to show
          \(\deepapl{\conf{\mem_{n},\reg_{n}}}{\buf} = \aconf_{\corr{\arun}{\mrun}(n)(|\buf|)}\) %
          (which concludes our goal \cref{lemma:corr:goal} for this case), it remains to show
          \[\deepapl{\aconf_{j}}{\pc \gets \slevel{\val{l'}}{\botsec} @ \val{l}} = \aconf_{j+1}\]
          From there, we have to consider several cases:
  \begin{enumerate}
    \item \(\prog{\val{l}} = \beqz{e}{\val{l_{br}}}\) and
          \(\semexpr{e}{\reg_{j}} = 0\). In this case, from \cref{def:deepapl},
          we have
          \(\deepapl{\aconf_{j}}{\pc \gets \slevel{\val{l'}}{\botsec} @ \val{l}} = \conf{\mem_{j},\reg_{j}\change{\pc}{\val{l_{br}}}}\).
          Moreover, from the evaluation rules of the sequential semantics
          (\cref{fig:sequential-semantics}, rule \textsc{branch}), we also have
          \(\aconf_{j+1} = \conf{\mem_{j},\reg_{j}\change{\pc}{\val{l_{br}}}}\),
          which concludes
          \(\deepapl{\aconf_{j}}{\pc \gets \slevel{\val{l'}}{\botsec} @ \val{l}} = \aconf_{j+1}\).
    \item \(\prog{\val{l}} = \beqz{e}{\val{l_{br}}}\) and
    \(\semexpr{e}{\reg_{j}} \neq 0\). In this case, from \cref{def:deepapl},
          we have
          \(\deepapl{\aconf_{j}}{\pc \gets \slevel{\val{l}}{\botsec} @ \val{l}} = \conf{\mem_{j},\reg_{j}\change{\pc}{\val{l+1}}}\).
          Moreover, from the evaluation rules of the sequential semantics
          (\cref{fig:sequential-semantics}, rule \textsc{branch}), we also have
          \(\aconf_{j+1} = \conf{\mem_{j},\reg_{j}\change{\pc}{\val{l+1}}}\), which
          concludes \(\deepapl{\aconf_{j}}{\pc \gets \slevel{\val{l'}}{\botsec} @ \val{l}} = \aconf_{j+1}\).
    \item \(\prog{\val{l}} = \jmp{e}\). In this case, from \cref{def:deepapl},
          we have
          \(\deepapl{\aconf_{j}}{\pc \gets \slevel{\val{l}}{\botsec} @ \val{l}} = \conf{\mem_{j},\reg_{j}\change{\pc}{\semexpr{e}{r_{j}}}}\).
          Moreover, from the evaluation rules of the sequential semantics
          (\cref{fig:sequential-semantics}, rule \textsc{jmp}), we also have
          \(\aconf_{j+1} = \conf{\mem_{j},\reg_{j}\change{\pc}{\semexpr{e}{r_{j}}}}\), which
          concludes \(\deepapl{\aconf_{j}}{\pc \gets \slevel{\val{l'}}{\botsec} @ \val{l}} = \aconf_{j+1}\).
  \end{enumerate}
  We have shown that for an arbitrary \(\buf \in \prefix(\buf_{n}, %
  \conf{\mem_{n}, \reg_{n}})\) we have \(\deepapl{\conf{\mem_{n},\reg_{n}}}{\buf} = \aconf_{\corr{\arun}{\mrun}(n)(|\buf|)}\), which concludes \cref{lemma:corr:goal}.
  \end{enumerate}
  \end{itemize}

  \newrule{fetch-others}: from the hardware evaluation rules (cf.\
  \cref{fig:fetch-rules}), we have
  \(\buf_{n} = \buf_{n-1} \cdot inst @ \varepsilon  \cdot \pc \gets \slevel{\val{l+1}}{\botsec} @ \varepsilon\)
  where
  \(\slevel{\val{l}}{\_} \mydef \semexpr{\pc}{\apl(\buf_{n-1}, \reg_{n-1})}\),
  \(inst = \prog{\val{l}}\) and \(inst \not\in \{\beqz{\_}{\_}, \jmp{\_}\}\). Additionally, we have
  \(\mem_{n} = \mem_{n-1}\) and \(\reg_{n} = \reg_{n-1}\). We have to consider
  two cases:
\begin{itemize}
  \item \(\transient(\mconf_{n-1})\). The proof is similar to the case
  \textsc{fetch-predict-branch-jmp}.
  \item  Because
        \(\conf{\mem_{n-1},\reg_{n-1}} = \conf{\mem_{n},\reg_{n}}\), we have
        \(\prefix(\buf_{n}, \conf{\mem_{n}, \reg_{n}}) = \prefix(\buf_{n}, \conf{\mem_{n-1}, \reg_{n-1}})\)
        From this, \cref{def:prefixes}, the definition of \(\buf_{n}\) and the
        fact that \(\buf_{n}\) is well formed (from \cref{lemma:wfbuf}),
        we get  %
        \[\prefix(\buf_{n}, \conf{\mem_{n}, \reg_{n}}) = \prefix(\buf_{n-1}, \conf{\mem_{n-1}, \reg_{n-1}}) \cup \{\buf_{n}\}\]
        Let \(\buf\) be an arbitrary prefix in
        \(\prefix(\buf_{n}, \conf{\mem_{n}, \reg_{n}})\). We can consider two
        cases:
\begin{enumerate}
  \item \(\buf \in \prefix(\buf_{n-1}, \conf{\mem_{n-1}, \reg_{n-1}})\). The
        proof is similar to the case \textsc{fetch-predict-branch-jmp}.
  \item \(\buf \not\in \prefix(\buf_{n-1}, \conf{\mem_{n-1}, \reg_{n-1}})\), in
        which case
        \(\buf = \buf_{n} = \buf_{n-1} \cdot inst @ \varepsilon \cdot \pc \gets \slevel{\val{l+1}}{\botsec} @ \varepsilon\).
        Because \(\neg\transient(\mconf_{n-1})\) and because reorder buffers are
        well-formed (\cref{lemma:wfbuf}), we have from \cref{def:prefixes} that
        \(\buf_{n-1} \in \prefix(\buf_{n-1}, \conf{\mem_{n-1}, \reg_{n-1}})\). %
        Hence, from \cref{lemma:corr:IH}, we have
        \[\deepapl{\conf{\mem_{n-1},\reg_{n-1}}}{\buf_{n-1}} = \aconf_{j}
        \text{ where } j = \corr{\arun}{\mrun}(n-1)(|buf_{n-1}|)\] %
        In the following, we let \(\aconf_{j} = \conf{\reg_{j}, \mem_{j}}\).
        From \cref{lemma:deepapl_apl_expr} and
        \(\slevel{\val{l}}{\_} = \semexpr{\pc}{\apl(\buf_{n-1}, \reg_{n-1})}\),
        we have \(\reg_{j}(\pc) = \slevel{\val{l}}{\_}\) and hence
        \(\lstpc(\mconf_{n-1}) = \val{l}\), meaning that
        \(\prognott{\lstpc(\mconf_{n-1})} \not\in \{\beqz{\_}{\_}, \jmp{\_}\}\).
        From this, \cref{def:corr}, and \(|\buf| = |\buf_{n-1}| + 2\), we get
        \(\corr{\arun}{\mrun}(n)(|\buf|) = \corr{\arun}{\mrun}(n-1)(|buf_{n-1}|) + 1 = j + 1\). %

        Additionally, from \cref{def:deepapl}, we have that
        \(\deepapl{\conf{\mem_{n},\reg_{n}}}{\buf} = \deepapl{(\deepapl{(\deepapl{\conf{\mem_{n},\reg_{n}}}{\buf_{n-1}})}{inst @ \varepsilon})}{\pc \gets \slevel{\val{l+1}}{\botsec} @ \varepsilon}\).
        From \(\conf{\mem_{n-1},\reg_{n-1}} = \conf{\mem_{n},\reg_{n}}\), and
        the definition of \(\aconf_{j}\), this gives us
        \(\deepapl{\conf{\mem_{n},\reg_{n}}}{\buf} = %
        \deepapl{(\deepapl{\arun_{j}}{inst @ \varepsilon})}{\pc \gets \slevel{\val{l+1}}{\botsec} @ \varepsilon}\).
        Hence, in order to show
        \(\deepapl{\conf{\mem_{n},\reg_{n}}}{\buf} = \aconf_{\corr{\arun}{\mrun}(n)(|\buf|)}\)
        (which concludes our goal \cref{lemma:corr:goal} for this case), it
        remains to show
        \[\deepapl{(\deepapl{\arun_{j}}{inst @ \varepsilon})}{\pc \gets \slevel{\val{l+1}}{\botsec} @ \varepsilon} = \aconf_{j+1}\]
        The proof follows by case analysis on the instruction \(inst\):
\begin{enumerate}
  \item \(inst = \mov{x}{e}\). In this case, from \cref{def:deepapl}, we have
        \(\deepapl{(\deepapl{\arun_{j}}{\mov{x}{e} @ \varepsilon})}{\pc \gets \slevel{\val{l+1}}{\botsec} @ \varepsilon} = \conf{\mem_{j}, \reg_{j}\change{\register{x}}{\semexpr{e}{\reg_{j}}}\change{\pc}{\val{l+1}}}\).
        Moreover, from the evaluation rules of the
        sequential semantics (\cref{fig:sequential-semantics}, rule
        \textsc{assign}), we also have
        \(\aconf_{j+1} = \conf{\mem_{j}, \reg_{j}\change{\register{x}}{\semexpr{e}{\reg_{j}}}\change{\pc}{\val{l+1}}}\),
        which concludes
        \(\deepapl{(\deepapl{\arun_{j}}{inst @ \varepsilon})}{\pc \gets \slevel{\val{l+1}}{\botsec} @ \varepsilon} = \aconf_{j+1}\).
  \item \(inst = \load{x}{e}\) and \(inst = \store{e_{a}}{e_{v}}\). Similar to
        the previous case, these follow by \cref{def:deepapl} and by inspection
        of the evaluation rules of the sequential semantics
        (\cref{fig:sequential-semantics}, rule \textsc{load} and
        \textsc{store}).
  \item Other cases are not possible from
        \(inst \not\in \{\beqz{\_}{\_}, \jmp{\_}\}\) and because the program
        does not contain direct assignments to \(\pc\) (cf. \cref{prop:no_pc_assign}).
  \end{enumerate}
\end{enumerate}
\end{itemize}

\proofcase{Execute directive.} In this case, we
have \(\directive = \execute{j}\) and from \cref{def:corr}
\(\corr{\arun}{\mrun}(n) = \corr{\arun}{\mrun}(n-1)\). The proof proceeds by
case analysis on the evaluation rule
(\cref{fig:execute-jmp-rules,fig:execute-mem-rules}). Notice that in all rules
for the execute directive, we have \(\mem_{n} = \mem_{n-1}\) and
\(\reg_{n} = \reg_{n-1}\).

\newrule{branch-commit}: from the evaluation rules (cf.\
\cref{fig:execute-jmp-rules}), we have:
\begin{align}
  \buf_{n-1} =~& \buf_{n-1}\bufrange{1}{j-1} \cdot
               \mov{\pc}{\slevel{\val{l}}{\_{}}} @ \val{l_{0}} \cdot
               \buf_{n-1}\bufrange{j+1}{|buf_{n-1}|} \tag{Hbufn-1}\label{lemma:corr:execute-branch-commit:hubufi1}\\
  \buf_{n}   =~& \buf_{n-1}\bufrange{1}{j-1} \cdot
               \mov{\pc}{\slevel{\val{l}}{\botsec}} @ \varepsilon \cdot
               \buf_{n-1}\bufrange{j+1}{|buf_{n-1}|} \tag{Hbufi}\label{lemma:corr:execute-branch-commit:hubufi}
\end{align}
with \(\prog{l_{0}} = \beqz{e}{\val{l'}}\). Furthermore, we have
\(\val{l} = (\ite{\val{c} = 0}{\val{l'}}{\val{l_0} + 1})\) where
\(\slevel{\val{c}}{\_} = \semexpr{e}{\aplsan(\buf_{n-1}\bufrange{0}{j-1}, \reg_{n-1})}\), meaning that:
\begin{align}
 \goodpred(\buf_{n-1}\bufindex{j}, \val{l_{0}}, \conf{\mem_{n-1},\reg_{n-1}})
 \tag{Hpred}\label{lemma:corr:execute-branch-commit:pred}
\end{align}

Let \(\buf\) be an arbitrary prefix in
\(\prefix(\buf_{n}, \conf{\mem_{n}, \reg_{n}})\). We can consider two cases:
\begin{itemize}
  \item \(|buf| < j\). In this case, by
        \cref{def:prefixes,lemma:corr:execute-branch-commit:hubufi1,lemma:corr:execute-branch-commit:hubufi},
        and because
        \(\conf{\mem_{n}, \reg_{n}} = \conf{\mem_{n-1}, \reg_{n-1}}\), we also
        have that
        \(\buf \in \prefix(\buf_{n-1}, \conf{\mem_{n-1}, \reg_{n-1}})\). Hence,
        from \cref{lemma:corr:IH}, we have that
        \(\deepapl{\conf{\mem_{n-1},\reg_{n-1}}}{\buf} = \aconf_{\corr{\arun}{\mrun}(n-1)(|\buf|)}\).
        From \(\conf{\mem_{n}, \reg_{n}} = \conf{\mem_{n-1}, \reg_{n-1}}\), and
        \(\corr{\arun}{\mrun}(n) = \corr{\arun}{\mrun}(n-1)\), we get that
        \(\deepapl{\conf{\mem_{n},\reg_{n}}}{\buf} = \aconf_{\corr{\arun}{\mrun}(n)(|\buf|)}\).
  \item \(|buf| \geq j\). Notice that, by \cref{def:prefixes} and
        \(\buf \in \prefix(\buf_{n}, \conf{\mem_{n}, \reg_{n}})\), there is no
        mispredicted instruction in \(\buf\) (except possibly the last one).
        Therefore, it follows from
        \(\conf{\mem_{n}, \reg_{n}} = \conf{\mem_{n-1}, \reg_{n-1}}\), that
        \(\buf' = \buf\bufrange{1}{j-1} \cdot \mov{\pc}{\slevel{\val{l}}{\_}} @ \val{l_{0}} \cdot \buf\bufrange{j+1}{|buf|}\)
        belongs to \(\prefix(\buf_{n-1}, \conf{\mem_{n-1}, \reg_{n-1}})\). %
        Hence, from \cref{lemma:corr:IH}, we have that
        \(\deepapl{\conf{\mem_{n-1},\reg_{n-1}}}{\buf'} = \aconf_{\corr{\arun}{\mrun}(n-1)(|\buf'|)}\).
        From \(|\buf'| = |\buf|\),
        \(\conf{\mem_{n}, \reg_{n}} = \conf{\mem_{n-1}, \reg_{n-1}}\), and
        \(\corr{\arun}{\mrun}(n) = \corr{\arun}{\mrun}(n-1)\), we get that
        \(\deepapl{\conf{\mem_{n},\reg_{n}}}{\buf'} = \aconf_{\corr{\arun}{\mrun}(n)(|\buf|)}\).
        Finally, from \cref{lemma:corr:execute-branch-commit:pred}, we have that
        \(\deepapl{\conf{\mem_{n},\reg_{n}}}{\buf'} = \deepapl{\conf{\mem_{n},\reg_{n}}}{\buf}\),
        which concludes
        \(\deepapl{\conf{\mem_{n},\reg_{n}}}{\buf} = \aconf_{\corr{\arun}{\mrun}(n)(|\buf|)}\).
\end{itemize}

\newrule{branch-rollback} from the evaluation rules (cf.\
\cref{fig:execute-jmp-rules}), we have:
\begin{align}
  \buf_{n-1} =~& \buf_{n-1}\bufrange{1}{j-1} \cdot
               \mov{\pc}{\slevel{\val{l}}{\_{}}} @ \val{l_{0}} \cdot
               \buf_{n-1}\bufrange{j+1}{|buf_{n-1}|} \tag{Hbufn-1}\label{lemma:corr:execute-branch-rollback:hubufi1}\\
  \buf_{n}   =~& \buf_{n-1}\bufrange{1}{j-1} \cdot
               \mov{\pc}{\slevel{\val{l''}}{\botsec}} @ \varepsilon \tag{Hbufi}\label{lemma:corr:execute-branch-rollback:hubufi}
\end{align}
with \(\prog{l_{0}} = \beqz{e}{\val{l'}}\).

Let \(\buf\) be an arbitrary prefix in
\(\prefix(\buf_{n}, \conf{\mem_{n}, \reg_{n}})\). We can consider two cases:
\begin{itemize}
  \item \(|buf| < j\). The proof for this case is similar to the proof for the
        rule \textsc{branch-commit}.
  \item \(|buf| = j\). Notice that, by \cref{def:prefixes} and
        \(\buf \in \prefix(\buf_{n}, \conf{\mem_{n}, \reg_{n}})\), there is no
        mispredicted instruction in \(\buf\) (except the last one).
        Therefore, it follows from
        \(\conf{\mem_{n}, \reg_{n}} = \conf{\mem_{n-1}, \reg_{n-1}}\), that
        \(\buf' = \buf\bufrange{1}{j-1} \cdot \mov{\pc}{\slevel{\val{l}}{\_}} @ \val{l_{0}}\)
        belongs to \(\prefix(\buf_{n-1}, \conf{\mem_{n-1}, \reg_{n-1}})\). %
        Hence, from \cref{lemma:corr:IH}, we have that
        \(\deepapl{\conf{\mem_{n-1},\reg_{n-1}}}{\buf'} = \aconf_{\corr{\arun}{\mrun}(n-1)(|\buf'|)}\).
        From \(|\buf'| = |\buf|\),
        \(\conf{\mem_{n}, \reg_{n}} = \conf{\mem_{n-1}, \reg_{n-1}}\), and
        \(\corr{\arun}{\mrun}(n) = \corr{\arun}{\mrun}(n-1)\), we get that
        \(\deepapl{\conf{\mem_{n},\reg_{n}}}{\buf'} = \aconf_{\corr{\arun}{\mrun}(n)(|\buf|)}\).
        Finally, from \cref{def:deepapl}, we have that
        \(\deepapl{\conf{\mem_{n},\reg_{n}}}{\buf'} = \deepapl{\conf{\mem_{n},\reg_{n}}}{\buf}\),
        which concludes
        \(\deepapl{\conf{\mem_{n},\reg_{n}}}{\buf} = \aconf_{\corr{\arun}{\mrun}(n)(|\buf|)}\).
\end{itemize}

\newrule{jmp-commit} The proof for this case is similar to the proof for the
case \textsc{branch-commit}.

\newrule{jmp-rollback} The proof for this case is similar to the proof for the
case \textsc{branch-rollback}.

\newrule{execute-load-predict}: from the evaluation rules (cf.\
\cref{fig:execute-mem-rules}), we have:
\begin{align}
  \buf_{n-1} =~& \buf_{n-1}\bufrange{1}{j-1} \cdot
               \load{x}{e} @ T \cdot
               \buf_{n-1}\bufrange{j+1}{|buf_{n-1}|} \tag{Hbufn-1}\label{lemma:corr:execute-load-predict:hubufi1}\\
  \buf_{n}   =~& \buf_{n-1}\bufrange{1}{j-1} \cdot
               \mov{x}{\slevel{\val{v}}{\botsec} @ \val{l_{0}}} \cdot
               \buf_{n-1}\bufrange{j+1}{|buf_{n-1}|} \tag{Hbufi}\label{lemma:corr:execute-load-predict:hubufi}
\end{align}
with
\(\slevel{\val{l_{0}}}{\_} = \semexpr{\pc}{\aplsan(\buf_{n-1}\bufrange{0}{j-1}, \reg_{n-1})}\).
Notice that from \cref{def:wfbuf}, we also have that
\(\prog{l_{0}} = \load{\register{x}}{e}\).

Let \(\buf\) be an arbitrary prefix in
\(\prefix(\buf_{n}, \conf{\mem_{n}, \reg_{n}})\). We can consider two cases:
\begin{itemize}
  \item \(|buf| < j\). The proof for this case is similar to the proof for the
        rule \textsc{branch-commit}.
  \item \(|buf| \geq j\). Notice that, by \cref{def:prefixes} and
        \(\buf \in \prefix(\buf_{n}, \conf{\mem_{n}, \reg_{n}})\), there is no
        mispredicted instruction in \(\buf\) (except possibly the last one).
        Therefore, it follows from
        \(\conf{\mem_{n}, \reg_{n}} = \conf{\mem_{n-1}, \reg_{n-1}}\), that
        \(\buf' = \buf\bufrange{1}{j-1} \cdot \load{x}{e} @ T \cdot \buf\bufrange{j+1}{|buf|}\)
        belongs to \(\prefix(\buf_{n-1}, \conf{\mem_{n-1}, \reg_{n-1}})\). %
        Hence, from \cref{lemma:corr:IH}, we have that
        \(\deepapl{\conf{\mem_{n-1},\reg_{n-1}}}{\buf'} = \aconf_{\corr{\arun}{\mrun}(n-1)(|\buf'|)}\).
        From \(|\buf'| = |\buf|\),
        \(\conf{\mem_{n}, \reg_{n}} = \conf{\mem_{n-1}, \reg_{n-1}}\), and
        \(\corr{\arun}{\mrun}(n) = \corr{\arun}{\mrun}(n-1)\), we get that
        \(\deepapl{\conf{\mem_{n},\reg_{n}}}{\buf'} = \aconf_{\corr{\arun}{\mrun}(n)(|\buf|)}\).
        Finally, because \(\prog{\val{l_{0}}} = \load{\register{x}}{e}\), we
        have that
        \(\deepapl{\conf{\mem_{n},\reg_{n}}}{\buf'} = \deepapl{\conf{\mem_{n},\reg_{n}}}{\buf}\),
        which concludes
        \(\deepapl{\conf{\mem_{n},\reg_{n}}}{\buf} = \aconf_{\corr{\arun}{\mrun}(n)(|\buf|)}\).
\end{itemize}

\newrule{execute-load-commit}: from the evaluation rules (cf.\
\cref{fig:execute-mem-rules}), we have:
\begin{align}
  \buf_{n-1} =~& \buf_{n-1}\bufrange{1}{j-1} \cdot
                 \mov{x}{\slevel{\mem_{n-1}(\val{a})}{\level{s}} @ \val{l_{0}}} \cdot
                 \buf_{n-1}\bufrange{j+1}{|buf_{n-1}|} \tag{Hbufn-1}\label{lemma:corr:execute-load-commit:hubufi1}\\
  \buf_{n}   =~& \buf_{n-1}\bufrange{1}{j-1} \cdot
                 \mov{x}{\slevel{\mem_{n-1}(\val{a})}{\memsec(\val{a})} @ \varepsilon} \cdot
                 \buf_{n-1}\bufrange{j+1}{|buf_{n-1}|} \tag{Hbufi}\label{lemma:corr:execute-load-commit:hubufi}
\end{align}
with \(\prog{l_{0}} = \load{\register{x}}{e}\), \(\val{a} = \semexpr{e}{\aplsan(\buf_{n-1}\bufrange{1}{j-1}, \reg_{n-1})}\), and \(\store{\_}{\_} \not\in \buf_{n-1}\bufrange{1}{j-1}\).

Let \(\buf\) be an arbitrary prefix in
\(\prefix(\buf_{n}, \conf{\mem_{n}, \reg_{n}})\). We can consider two cases:
\begin{itemize}
  \item \(|buf| < j\). The proof for this case is similar to the proof for the
        rule \textsc{branch-commit}.
  \item \(|buf| \geq j\). Notice that, by \cref{def:prefixes} and
        \(\buf \in \prefix(\buf_{n}, \conf{\mem_{n}, \reg_{n}})\), there is no
        mispredicted instruction in \(\buf\) (except possibly the last one).
        Therefore, it follows from
        \(\conf{\mem_{n}, \reg_{n}} = \conf{\mem_{n-1}, \reg_{n-1}}\), that
        \(\buf' = \buf\bufrange{1}{j-1} \cdot \mov{x}{\slevel{\val{v}}{\level{s}} @ \val{l_{0}}} \cdot \buf\bufrange{j+1}{|buf|}\)
        belongs to \(\prefix(\buf_{n-1}, \conf{\mem_{n-1}, \reg_{n-1}})\). %
        Hence, from \cref{lemma:corr:IH}, we have that
        \(\deepapl{\conf{\mem_{n-1},\reg_{n-1}}}{\buf'} = \aconf_{\corr{\arun}{\mrun}(n-1)(|\buf'|)}\).
        From \(|\buf'| = |\buf|\),
        \(\conf{\mem_{n}, \reg_{n}} = \conf{\mem_{n-1}, \reg_{n-1}}\), and
        \(\corr{\arun}{\mrun}(n) = \corr{\arun}{\mrun}(n-1)\), we get that
        \(\deepapl{\conf{\mem_{n},\reg_{n}}}{\buf'} = \aconf_{\corr{\arun}{\mrun}(n)(|\buf|)}\).
        Let
        \(\conf{\mem', \reg'} = \deepapl{\conf{\mem_{n},\reg_{n}}}{\buf_{n-1}\bufrange{1}{j-1}}\).
        Notice that from \cref{lemma:deepapl_apl}, \(\reg_{n} = \reg_{n-1}\) and
        \(\val{a} = \semexpr{e}{\aplsan(\buf_{n-1}\bufrange{1}{j-1}, \reg_{n-1})}\)
        we also have that
        \(\semexpr{e}{\reg'}\).
         Moreover, because
        \(\store{\_}{\_} \not\in \buf_{n-1}\bufrange{1}{j-1}\) and \(\mem_{n} = \mem_{n-1}\), we have \(\mem_{n-1} = \mem'\), which implies
        \(\mem_{n-1}(\val{a}) = \mem'(\val{a})\).
        Hence, from this and \cref{def:deepapl}, we get that
        \(\deepapl{\conf{\mem_{n},\reg_{n}}}{\buf'} = \deepapl{\conf{\mem_{n},\reg_{n}}}{\buf}\)
        which concludes
        \(\deepapl{\conf{\mem_{n},\reg_{n}}}{\buf} = \aconf_{\corr{\arun}{\mrun}(n)(|\buf|)}\).
        \lestodo{More details would not hurt.}
\end{itemize}

\newrule{execute-load-rollback} from the evaluation rules (cf.\
\cref{fig:execute-mem-rules}), we have:
\begin{align*}
  \buf_{n-1} =~& \buf_{n-1}\bufrange{1}{j-1} \cdot
                 \mov{x}{\slevel{\val{v}}{\level{s}} @ \val{l_{0}}} \cdot
                 \incrpc{\slevel{l}{s}} @\varepsilon \cdot
                 \buf_{n-1}\bufrange{j+2}{|buf_{n-1}|}\\
  \buf_{n}   =~& \buf_{n-1}\bufrange{1}{j-1} \cdot
                 \mov{x}{\slevel{\mem_{n-1}(\val{a})}{\memsec(\val{a})} @ \varepsilon} \cdot
                 \incrpc{\slevel{l}{s}} @\varepsilon
\end{align*}
with \(\prog{l_{0}} = \load{\register{x}}{e}\),
\(\val{a} = \semexpr{e}{\aplsan(\buf_{n-1}\bufrange{1}{j-1}, \reg_{n-1})}\) and \(\store{\_}{\_} \not\in \buf_{n-1}\bufrange{1}{j-1}\).

Let \(\buf\) be an arbitrary prefix in
\(\prefix(\buf_{n}, \conf{\mem_{n}, \reg_{n}})\). We can consider two cases:
\begin{itemize}
  \item \(|buf| < j\). The proof for this case is similar to the proof for the
        rule \textsc{branch-commit}.
  \item \(|buf| \geq j\). Notice that, by \cref{def:prefixes},
        \cref{lemma:wfbuf} and
        \(\buf \in \prefix(\buf_{n}, \conf{\mem_{n}, \reg_{n}})\), we have
        \(\buf = \buf_{n}\) and there is no mispredicted instruction in
        \(\buf\). Therefore, it follows from
        \(\conf{\mem_{n}, \reg_{n}} = \conf{\mem_{n-1}, \reg_{n-1}}\), that
        \(\buf' = \buf\bufrange{1}{j-1} \cdot \mov{x}{\slevel{\val{v}}{\level{s}} @ \val{l_{0}}} \cdot \incrpc{\slevel{l}{s}} @\varepsilon\)
        belongs to \(\prefix(\buf_{n-1}, \conf{\mem_{n-1}, \reg_{n-1}})\). %
        Hence, from \cref{lemma:corr:IH}, we have that
        \(\deepapl{\conf{\mem_{n-1},\reg_{n-1}}}{\buf'} = \aconf_{\corr{\arun}{\mrun}(n-1)(|\buf'|)}\).
        From \(|\buf'| = |\buf|\),
        \(\conf{\mem_{n}, \reg_{n}} = \conf{\mem_{n-1}, \reg_{n-1}}\), and
        \(\corr{\arun}{\mrun}(n) = \corr{\arun}{\mrun}(n-1)\), we get that
        \(\deepapl{\conf{\mem_{n},\reg_{n}}}{\buf'} = \aconf_{\corr{\arun}{\mrun}(n)(|\buf|)}\).

        From there, it sufficices to show \(\deepapl{\conf{\mem_{n},\reg_{n}}}{\buf'} = \deepapl{\conf{\mem_{n},\reg_{n}}}{\buf}\)
        to get to our goal
        \(\deepapl{\conf{\mem_{n},\reg_{n}}}{\buf} = \aconf_{\corr{\arun}{\mrun}(n)(|\buf|)}\).

        Let
        \(\conf{\mem', \reg'} = \deepapl{\conf{\mem_{n},\reg_{n}}}{\buf_{n-1}\bufrange{1}{j-1}}\).
        By \cref{def:deepapl} and definitions of \(\buf\) and \(\buf'\), we have:
        \(\deepapl{\conf{\mem_{n},\reg_{n}}}{\buf} = \deepapl{\conf{\mem',\reg'}}{\mov{x}{\slevel{\mem_{n-1}(\val{a})}{\memsec(\val{a})} @ \varepsilon} \cdot \incrpc{\slevel{l}{s}} @\varepsilon}\)
        and
        \(\deepapl{\conf{\mem_{n},\reg_{n}}}{\buf'} = \deepapl{\conf{\mem',\reg'}}{\mov{x}{\slevel{\val{v}}{\level{s}} @ \val{l_{0}}} \cdot \incrpc{\slevel{l}{s}} @\varepsilon}\).

        Because \(\store{\_}{\_} \not\in \buf_{n-1}\bufrange{1}{j-1}\) and
        \(\mem_{n} = \mem_{n-1}\), we have \(\mem' = \mem_{n-1}\). From this and
        \cref{def:deepapl}, we get
        \(\deepapl{\conf{\mem_{n},\reg_{n}}}{\buf} = \deepapl{\conf{\mem_{n-1},\reg'\change{\register{x}}{\slevel{\mem_{n-1}(\val{a})}{\memsec(\val{a})}}}}{\incrpc{\slevel{l}{s}} @\varepsilon}\) %
        and %
        \(\deepapl{\conf{\mem_{n},\reg_{n}}}{\buf'} = \deepapl{\conf{\mem_{n-1},\reg'\change{\register{x}}{\slevelnormal{\mem_{n-1}(\semexpr{e}{\reg'})}{\memsec(\semexpr{e}{\reg'})}}}}{\incrpc{\slevel{l}{s}} @\varepsilon}\). %
        Hence, we simply have to show \(\semexpr{e}{\reg'} = \val{a}\), which follows from \cref{lemma:deepapl_apl}, and definition of \(\val{a}\).
\end{itemize}

\newrule{execute-assign}: from the hardware evaluation rules (cf.\
\cref{fig:execute-mem-rules}), we have:
\begin{align*}
  \buf_{n-1} =~& \buf_{n-1}\bufrange{1}{j-1} \cdot
                 \mov{\register{r}}{e} @ T \cdot
                 \buf_{n-1}\bufrange{j+1}{|buf_{n-1}|}\\
  \buf_{n}   =~& \buf_{n-1}\bufrange{1}{j-1} \cdot
                 \mov{\register{r}}{\slevel{\val{v}}{s}} @ T \cdot
                 \buf_{n-1}\bufrange{j+1}{|buf_{n-1}|}
\end{align*}
with \(e \not\in \SVals\) and
\(\slevel{\val{v}}{s} = \semexpr{e}{\apl(\buf_{n-1}\bufrange{1}{j-1}, \reg_{n-1})}\).

Let \(\buf\) be an arbitrary prefix in
\(\prefix(\buf_{n}, \conf{\mem_{n}, \reg_{n}})\). We can consider two cases:
\begin{itemize}
  \item \(|buf| < j\). The proof for this case is similar to the proof for the
        rule \textsc{branch-commit}.
  \item \(|buf| \geq j\). Notice that, by \cref{def:prefixes} and
        \(\buf \in \prefix(\buf_{n}, \conf{\mem_{n}, \reg_{n}})\), there is no
        mispredicted instruction in \(\buf\) (except possibly the last one).
        Therefore, it follows from
        \(\conf{\mem_{n}, \reg_{n}} = \conf{\mem_{n-1}, \reg_{n-1}}\), that
        \(\buf' = \buf\bufrange{1}{j-1} \cdot \mov{\register{r}}{e} @ T \cdot \buf\bufrange{j+1}{|buf|}\)
        belongs to \(\prefix(\buf_{n-1}, \conf{\mem_{n-1}, \reg_{n-1}})\). %
        Hence, from \cref{lemma:corr:IH}, we have that
        \(\deepapl{\conf{\mem_{n-1},\reg_{n-1}}}{\buf'} = \aconf_{\corr{\arun}{\mrun}(n-1)(|\buf'|)}\).
        From this, \(|\buf'| = |\buf|\),
        \(\conf{\mem_{n}, \reg_{n}} = \conf{\mem_{n-1}, \reg_{n-1}}\), and
        \(\corr{\arun}{\mrun}(n) = \corr{\arun}{\mrun}(n-1)\), we get that
        \(\deepapl{\conf{\mem_{n},\reg_{n}}}{\buf'} = \aconf_{\corr{\arun}{\mrun}(n)(|\buf|)}\).
        Let
        \(\conf{\mem', \reg'} = \deepapl{\conf{\mem_{n},\reg_{n}}}{\buf_{n-1}\bufrange{1}{j-1}}\).
        Notice that from \cref{lemma:deepapl_apl}, \(\reg_{n} = \reg_{n-1}\) and
        \(\slevel{\val{v}}{s} = \semexpr{e}{\apl(\buf_{n-1}\bufrange{1}{j-1}, \reg_{n-1})}\),
        we get \(\slevel{\val{v}}{s} = \semexpr{e}{\reg'}\). Moreover, because
        \(\buf_{n-1}\) is well formed (cf.\cref{lemma:wfbuf}), from
        \cref{def:wfbuf} and \(e \in \SVals\), we know that \(T = \varepsilon\).
        Therefore, by \cref{def:deepapl} and rewriting we get
        \begin{align*}
        \deepapl{\conf{\mem_{n},\reg_{n}}}{\buf'} =~ %
        &\deepapl{\conf{\mem', \reg'}}{\mov{\register{r}}{e} @ T \cdot \buf\bufrange{j+1}{|buf|}}        \\
          =~& \deepapl{\conf{\mem', \reg'}}{\mov{\register{r}}{\slevel{\val{v}}{s}} @ \varepsilon \cdot \buf\bufrange{j+1}{|buf|}} \\
          =~& \deepapl{\conf{\mem_{n},\reg_{n}}}{\buf}
        \end{align*}
        which by \(\deepapl{\conf{\mem_{n},\reg_{n}}}{\buf} = \aconf_{\corr{\arun}{\mrun}(n)(|\buf|)}\) concludes \cref{lemma:corr:goal}.
\end{itemize}

\newrule{execute-store}: The proof is similar to the case
\textsc{execute-assign}.

\proofcase{Retire directive.} In this case, from \cref{def:corr},
we have \(\corr{\arun}{\mrun}(n) = \shift(\corr{\arun}{\mrun}(n-1))\). The proof
proceeds by case analysis on the evaluation rule.

\newrule{retire-assign}: In this case, from the hardware evaluation rules in
\cref{fig:retire-rules}, we have
\(\buf_{n-1} = \mov{\register{r}}{\slevel{\val{v}}{s}} @ \varepsilon \cdot \buf_{n}\),
\(\mem_{n} = \mem_{n-1}\) and
\(\reg_{n} = \reg_{n-1}\change{\register{r}}{\slevel{\val{v}}{s}}\). We consider
two cases:
\begin{itemize}
  \item \(\register{r} \neq \pc\). In this case, because \(\buf_{n-1}\) is
        well-formed (cf. \cref{lemma:wfbuf}) and from the definition of
        well-formed buffers (\cref{def:wfbuf}), we have that
        \(\buf_{n} = \incrpc{\slevel{\val{l}}{\botsec}} \cdot \buf_{n}'\).
        Hence, from \cref{def:prefixes}, we have
        \begin{align*}
          \prefix(\buf_{n-1}, \conf{\mem_{n-1}, \reg_{n-1}}) =~& \{\varepsilon \} \cup %
                                  \{\mov{\register{r}}{\slevel{\val{v}}{s}} @ \varepsilon \cdot \pc \gets \slevel{\val{l}}{\botsec} @ \varepsilon \cdot \buf' ~|~\\ %
           &\buf' \in \prefix(\buf_{n}', \deepapl{\conf{\mem_{n-1}, \reg_{n-1}}}{\mov{\register{r}}{\slevel{\val{v}}{s}} @ \varepsilon \cdot \pc \gets \slevel{\val{l}}{\botsec} @ \varepsilon})\}\\
          \prefix(\buf_{n}, \conf{\mem_{n}, \reg_{n}}) =~& \{\varepsilon \} \cup %
                                \{\pc \gets \slevel{\val{l}}{\botsec} @ \varepsilon \cdot \buf' ~|~ %
                                \buf' \in \prefix(\buf_{n}', \deepapl{\conf{\mem_{n}, \reg_{n}}}{\pc \gets \slevel{\val{l}}{\botsec} @ \varepsilon})\}\\
        \end{align*}
        Let \(\buf\) be an arbitrary prefix in \(\buf_{n}\). By \cref{def:prefixes}
        and the definition of \(\buf_{n}\) and \(\buf_{n-1}\), we have that
        \(\mov{\register{r}}{\slevel{\val{v}}{s}} @ \varepsilon \cdot \buf\) is
        a prefix in \(\prefix(\buf_{n-1}, \conf{\mem_{n-1}, \reg_{n-1}})\). Hence, from \cref{lemma:corr:IH}, we have
        \(\deepapl{\conf{\mem_{n-1},\reg_{n-1}}}{\mov{\register{r}}{\slevel{\val{v}}{s}} @ \varepsilon \cdot \buf} = \aconf_{\corr{\arun}{\mrun}(n-1)(|\buf| + 1)}\). Hence, by \cref{def:deepapl}, we get
        \(\deepapl{\conf{\mem_{n-1},\reg_{n-1}\change{\register{r}}{\slevel{\val{v}}{s}}}}{\buf} = \aconf_{\corr{\arun}{\mrun}(n-1)(|\buf|+1)}\).
        By \(\mem_{n} = \mem_{n-1}\) and
        \(\reg_{n} = \reg_{n-1}\change{\register{r}}{\slevel{\val{v}}{s}}\), this gives us \(\deepapl{\conf{\mem_{n},\reg_{n}}}{\buf} = \aconf_{\corr{\arun}{\mrun}(n-1)(|\buf|+1)}\).
        Finally, by the definition of \(\shift\), we get
        \(\deepapl{\conf{\mem_{n},\reg_{n}}}{\buf} = \aconf_{\shift(\corr{\arun}{\mrun}(n-1)(|\buf|))}\), which from
    \(\corr{\arun}{\mrun}(n) = \shift(\corr{\arun}{\mrun}(n-1))\),
    entails
    \(\deepapl{\conf{\mem_{n},\reg_{n}}}{\buf} = \aconf_{\corr{\arun}{\mrun}(n)(|\buf|)}\) and concludes \cref{lemma:corr:goal}.

  \item \(\register{r} = \pc\). In this case, from \cref{def:prefixes}, we have
  \begin{align*}
    \prefix(\buf_{n-1}) =~& \{\varepsilon \} \cup %
    \{\pc \gets \slevel{\val{v}}{s} @ \varepsilon \cdot \buf' ~|~ %
    \buf' \in \prefix(\buf_{n}, \deepapl{\conf{\mem, \reg}}{\pc \gets \slevel{\val{v}}{s} @ \varepsilon})\}\\
  \end{align*}
  Let \(\buf\) be an arbitrary prefix in \(\buf_{n}\). By \cref{def:prefixes}
  and the definition of \(\buf_{n-1}\), we have that
  \(\mov{\pc}{\slevel{\val{v}}{s}} @ \varepsilon \cdot \buf\) is
  a prefix in \(\prefix(\buf_{n-1})\). Hence, from \cref{lemma:corr:IH}, we have
  \(\deepapl{\conf{\mem_{n-1},\reg_{n-1}}}{\mov{\pc}{\slevel{\val{v}}{s}} @ \varepsilon \cdot \buf} = \aconf_{\corr{\arun}{\mrun}(n-1)(|\buf| + 1)}\). Hence, by \cref{def:deepapl}, we get
  \(\deepapl{\conf{\mem_{n-1},\reg_{n-1}\change{\pc}{\slevel{\val{v}}{s}}}}{\buf} = \aconf_{\corr{\arun}{\mrun}(n-1)(|\buf|+1)}\).
  By \(\mem_{n} = \mem_{n-1}\) and
  \(\reg_{n} = \reg_{n-1}\change{\pc}{\slevel{\val{v}}{s}}\), this gives us \(\deepapl{\conf{\mem_{n},\reg_{n}}}{\buf} = \aconf_{\corr{\arun}{\mrun}(n-1)(|\buf|+1)}\).
  Finally, by the definition of \(\shift\), we get
        \(\deepapl{\conf{\mem_{n},\reg_{n}}}{\buf} = \aconf_{\shift(\corr{\arun}{\mrun}(n-1)(|\buf|))}\), which from
    \(\corr{\arun}{\mrun}(n) = \shift(\corr{\arun}{\mrun}(n-1))\),
    entails
    \(\deepapl{\conf{\mem_{n},\reg_{n}}}{\buf} = \aconf_{\corr{\arun}{\mrun}(n)(|\buf|)}\) and concludes \cref{lemma:corr:goal}.
\end{itemize}

\newrule{retire-assign-marked}: The proof for this case is similar to the proof for the rule \textsc{retire-assign}, case \(\register{r} = \pc\).

\newrule{retire-store-low} and \textsc{retire-store-high} The proof for this case is similar to the proof for the rule \textsc{retire-assign}.

\medskip%
\noindent%
We have shown for any evaluation rule in the hardware semantics that if
\cref{lemma:corr:goal} holds at step \(n-1\), then it also holds at step \(i\).
This concludes the proof of \cref{lemma:corr}.
\end{proof}

\subsection{Low-equivalence relations between pairs of executions}\label{sec:proof-low-eq-lemmas}
For the proofs, we refine the low-projection in order to distinguish secret
values in \(\SVals\) from undefined values (\(\botval\))\footnote{In particular,
  this is important to prove \cref{lemma:rloweq_svals} and guarantee that two
  low-equivalent configuration can make progress at the same time}. This refined
low-projection discloses public values and undefined values (\(\botval\)), and
replaces secret with a special symbol \(\dummy\).
\begin{definition}[Refined low-projection]\label{def:rsproj}
  \begin{mathpar}
  \rsproj{\slevel{\val{v}}{\botsec}} = \slevel{\val{v}}{\botsec}\and
  \rsproj{\slevel{\val{v}}{\topsec}} = \dummy\and
  \rsproj{\botval} = \botval
\end{mathpar}
\end{definition}
We let \(\sproj{\reg}\) be the point-wise extension of \(\sproj{\cdot}\) to
register maps.

Similarly, we define a refined low buffer projection that replaces secrets with \(\dummy\) instead of \(\botval\):
\begin{definition}[Refined low buffer projection \(\rsbufproj{\buf}\)]\label{def:rsbufproj}
For all reorder buffer \(\buf\) its refined low-projection \(\rsbufproj{\buf}\) is given by:
  \begin{equation*}
    \begin{array}{rl}
      \rsbufproj{e} = &
                      \begin{cases}
                        \dummy & {\sf if~} e = \slevel{\val{v}}{\topsec} \\
                        e & {\sf otherwise} \\
                      \end{cases}
      \\
      \rsbufproj{\varepsilon} = & \varepsilon \\
      \rsbufproj{\mov{r}{e}} = & \mov{r}{\rsbufproj{e}} \\
      \rsbufproj{\load{x}{e}} = & \load{x}{\rsbufproj{e}} \\
      \rsbufproj{\store{e_{a}}{e_{v}}} = & \store{\rsbufproj{e_{a}}}{\rsbufproj{e_{v}}} \\
      \rsbufproj{inst @ T \cdot \buf} = & \rsbufproj{inst} @ T \cdot \rsbufproj{\buf}
  \end{array}
\end{equation*}
\end{definition}

The following lemma expresses that low-projections and refined
low-projection for values in \(\SVals\) are equivalent.
\begin{lemma}\label{lemma:sproj_rsproj_vals}
  For all pairs of labeled values \(\slevel{v}{s}\) and \(\slevel{v'}{s'}\),
  \begin{equation*}
    \sproj{\slevel{v}{s}} = \sproj{\slevel{v'}{s'}} \iff \rsproj{\slevel{v}{s}} = \rsproj{\slevel{v'}{s'}}
  \end{equation*}
\end{lemma}
\begin{proof}
  The proof directly follows from \cref{def:sproj} and
  \cref{def:rsproj}.
\end{proof}

\begin{corollary}\label{lemma:sproj_rsproj_reg}
  For all register maps \(\reg\) and \(\reg'\) such that
  \(\sproj{\reg} = \sproj{\reg'}\) and that do not contain unresolved values
  (i.e.,\ for all register \register{r}, \(\reg(\register{r}) \in \SVals\)), we
  have that \(\rsproj{\reg} = \rsproj{\reg'}\).
\end{corollary}

The following lemma expresses that equality of refined low-projections
implies and equality of low-projections.
\begin{lemma}\label{lemma:rsproj_sproj}
  For all pairs of (possibly undefined) value \(u, u' \in \SVals \cup \{\botval\}\),
  \begin{equation*}
    \rsproj{u} = \rsproj{u'} \implies \sproj{u} = \sproj{u'}
  \end{equation*}
\end{lemma}
\begin{proof}
  The proof goes by case analysis on \(u\):
  \begin{itemize}
    \item Case \(u = \slevel{v}{\botsec}\): from \cref{def:rsproj} we have
          \(\rsproj{u} = \rsproj{u'} = \slevel{v}{\botsec}\), meaning that
          \(u = u' = \slevel{v}{\botsec}\). Therefore from \cref{def:sproj} we
          have \(\sproj{u} = \sproj{u'} = \slevel{v}{\botsec}\);
    \item Case \(u = \slevel{v}{\topsec}\): from \cref{def:rsproj} we have
          \(\rsproj{u} = \rsproj{u'} = \dummy\), meaning that
          \(u = \slevel{\_}{\topsec}\) and \(u' = \slevel{\_}{\topsec}\). %
          Therefore from \cref{def:sproj} we have
          \(\sproj{u} = \sproj{u'} = \botval\);
    \item Case \(u = \botval\): from \cref{def:rsproj} we have
          \(\rsproj{u} = \rsproj{u'} = \botval\), meaning that
          \(u = u' = \botval\). %
          Therefore from \cref{def:sproj} we have
          \(\sproj{u} = \sproj{u'} = \botval\);
  \end{itemize}
\end{proof}

\begin{corollary}\label{lemma:rsproj_sproj_reg}
  For all register maps \(\reg\) and \(\reg'\):
  \begin{equation*}
    \rsproj{\reg} = \rsproj{\reg'} \implies \sproj{\reg} = \sproj{\reg'}
  \end{equation*}
\end{corollary}
\begin{proof}
  The proof directly follows from \cref{def:rsproj}, \cref{def:sproj}, and
  \cref{lemma:rsproj_sproj}.
\end{proof}

The following lemma expresses that for two reorder buffer, if their refined low-projections are equal, then their low-projections are equal:
\begin{lemma}\label{lemma:rsbufproj_bufsproj_buf}
  For all reorder buffers \(\buf\) and \(\buf'\):
  \begin{equation*}
    \rsbufproj{\buf} = \rsbufproj{\buf'} \implies \sbufproj{\buf} = \sbufproj{\buf'}
  \end{equation*}
\end{lemma}
\begin{proof} Notice that from \(\rsbufproj{\buf} = \rsbufproj{\buf'}\) and
  \cref{def:rsbufproj}, \(\buf\) and \(\buf'\) have the same size. The proof
  goes by induction on the size of \(\buf\).

  \proofcase{Base case} (\(\buf = \buf' = \varepsilon\)) is trivial.

  \proofcase{Inductive case.} Consider that \cref{lemma:rsbufproj_bufsproj_buf}
  holds for reorder buffers of size \(n-1\), \(\buf_{n-1}\) and \(\buf_{n-1}'\).
  We show that \cref{lemma:rsbufproj_bufsproj_buf} still holds for buffers
  of size \(n\):
  \begin{gather}
    \rsbufproj{inst @ T \cdot \buf_{n-1}} = \rsbufproj{inst' @ T' \cdot \buf'_{n-1}} \implies\\
    \sbufproj{inst @ T \cdot \buf_{n-1}} = \sbufproj{inst' @ T \cdot \buf'_{n-1}}
  \end{gather}
  By the induction hypothesis, \cref{def:rsbufproj} and \cref{def:sbufproj}, we
  this amounts to show that
  \begin{equation*}
    \rsbufproj{inst @ T} = \rsbufproj{inst' @ T'} \implies
    \sbufproj{inst @ T} = \sbufproj{inst' @ T}
  \end{equation*}
  The proof continues by case analysis on \(inst\). Because
  \(\rsbufproj{inst @ T}\) is strictly equivalent to \(\sbufproj{inst @ T}\)
  except for expressions, it is sufficient to show that for any expressions \(e, e'\),
  \begin{equation*}
    \rsbufproj{e} = \rsbufproj{e'} \implies
    \sbufproj{e} = \sbufproj{e'}
  \end{equation*}
  The proof proceeds by case analysis of \(e\):
  \begin{itemize}
    \item Case \(e = \slevel{v}{\topsec}\): from \cref{def:rsbufproj}, we have that
          \(\rsbufproj{e} = \dummy\). %
          From the hypothesis \(\rsbufproj{e} = \rsbufproj{e'}\), we also have
          \(\rsbufproj{e'} = \dummy\), meaning that
          \(e' = \slevel{v'}{\topsec}\). Finally, from \cref{def:sbufproj}, we have
          \(\sbufproj{e} = \sbufproj{e'} = \botval\) which concludes our goal.
    \item Otherwise, from \cref{def:rsbufproj}, we have that
          \(\rsbufproj{e} = e\). %
          From the hypothesis \(\rsbufproj{e} = \rsbufproj{e'}\), we also have
          \(\rsbufproj{e'} = e\), meaning that \(e' = e\). Finally, from
          \cref{def:sbufproj}, we have \(\sbufproj{e} = \sbufproj{e'} = e\)
          which concludes our goal.
  \end{itemize}
\end{proof}

The following lemma expresses that for values, the refined low-buffer projection
and the refined low-equivalence relation are equivalent.
\begin{lemma}\label{lemma:val_rsbufproj_rsproj}
  For all values \(\slevel{v}{s}, \slevel{v'}{s'} \in \SVals\):
  \begin{equation*}
    \rsbufproj{\slevel{v}{s}} = \rsbufproj{\slevel{v'}{s'}} \iff \rsproj{\slevel{v}{s}} = \rsproj{\slevel{v'}{s'}}
  \end{equation*}
\end{lemma}
\begin{proof} We consider both sides of the equivalence separately:
  \begin{itemize}
    \item[\(\implies\)] We consider the two cases \(\level{s} = \topsec\) and
          \(\level{s} = \botsec\):
    \begin{itemize}
      \item In case \(\level{s} = \topsec\), from \cref{def:rsbufproj}, we have
            \(\rsbufproj{\slevel{v}{s}} = \dummy\). Therefore, from
            \(\rsbufproj{\slevel{v}{s}} = \rsbufproj{\slevel{v}{\level{s'}}}\), we
            have \(\rsbufproj{\slevel{v'}{\level{s'}}} = \dummy\), meaning that
            \(\level{s'} = \topsec\). Finally,
            \(\rsproj{\slevel{v}{\topsec}} = \dummy\) and
            \(\rsproj{\slevel{v'}{\topsec}} = \dummy\), therefore
            \(\rsproj{\slevel{v}{s}} = \rsproj{\slevel{v'}{\level{s'}}}\).
      \item In case \(\level{s} = \botsec\), from \cref{def:rsbufproj}, we have
            \(\rsbufproj{\slevel{v}{s}} = \slevel{v}{s}\). Therefore, from
            \(\rsbufproj{\slevel{v}{s}} = \rsbufproj{\slevel{v}{\level{s'}}}\), we
            have \(\rsbufproj{\slevel{v'}{\level{s'}}} = \slevel{v}{s}\), meaning
            that \(\slevel{v'}{\level{s'}} = \slevel{v}{s}\), which entails
            \(\rsproj{\slevel{v}{s}} = \rsproj{\slevel{v'}{\level{s'}}}\).
    \end{itemize}
    \item[\(\Longleftarrow\)] We consider the two cases \(\level{s} = \topsec\) and
    \(\level{s} = \botsec\):
    \begin{itemize}
      \item In case \(\level{s} = \topsec\), by \cref{def:rsproj}, we have
            \(\rsproj{\slevel{v}{\topsec}} = \dummy\). Therefore, from
            \(\rsproj{\slevel{v}{s}} = \rsproj{\slevel{v'}{\level{s'}}}\), we have
            \(\rsproj{\slevel{v'}{\level{s'}}} = \dummy\), meaning that
            \(\level{s'} = \topsec\). Finally, by \cref{def:rsbufproj}, we have
            \(\rsbufproj{\slevel{v}{s}} = \rsbufproj{\slevel{v'}{\level{s'}}} = \dummy\).
      \item In case \(\level{s} = \botsec\), by \cref{def:rsproj}, we have
            \(\rsproj{\slevel{v}{\botsec}} = \slevel{v}{\botsec}\). Therefore,
            from \(\rsproj{\slevel{v}{s}} = \rsproj{\slevel{v'}{\level{s'}}}\), we
            have \(\rsproj{\slevel{v'}{\level{s'}}} = \slevel{v}{\botsec}\),
            meaning that \(\val{v} = \val{v'}\) and \(\level{s} = \level{s'}\),
            which entails
            \(\rsbufproj{\slevel{v}{s}} = \rsbufproj{\slevel{v'}{\level{s'}}}\).
    \end{itemize}
  \end{itemize}
\end{proof}

The following lemma expresses that speculations are disclosed in the
refined low-projections of reorder buffers, meaning that two low-equivalent
reorder buffers \(\buf\) and \(\buf'\) correspond both to sequential executions
or both to speculative execution.
\begin{lemma}\label{lemma:speculations}
  For all reorder buffers \(buf\) and \(\buf'\) such that
  \(\rsbufproj{\buf} = \rsbufproj{\buf'}\): %
  \begin{equation*}
  \exists inst @T \in \buf.\ T \neq \varepsilon \iff
  \exists inst' @T' \in \buf'.\ T' \neq \varepsilon
  \end{equation*}
\end{lemma}
\begin{proof}
  Notice that from \cref{def:rsbufproj}, \(\rsbufproj{\buf} = \rsbufproj{\buf'}\),
  implies the pointwise equality of \(\rsbufproj{\buf}\) and
  \(\rsbufproj{\buf'}\).
  Therefore for each \(i^{th}\) instructions \({inst @ T}\) in \(\buf\), the
  corresponding \(i^{th}\) instructions \({inst' @ T'}\) in \(\buf'\) satisfies
  \(\rsbufproj{inst @ T} = \rsbufproj{inst' @ T'}\). Moreover, from
  \cref{def:rsbufproj}, we have that \(T = T'\). Therefore,
  \(T \neq \varepsilon \iff T' \neq \varepsilon\).
\end{proof}

The following proposition expresses that a register that is public in a reorder
buffer \(\buf\) and a register map \(\reg\), is also public in
\(\apl(\buf, \reg)\). %
\begin{lemma}[\(\apl\) preserves security level]\label{lemma:apl_low_var}
  For all register \(\register{r}\), register map \(\reg\) and reorder buffer \(\buf\):
  \begin{gather}
    (\reg(\register{r}) = \slevel{\_}{\botsec} \vee \reg(\register{r}) = \botval) ~\wedge~\tag{Hreg}\label{apl_low_var:h:reg}\\
    (\mov{\register{r}}{\slevel{\_}{s}} @ T \in \buf_{n} \implies \level{s} = \botsec\tag{Hbuf}\label{apl_low_var:h:buf})\\     %
    \implies apl(\buf, \reg)(\register{r}) = \slevel{\_}{\botsec} \vee apl(\buf, \reg)(\register{r}) = \botval \notag
  \end{gather}
\end{lemma}
\begin{proof} The proof that
  \(apl(\buf, \reg)(\register{r}) = \slevel{\_}{\botsec} \vee apl(\buf, \reg)(\register{r}) = \botval\)
  goes by induction on the size of \(\buf\):

  \proofcase{Base case:} \(\buf = \varepsilon\). Consider a register map
  \(\reg\) such that \cref{apl_low_var:h:reg} hold. By \cref{def:apl},
  \(\apl(\varepsilon, \reg) = \reg\). Hence,
  \(apl(\varepsilon, \reg)(\register{r}) = \slevel{\_}{\botsec} \vee apl(\varepsilon, \reg)(\register{r}) = \botval\)
  directly follows from \cref{apl_low_var:h:reg}.

  \proofcase{Inductive case.} Consider that \cref{lemma:apl_low_var} holds for a
  reorder buffer \(\buf_{n-1}\) of size \(n-1\). We show that it still holds for
  \(\buf_{n} = inst @ T \cdot \buf_{n-1}\). That is for a register map \(\reg\)
  such that \cref{apl_low_var:h:reg} hold and \(\buf_{n}\) such that
  \cref{apl_low_var:h:buf} hold, then
  \(apl(\buf_{n}, \reg)(\register{r}) = \slevel{\_}{\botsec} \vee apl(\buf_{n}, \reg)(\register{r}) = \botval\).
  The proof continues by case analysis on \(inst\):
  \begin{itemize}
    \item Case \(inst = \mov{\register{r}}{\slevel{v}{s}}\). From
          \cref{apl_low_var:h:buf}, we have that \(\level{s} = \botsec\). From
          \cref{def:apl}, we have that
          \(\apl(inst @ T \cdot \buf_{n}, \reg) = \apl(\buf_{n-1}, \reg\change{\register{r}}{\slevel{v}{\botsec}})\).
          Because \(\reg\change{\register{r}}{\slevel{v}{\botsec}}\) satisfies
          \cref{apl_low_var:h:reg}, we can apply the induction hypothesis, which
          gives us
          \(\apl(\buf_{n-1}, \reg\change{\register{r}}{\slevel{v}{\botsec}})(\register{r})\)
          equals \(\slevel{\_}{\botsec}\) or \(\botval\), hence
          \(\apl(\buf_{n}, \reg)(\register{r})\) equals \(\slevel{\_}{\botsec}\)
          or \(\botval\).
    \item Case \(inst = \mov{\register{r'}}{\slevel{v}{s}}\) with
          \(\register{r} \neq \register{r'}\). From \cref{def:apl}, we have that
          \(\apl(inst @ T \cdot \buf_{n}, \reg) = \apl(\buf_{n-1}, \reg\change{\register{r'}}{\slevel{v}{s}})\).
          Because \(\reg\change{\register{r'}}{\slevel{v}{s}}\) satisfies
          \cref{apl_low_var:h:reg}, we can apply the induction hypothesis, which
          gives us
          \(\apl(\buf_{n-1}, \reg\change{\register{r}}{\slevel{v}{\botsec}})(\register{r})\)
          equals \(\slevel{\_}{\botsec}\) or \(\botval\), hence
          \(\apl(\buf_{n}, \reg)(\register{r})\) equals \(\slevel{\_}{\botsec}\)
          or \(\botval\).
    \item Case \(inst = \mov{r'}{e}\) where \(e \not\in \SVals\). From
          \cref{def:apl}, we have that
          \(\apl(inst @ T \cdot \buf_{n-1}, \reg) = \apl(\buf_{n-1}, \reg\change{\register{r'}}{\botval})\).
          Because \(\reg\change{\register{r'}}{\botsec}\) satisfies
          \cref{apl_low_var:h:reg}, we can apply the induction hypothesis, which
          gives us
          \(\apl(\buf_{n-1}, \reg\change{\register{r'}}{\slevel{v}{\botsec}})(\register{r})\)
          equals \(\slevel{\_}{\botsec}\) or \(\botval\), hence
          \(\apl(\buf_{n}, \reg)(\register{r})\) equals \(\slevel{\_}{\botsec}\)
          or \(\botval\).
    \item Case \(inst = \load{r'}{\_}\): the proof is similar to the previous
          case.
    \item Case \(inst = \store{\_}{\_}\): from \cref{def:apl}, we have that
          \(\apl(inst @ T \cdot \buf_{n-1}, \reg) = \apl(\buf_{n-1}, \reg)\).
          Hence,
          \(\apl(\buf_{n}, \reg)(\register{r}) = \slevel{\_}{\botsec} \vee \apl(\buf_{n}, \reg)(\register{r}) = \botval\)
          directly follows from the induction hypothesis.
  \end{itemize}
\end{proof}

The following lemma expresses that the function \(\apl\) applied to two
low-equivalent buffers and two low-equivalent register maps returns two low
equivalent register maps:
\begin{lemma}\label{lemma:apl_rloweq}
  For all register maps \(\reg\) and \(\reg'\) and buffers \(\buf\) and
  \(\buf'\): %
  \begin{equation*}
  \rsproj{\reg} = \rsproj{\reg'} \wedge \rsbufproj{\buf} = \rsbufproj{\buf'} \implies
  \rsproj{\apl(\buf, \reg)} = \rsproj{\apl(\buf', \reg')}
  \end{equation*}
\end{lemma}
\begin{proof}
  First, notice that from the hypothesis \(\rsbufproj{\buf} = \rsbufproj{\buf'}\)
  and \cref{def:rsbufproj}, we have, \(|\buf| = |\buf'|\). The proof goes by
  induction on the size of \(\buf\) and \(\buf'\).

  \proofcase{Base case:} \(\buf = \buf' = \varepsilon\). By
  \cref{def:apl}, \(\apl(\varepsilon, \reg) = \reg\) and
  \(\apl(\varepsilon, \reg') = \reg'\). Hence,
  \(\rsproj{\apl(\buf, \reg)} = \rsproj{\apl(\buf', \reg')}\) directly follows
  from the hypothesis \(\rsproj{\reg} = \rsproj{\reg'}\).

  \proofcase{Inductive case.} Consider that \cref{lemma:apl_rloweq} holds for
  reorder buffers of size \(n-1\), \(\buf_{n-1}\) and \(\buf_{n-1}'\), meaning that
  for all register maps \(\reg\), \(\reg'\):
  \begin{equation}
    \rsproj{\reg} = \rsproj{\reg'} \wedge %
    \rsbufproj{\buf_{n-1}} = \rsbufproj{\buf_{n-1}'} \implies
    \rsproj{\apl(\buf_{n-1}, \reg)} = \rsproj{\apl(\buf_{n-1}', \reg')}
    \tag{IH}\label{lemma:apl_loweq:ih}
  \end{equation}
  We show that \cref{lemma:apl_rloweq} still holds for buffers of size \(n\),
  meaning that for all register maps \(\reg\), \(\reg'\):
  \begin{gather}
    \rsproj{\reg} = \rsproj{\reg'} ~\wedge \tag{Hreg}\label{lemma:apl_loweq:h:reg} \\
    \rsbufproj{inst @ T \cdot \buf_{n-1}} = \rsbufproj{inst' @ T \cdot \buf_{n-1}'}
    \implies %
    \tag{Hbufn}\label{lemma:apl_loweq:h:buf} \\
    \rsproj{\apl(inst @ T \cdot \buf_{n-1}, \reg)} = \rsproj{\apl(inst' @ T \cdot \buf_{n-1}, \reg')} \tag{G}\label{lemma:apl_loweq:goal}
  \end{gather}
  From \cref{lemma:apl_loweq:h:buf} and \cref{def:rsbufproj}, we have:
  \begin{equation*}
    \rsbufproj{inst} = \rsbufproj{inst'}
    \tag{Hinst}\label{lemma:apl_loweq:h:instr}
  \end{equation*}
  and
  \begin{equation*}
    \rsbufproj{buf_{n-1}} = \rsbufproj{buf_{n-1}'}
    \tag{Hbufn-1}\label{lemma:apl_loweq:h:bufn}
  \end{equation*}
  The proof continues by case analysis on \(inst\):
  \begin{itemize}
    \item Case \(inst = \store{\_}{\_}\): from \cref{def:rsbufproj}, we have
          that \(\rsbufproj{inst} = \store{\_}{\_}\). %
          From \cref{lemma:apl_loweq:h:instr}, we also have
          \(\rsbufproj{inst'} = \store{\_}{\_}\), meaning that
          \(inst' = \store{\_}{\_}\). %
          From \cref{def:apl}, we have:
          \begin{gather*}
            \apl(inst @ T \cdot \buf_{n-1}, \reg) = apl(\buf_{n-1}, \reg) \text{ and }\\ %
            \apl(inst' @ T \cdot \buf_{n-1}', \reg) = apl(\buf_{n-1}', \reg')
          \end{gather*}
          Therefore, \cref{lemma:apl_loweq:goal} follows from the application of
          \cref{lemma:apl_loweq:ih} with \cref{lemma:apl_loweq:h:reg} and
          \cref{lemma:apl_loweq:h:bufn}.

    \item Case \(inst = \load{x}{\_}\): from \cref{def:rsbufproj}, we have that
          \(\rsbufproj{inst} = \load{x}{\_}\). %
          From \cref{lemma:apl_loweq:h:instr}, we also have
          \(\rsbufproj{inst'} = \load{x}{\_}\), meaning that
          \(inst' = \load{x}{\_}\). %
          From \cref{def:apl}, we have:
          \begin{gather*}
            \apl(inst @ T \cdot \buf_{n-1}, \reg) = apl(\buf_{n-1}, \reg\change{x}{\botval}) \text{ and }\\ %
            \apl(inst' @ T \cdot \buf_{n-1}', \reg) = apl(\buf_{n-1}', \reg'\change{x}{\botval})
          \end{gather*}
          Finally, from \cref{def:rsproj} and \cref{lemma:apl_loweq:h:reg}, we
          have
          \(\rsproj{\reg\change{x}{\botval}} = \rsproj{\reg'\change{x}{\botval}}\).
          Therefore, \cref{lemma:apl_loweq:goal} follows from the application of
          \cref{lemma:apl_loweq:ih} with \cref{lemma:apl_loweq:h:bufn} and
          \(\rsproj{\reg\change{x}{\botval}} = \rsproj{\reg'\change{x}{\botval}}\).

    \item Case \(inst = \mov{r}{e}\) where \(e \not\in \SVals\):
          from \cref{def:rsbufproj}, \(\rsbufproj{inst} = \mov{r}{e}\). From
          \cref{lemma:apl_loweq:h:instr}, we also have
          \(\rsbufproj{inst'} = \mov{r}{e}\), meaning that
          \(inst' = \mov{r}{e}\). From \cref{def:apl}, we have:
          \begin{gather*}
            \apl(inst @ T \cdot \buf_{n-1}, \reg) = apl(\buf_{n-1}, \reg\change{x}{\botval}) \text{ and }\\ %
            \apl(inst' @ T \cdot \buf_{n-1}', \reg) = apl(\buf_{n-1}', \reg'\change{x}{\botval})
          \end{gather*}
          The rest of the proof is similar to case \(inst = \load{x}{\_}\).

    \item Case \(inst = \mov{r}{\slevel{v}{\botsec}}\): from \cref{def:rsbufproj},
          \(\rsbufproj{inst} = \slevel{v}{\botsec}\). From
          \cref{lemma:apl_loweq:h:instr}, we also have
          \(\rsbufproj{inst'} = \mov{r}{\slevel{v}{\botsec}}\), meaning that
          \(inst' = \mov{r}{\slevel{v}{\botsec}}\). From \cref{def:apl}, we have:
          \begin{gather*}
            \apl(inst @ T \cdot \buf_{n-1}, \reg) = apl(\buf_{n-1}, \reg\change{r}{\slevel{v}{\botsec}}) \text{ and }\\ %
            \apl(inst' @ T \cdot \buf_{n-1}', \reg) = apl(\buf_{n-1}', \reg'\change{r}{\slevel{v}{\botsec}})
          \end{gather*}
          Finally, from \cref{def:rsproj} and \cref{lemma:apl_loweq:h:reg}, we
          have
          \(\rsproj{\reg\change{r}{\slevel{v}{\botsec}}} = \rsproj{\reg'\change{r}{\slevel{v}{\botsec}}}\).
          Therefore, \cref{lemma:apl_loweq:goal} follows from the application of
          \cref{lemma:apl_loweq:ih} with \cref{lemma:apl_loweq:h:bufn} and
          \(\rsproj{\reg\change{r}{\slevel{v}{\botsec}}} = \rsproj{\reg'\change{r}{\slevel{v}{\botsec}}}\).

    \item Case \(inst = \mov{r}{\slevel{v}{\topsec}}\): from \cref{def:rsbufproj},
          \(\rsbufproj{inst} = \mov{r}{\dummy}\). From
          \cref{lemma:apl_loweq:h:instr}, we also have
          \(\rsbufproj{inst'} = \mov{r}{\dummy}\), meaning that
          \(inst' = \mov{r}{\slevel{v'}{\topsec}}\). From
          \cref{def:apl}, we have:
          \begin{gather*}
            \apl(inst @ T \cdot \buf_{n-1}, \reg) = apl(\buf_{n-1}, \reg\change{r}{\slevel{v}{\topsec}}) \text{ and }\\ %
            \apl(inst' @ T \cdot \buf_{n-1}', \reg) = apl(\buf_{n-1}', \reg'\change{r}{\slevel{v'}{\topsec}})
          \end{gather*}
          Finally, from \cref{def:rsproj} and \cref{lemma:apl_loweq:h:reg}, we
          have
          \(\rsproj{\reg\change{r}{\slevel{v}{\topsec}}} = \rsproj{\reg'\change{r}{\slevel{v'}{\topsec}}}\).
          Therefore, \cref{lemma:apl_loweq:goal} follows from the application of
          \cref{lemma:apl_loweq:ih} with \cref{lemma:apl_loweq:h:bufn} and
          \(\rsproj{\reg\change{r}{\slevel{v}{\topsec}}} = \rsproj{\reg'\change{r}{\slevel{v'}{\topsec}}}\).
        \end{itemize}
\end{proof}

The following lemma expresses that the function \(\aplsan\) applied to two
low-equivalent buffers and two low-equivalent register maps returns two low
equivalent register maps.
\begin{lemma}\label{lemma:raplsan}
  \(\rsproj{\reg} = \rsproj{\reg'} \wedge \rsbufproj{\buf} = \rsbufproj{\buf'}
  \implies %
  \rsproj{\aplsan(\buf, \reg)} = \rsproj{\aplsan(\buf',\reg')}\)
\end{lemma}
\begin{proof}
  We consider the two cases in \cref{def:aplsan}:
  \begin{itemize}
    \item \proofcase{Sequential execution
          (\(\forall inst @T \in \buf.\ T = \varepsilon\)):} %
          In this case, \(\aplsan(\buf, \reg) = \apl(\buf, \reg)\). %
          From the hypothesis \(\rsbufproj{\buf} = \rsbufproj{\buf'}\) and
          \cref{lemma:speculations}, we also have
          \(\forall inst @T \in \buf'.\ T = \varepsilon\). Therefore,
          \(\aplsan(\buf', \reg') = \apl(\buf', \reg')\). Consequently, to show
          \(\rsproj{\aplsan(\buf, \reg)} = \rsproj{\aplsan(\buf',\reg')}\), we
          have to show \(\rsproj{\apl(\buf, \reg)} = \rsproj{\apl(\buf',\reg')}\).
          This follows from the application of \cref{lemma:apl_rloweq} with
          \(\rsproj{\reg} = \rsproj{\reg'}\) and
          \(\rsbufproj{\buf} = \rsbufproj{\buf'}\).
    \item \proofcase{Speculative execution
          (\(\exists inst @T \in \buf.\ T \neq \varepsilon\)):} In this case,
          \(\aplsan(\buf, \reg) = \sproj{\apl(\buf, \reg)}\). %
          From the hypothesis \(\rsbufproj{\buf} = \rsbufproj{\buf'}\) and
          \cref{lemma:speculations}, we also have
          \(\exists inst @T \in \buf'.\ T \neq \varepsilon\). Therefore,
          \(\aplsan(\buf', \reg') = \sproj{\apl(\buf', \reg')}\). Consequently,
          to show
          \(\rsproj{\aplsan(\buf, \reg)} = \rsproj{\aplsan(\buf',\reg')}\), we
          have to show
          \(\rsproj{\sproj{\apl(\buf, \reg)}} = \rsproj{\sproj{\apl(\buf',\reg')}}\),
          which amounts to show
          \(\sproj{\apl(\buf, \reg)} = \sproj{\apl(\buf',\reg')}\). By
          \cref{lemma:rsproj_sproj}, we have that
          \(\sproj{\apl(\buf, \reg)} = \sproj{\apl(\buf',\reg')}\) follows from
          \(\rsproj{\apl(\buf, \reg)} = \rsproj{\apl(\buf',\reg')}\), which in
          turns follows from the application of \cref{lemma:apl_rloweq} with
          \(\rsproj{\reg} = \rsproj{\reg'}\) and
          \(\rsbufproj{\buf} = \rsbufproj{\buf'}\).
  \end{itemize}
\end{proof}

The following lemma expresses that refined low projection can distinguish values from non values.
\begin{lemma}\label{lemma:rloweq_svals} For all (possibly unresolved) values
  \(u, u' \in \SVals \cup \{\botval\}\) such that \(\rsproj{u} = \rsproj{u'}\). \(u \in \SVals \iff u' \in \SVals\)
\end{lemma}
\begin{proof}
  The proof goes by case analysis on \(u\):
  \begin{itemize}
    \item Case \(u = \slevel{v}{\botsec}\): we have
          \(\rsproj{u} = \rsproj{u'} = \slevel{v}{\botsec}\), therefore
          \(u' = \slevel{v}{\botsec}\), meaning that \(u, u' \in \SVals\);
    \item Case \(u = \slevel{v}{\topsec}\): we have
          \(\rsproj{u} = \rsproj{u'} = \dummy\), therefore
          \(u' = \slevel{v'}{\topsec}\), meaning that \(u, u' \in \SVals\);
    \item Case \(u = \botval\): we have \(\rsproj{u} = \rsproj{u'} = \botval\),
          therefore \(u' = \botval\), meaning that \(u, u' \not\in \SVals\).
  \end{itemize}
\end{proof}
\begin{corollary}\label{lemma:rloweq_svals_regs} %
  For all register maps \(\reg\) and \(\reg'\) such that
  \(\rsproj{\reg} = \rsproj{\reg'}\) and for all register \(\register{r}\).
  \(\reg(\register{r}) \in \SVals \iff \reg'(\register{r}) \in \SVals\)
\end{corollary}
\begin{proof}
  This directly follows from \cref{lemma:rloweq_svals} and the definition of the
  refined low-projection for register maps (\cref{def:rsproj}).
\end{proof}

The following lemma expresses that the evaluation of an expression with two
low-equivalent register maps returns low-equivalent values.
\begin{lemma}\label{lemma:expr_rloweq}
  For all register maps \(\reg\) and \(\reg'\) such that
  \(\rsproj{\reg} = \rsproj{\reg'}\) and for all expression \(e\), %
  either both \(\semexpr{e}{\reg}\) and \(\semexpr{e}{\reg'}\)
  are undefined, or \(\rsproj{\semexpr{e}{\reg}} = \rsproj{\semexpr{e}{\reg'}}\).
\end{lemma}
\begin{proof}
  The proof goes by induction on the expression evaluation rules (cf.\
  \cref{fig:eval_expr}).

  \proofcase{Base cases:}
  \begin{itemize}
    \item Case \(e = \slevel{v}{s}\): \(\semexpr{\slevel{v}{s}}{\reg}\) and
          \(\semexpr{\slevel{v}{s}}{\reg'}\) both evaluate to \(\slevel{v}{s}\),
          therefore
          \(\rsproj{\semexpr{\slevel{v}{s}}{\reg}} = \rsproj{\semexpr{\slevel{v}{s}}{\reg'}}\);
    \item Case \(e = \register{r}\) and \(\reg(\register{r}) \in \SVals\): From
          the hypothesis \(\rsproj{\reg} = \rsproj{\reg'}\) and
          \cref{lemma:rloweq_svals_regs}, we have
          \(\reg'(\register{r}) \in \SVals\). Therefore,
          \(\semexpr{\register{r}}{\reg} = \reg(\register{r})\) and
          \(\semexpr{\register{r}}{\reg'} = \reg'(\register{r})\). From the
          hypothesis \(\rsproj{\reg} = \rsproj{\reg'}\) and \cref{def:rsproj},
          we have \(\rsproj{\reg(\register{r})} = \rsproj{\reg'(\register{r})}\),
          which entails
          \(\rsproj{\semexpr{\register{r}}{\reg}} = \rsproj{\semexpr{\register{r}}{\reg'}}\);
    \item Case \(e = \register{r}\) and \(\reg(\register{r}) \not\in \SVals\):
          From the hypothesis \(\rsproj{\reg} = \rsproj{\reg'}\) and
          \cref{lemma:rloweq_svals_regs}, we have
          \(\reg'(\register{r}) \not\in \SVals\). Therefore, both
          \(\semexpr{\register{r}}{\reg}\) and \(\semexpr{\register{r}}{\reg}\)
          are undefined.
  \end{itemize}

  \proofcase{Inductive case:} Case \(e = e_1 \otimes e_2\). %
  Consider two expressions \(e_{1}\) and \(e_{2}\) such that
  \cref{lemma:expr_rloweq} holds, meaning that:
  \begin{equation}
    \tag{IH}\label{lemma:expr_loweq:ih}
    \begin{aligned}
    &\text{Either both } \semexpr{e_1}{\reg} \text{ and } \semexpr{e_1}{\reg'} %
    \text{ are undefined or }
    \rsproj{\semexpr{e_1}{\reg}} = \rsproj{\semexpr{e_1}{\reg'}} \text{ and} \\ %
    &\text{either both } \semexpr{e_2}{\reg} \text{ and } \semexpr{e_2}{\reg'} %
    \text{ are undefined or }
    \rsproj{\semexpr{e_2}{\reg}} = \rsproj{\semexpr{e_2}{\reg'}} \\ %
    \end{aligned}
  \end{equation}
  To show
  \(\rsproj{\semexpr{e_1 \otimes e_2}{\reg}} = \rsproj{\semexpr{e_1 \otimes e_2}{\reg'}}\),
  we have to consider the following cases:
  \begin{itemize}
    \item Case \(\semexpr{e_1}{\reg}\) or \(\semexpr{e_2}{\reg}\) is undefined.
          In this case, from \cref{lemma:expr_loweq:ih} we also have
          \(\semexpr{e_1}{\reg'}\) or \(\semexpr{e_2}{\reg'}\) is undefined.
          Therefore, both \(\semexpr{e_1 \otimes e_2}{\reg}\) and
          \(\semexpr{e_1 \otimes e_2}{\reg'}\) are undefined.
    \item Case \(\semexpr{e_1}{\reg} = \slevel{v_{1}}{\botsec}\) and
          \(\semexpr{e_2}{\reg} = \slevel{v_{2}}{\botsec}\). In this case, from
          \cref{lemma:expr_loweq:ih} and \cref{def:rsproj}, we also have %
          \(\semexpr{e_1}{\reg'} = \slevel{v_{1}}{\botsec}\) and
          \(\semexpr{e_2}{\reg'} = \slevel{v_{2}}{\botsec}\). %
          Therefore, both \(\semexpr{e_1 \otimes e_2}{\reg}\) and
          \(\semexpr{e_1 \otimes e_2}{\reg'}\) evaluate to
          \(\slevel{v_{1} \otimes v_{2}}{\botsec}\), which entails
          \(\rsproj{\semexpr{e_1 \otimes e_2}{\reg}} = \rsproj{\semexpr{e_1 \otimes e_2}{\reg'}}\).
    \item Case \(\semexpr{e_1}{\reg} = \slevel{v_{1}}{s_{1}}\) and
          \(\semexpr{e_2}{\reg} = \slevel{v_{2}}{s_{2}}\) such that
          \(\level{s_{1}} = \topsec\) or \(\level{s_{2}} = \topsec\). In this
          case, from \cref{lemma:expr_loweq:ih} and \cref{def:rsproj}, we also
          have %
          \(\semexpr{e_1}{\reg'} = \slevel{v_{1}'}{\level{s_{1}}'}\) and
          \(\semexpr{e_2}{\reg'} = \slevel{v_{2}'}{\level{s_{2}}'}\) such that
          \(\level{s_{1}}' = \topsec\) or \(\level{s_{2}}' = \topsec\). %
          Consequently,
          \(\semexpr{e_1 \otimes e_2}{\reg} = \slevel{v_{1} \otimes v_{2}}{\level{s_{1}} \sqcup \level{s_{2}}}\)
          with \(\level{s_{1}} \sqcup \level{s_{2}} = \topsec\) and
          \(\semexpr{e_1 \otimes e_2}{\reg'} = \slevel{v_{1}' \otimes v_{2}'}{\level{s_{1}}' \sqcup \level{s_{2}}'}\)
          with \(\level{s_{1}}' \sqcup \level{s_{2}}' = \topsec\). %
          Finally, because
          \(\rsproj{\slevel{v_{1} \otimes v_{2}}{\topsec}} = \rsproj{\slevel{v_{1}' \otimes v_{2}'}{\topsec}} = \dummy\),
          we have
          \(\rsproj{\semexpr{e_1 \otimes e_2}{\reg}} = \rsproj{\semexpr{e_1 \otimes e_2}{\reg'}}\).
  \end{itemize}
\end{proof}

\subsection{Proof of security for constant-time programs without declassification}\label{sec:proof-ct}
The following hypotheses are used to prove security for constant-time programs
without declassification:
\hypdeterministic*

\begin{restatable}{hypothesis}{hypct}\label{hyp:ct-no-decl} The program is constant-time and does not declassify secret data (according to \cref{def:ct}). In particular, it means that for all configurations
  \(\conf{\reg_{0},\mem_{0}}\) and \(\conf{\reg_{0}',\mem_{0}'}\) such that
  \(\sproj{\reg_{0}} = \sproj{\reg_{0}'}\) and
  \(\sproj{\mem_{0}} = \sproj{\mem_{0}'}\), and for all number of step \(n\),
  \begin{equation*}
    \sarch{n}{\declassified}{\conf{\mem_{0}, \reg_{0}}}{\conf{\mem_{n}, \reg_{n}}}{\obs}%
    \implies%
    \sarch{n}{\declassify{\declassified'}}{\conf{\mem_{0}', \reg_{0}'}}{\conf{\mem_{n}', \reg_{n}'}}{\obs'}%
    \wedge\leaked{\obs} = \leaked{\obs'} \wedge \declassified = \declassify{\declassified'}
  \end{equation*}
\end{restatable}

\newpar%
This is the main lemma:
\begin{lemma}\label{lemma:security0}
  Under \cref{hyp:deterministic,hyp:ct-no-decl}, for all program \(\prog{}\),
  number of steps \(n\), memories \(\mem_{0}\) and \(\mem_{0}'\), register maps
  \(\reg_{0}\) and \(\reg_{0}'\), and microarchitectural context \(\micro_{0}\),
  \begin{gather}
    \smemproj{\mem_{0}} = \smemproj{\mem_{0}'} ~\wedge~ \sproj{\reg_{0}'} = \sproj{\reg_{0}'} ~\wedge~ %
    \smicro{}{n}{\conf{\mem_{0}, \reg_{0}, \varepsilon, \micro_{0}}}{\conf{\mem_{n}, \reg_{n}, \buf_{n},  \micro_{n}}}{} \implies \notag \\
    \smicro{}{n}{\conf{\mem_{0}', \reg_{0}', \varepsilon, \micro_{0}}}{\conf{\mem_{n}', \reg_{n}', \buf_{n}', \micro_{n}'}}{} ~\wedge \tag{Gstepn}\label{security0:stepfull} \\
    \sbufproj{\buf_{n}} = \sbufproj{\buf_{n}'} ~\wedge \tag{Gbuf}\label{security0:buf} \\
    \rsproj{\reg_{n}} = \rsproj{\reg_{n}'} ~\wedge \tag{Greg}\label{security0:reg} \\
    \smemproj{\mem_{n}} = \smemproj{\mem_{n}'} ~\wedge \tag{Gmem}\label{security0:mem} \\
    \micro_{n} = \micro_{n}' \tag{Gmicro}\label{security0:micro} ~\wedge \\
    \corr{\arun}{\mrun}(n) = \corr{\arun'}{\mrun'}(n)\tag{Gcorr}\label{security0:corr} ~\wedge \\
    \forall 0 \leq i \leq |\buf_{n}|.\
    \transient(\conf{\mem_{n}, \reg_{n}, \buf_{n}\bufrange{0}{i}, \_}) = \transient(\conf{\mem_{n}', \reg_{n}', \buf_{n}'\bufrange{0}{i}, \_})\tag{Gtrans}\label{security0:trans}
  \end{gather}
  where \(\arun\) and \(\arun'\) are the longest sequential runs, respectively
  starting from \(\conf{\mem_{0},\reg_{0}}\) and \(\conf{\mem_{0}',\reg_{0}'}\);
  and \(\mrun\) and \(\mrun'\) are the longest \name{} runs, respectively
  starting from \(\conf{\mem_{0}, \reg_{0}, \varepsilon, \micro}\) and
  \(\conf{\mem_{0}', \reg_{0}', \varepsilon, \micro}\).
\end{lemma}

\begin{proof}
  Assume a program \(\prog{}\) satisfying \cref{hyp:ct-no-decl}, memories
  \(\mem_{0}\) and \(\mem_{0}'\), register maps \(\reg_{0}\) and \(\reg_{0}'\),
  and a microarchitectural context \(\micro_{0}\), %
  such that:
  \begin{gather}
    \tag{Hloweq} \smemproj{\mem_{0}} = \smemproj{\mem_{0}'} \text{ and } \sproj{\reg_{0}} = \sproj{\reg_{0}'} \label{security0:h:loweq} \\
    \tag{Hstepn} \smicro{}{n}{\conf{\mem_{0}, \reg_{0}, \varepsilon, \micro_{0}}}{\conf{\mem_{n}, \reg_{n}, \buf_{n}, \micro_{n}}}{} \label{security0:h:stepfull}
  \end{gather}
  Under these hypothesis, we show that \cref{security0:stepfull},
  \cref{security0:buf}, \cref{security0:mem}, \cref{security0:micro}, \cref{security0:corr}, and \cref{security0:trans} The proof goes by induction on the number of steps \(n\).

  \proofcase{Base case \((n = 0)\):}
  \begin{itemize}
    \item[\eqref{security0:stepfull}] \(\smicro{}{0}{\conf{\mem_{0}', \reg_{0}', \buf_{0}', \micro_{0}'}}{\conf{\mem_{0}', \reg_{0}', \buf_{0}', \micro_{0}'}}{}\)
    \item[\eqref{security0:buf}] \(\sbufproj{\buf_{0}} = \sbufproj{\buf_{0}'} = \varepsilon\)
    \item[\eqref{security0:reg}]
          \(\rsproj{\reg_{0}} = \rsproj{\reg_{0}'}\) follows from the
          application of \cref{lemma:sproj_rsproj_reg} with
          \cref{security_patched:h:loweq} and \cref{lemma:reg_sval},
    \item[\eqref{security0:mem}] \(\smemproj{\mem_{0}} = \smemproj{\mem_{0}'}\)
          directly follows from \cref{security0:h:loweq},
    \item[\eqref{security0:micro}] \(\micro_{0} = \micro_{0}'\) because the
          initial microarchitectural context is the same in both initial
          configurations,
    \item[\eqref{security0:corr}]
          \(\corr{\arun}{\mrun}(0) = \corr{\arun'}{\mrun'}(0) = \{0 \mapsto 0\}\) by \cref{def:corr},
    \item[\eqref{security0:trans}]
          \(\forall 0 \leq i \leq |\buf_{0}|.\ \transient(\conf{\mem_{0}, \reg_{0}, \buf_{0}\bufrange{0}{i}, \micro_{0}}) = \transient(\conf{\mem_{0}', \reg_{0}', \buf_{0}'\bufrange{0}{i}, \micro_{0}'})\)
          follows directly by definition of \(\transient\) (cf.
          \cref{def:trans}) and \(\buf_{0} = \buf_{0}' = \varepsilon\).
  \end{itemize}

  \proofcase{Inductive case:}
  We assume that \cref{lemma:security0} holds at step \(n - 1\), meaning that:
  \begin{gather}
    \tag{IHstepn} \smicro{}{n-1}{\conf{\mem_{0}', \reg_{0}', \varepsilon, \micro_{0}}}{\conf{\mem_{n-1}', \reg_{n-1}', \buf_{n-1}', \micro_{n-1}'}}{} \label{security0:ih:stepfull} \\
    \tag{IHbuf} \rsbufproj{\buf_{n-1}} = \rsbufproj{\buf_{n-1}'} \label{security0:ih:buf} \\
    \tag{IHreg} \rsproj{\reg_{n-1}} = \rsproj{\reg_{n-1}'} \label{security0:ih:reg} \\
    \tag{IHmem} \smemproj{\mem_{n-1}} = \smemproj{\mem_{n-1}'} \label{security0:ih:mem} \\
    \tag{IHmicro} \micro_{n-1} = \micro_{n-1}' \label{security0:ih:micro} \\
    \tag{IHcorr} \corr{\arun}{\mrun}(n-1) = \corr{\arun'}{\mrun'}(n-1) \label{security0:ih:corr} \\
    \forall 0 \leq i \leq |\buf_{n-1}|.\
    \transient(\conf{\mem_{n-1}, \reg_{n-1}, \buf_{n-1}\bufrange{0}{i}, \_}) = \transient(\conf{\mem_{n-1}', \reg_{n-1}', \buf_{n-1}'\bufrange{0}{i}, \_})\tag{IHtrans}\label{security0:ih:trans}
  \end{gather}
  Under these hypotheses, we show that \cref{lemma:security0} still holds at
  step \(n\).

  \proofcase{Simplify goal \cref{security0:stepfull}:}
  Notice that by \cref{def:n-step} and \cref{security0:h:stepfull}, we have that:
  \begin{equation}
    \smicro{}{}{\conf{\mem_{n-1}, \reg_{n-1}, \buf_{n-1}, \micro_{n-1}}}{\conf{\mem_{n}, \reg_{n}, \buf_{n}, \micro_{n}}}{} \tag{IHstep}\label{security0:ih:step}
  \end{equation}
  which together with \cref{security0:h:loweq}, gives us that
  \cref{security0:ih:stepfull}, \cref{security0:ih:buf},
  \cref{security0:ih:reg} \cref{security0:ih:mem} and
  \cref{security0:ih:micro} hold at step \(n-1\).
  Additionally, from \cref{def:n-step} and \cref{security0:ih:stepfull}, proving
  \cref{security0:stepfull} amounts to show:
  \begin{equation}
    \smicro{}{}{\conf{\mem_{n-1}', \reg_{n-1}', \buf_{n-1}', \micro_{n-1}'}}{\conf{\mem_{n}', \reg_{n}', \buf_{n}', \micro_{n}'}}{} \tag{Gstep}\label{security0:step}
  \end{equation}
  Consequently, we have to show that \cref{security0:step},
  \cref{security0:buf}, \cref{security0:reg} \cref{security0:mem},
  \cref{security0:micro}, \cref{security0:corr} and
  \cref{security0:trans} hold at step \(n\).

  \proofcase{Transform \cref{security0:ih:corr}:}
  First, observe that because the semantics is deterministic (cf
  \cref{hyp:deterministic}), we have that:
  \begin{gather}
    \mrun_{n-1} = \conf{\mem_{n-1}, \reg_{n-1}, \buf_{n-1}, \micro_{n-1}} \text{ and } %
    \mrun_{n-1}' = \conf{\mem_{n-1}', \reg_{n-1}', \buf_{n-1}', \micro_{n-1}'} %
    \tag{H$\mrun_{n-1}$}\label{security0:HmrunA}
  \end{gather}
  From this, \cref{lemma:corr} and \cref{security0:ih:corr}, we have that for all
  \(\buf \in \prefix(\buf_{n-1}, \conf{\mem_{n-1},\reg_{n-1}})\) and
  \(\buf' \in \prefix(\buf_{n-1}', \conf{\mem_{n-1}',\reg_{n-1}'})\) %
  such that %
  \(|\buf| = |\buf'|\),
  \begin{equation}
    \deepapl{\conf{\mem_{n-1},\reg_{n-1}}}{\buf} = \aconf_{j} %
    \text{ and } \deepapl{\conf{\mem_{n-1}',\reg_{n-1}'}}{\buf} = \aconf_{j}' %
    \text{ where } j = \corr{\arun}{\mrun}(n-1)(|\buf|) \tag{IHarch}\label{security0:ih:arch} %
  \end{equation}

  \proofcase{Step rule:} The step rule updates the microarchitectural context such
  that:
  \begin{align*}
    \micro_{aux} &\mydef \update(\micro_{n-1}, \smemproj{\mem_{n-1}}, \sproj{\reg_{n-1}}, \sbufproj{\buf_{n-1}}) \\
    \micro_{aux}' &\mydef \update(\micro_{n-1}', \smemproj{\mem_{n-1}'}, \sproj{\reg_{n-1}'}, \sbufproj{\buf_{n-1}'})
  \end{align*}
  We start by showing that \(\micro_{aux} = \micro_{aux}'\). Because \(\update\)
  is deterministic, this follows from:
  \begin{itemize}
    \item \(\micro_{n-1} = \micro_{n-1}'\), which directly follows from \cref{security0:ih:micro},
    \item \(\smemproj{\mem_{n-1}} = \smemproj{\mem_{n-1}'}\), which directly follows from \cref{security0:ih:mem},
    \item \(\sproj{\reg_{n-1}} = \sproj{\reg_{n-1}'}\), which follows from
          \cref{security0:ih:reg}, \cref{lemma:sproj_rsproj_reg} and \cref{lemma:reg_sval},
    \item \(\sbufproj{\buf_{n-1}} = \sbufproj{\buf_{n-1}'}\), which follows from
          \cref{security0:ih:buf} and \cref{lemma:rsbufproj_bufsproj_buf}.
  \end{itemize}

  Next, the rule chooses a directive and applies the corresponding evaluation
  rule on the current configuration. By \cref{security0:ih:step}, in the first
  configuration, the rule selects a directive
  \(\directive \mydef\nextdir(\micro_{aux})\) and updates the current
  configuration such that:
  \begin{equation*}
    \smicro{\directive}{}{\conf{\mem_{n-1}, \reg_{n-1}, \buf_{n-1}, \micro_{aux}}}{\conf{\mem_{n}, \reg_{n}, \buf_{n}, \micro_{n}}}{}
  \end{equation*}
  We show that in the second configuration, the \textsc{step} rule also applies
  the same directive \(\directive\).

  In the second configuration, the next directive to apply is
  \(\directive' \mydef \nextdir(\micro_{aux}')\). By
  \(\micro_{aux} = \micro_{aux}'\) and because \(\nextdir\) is deterministic (from
  \cref{hyp:deterministic}), we have \(\directive = \directive'\) which means
  that both configurations evaluate the same directive at step \(n-1\).

  \proofcase{Proof of \cref{security0:corr}:} Observe that to show
  \cref{security0:corr}, we have to show that
  \(\corr{\arun}{\mrun}(n) = \corr{\arun'}{\mrun'}(n)\). By
  \cref{security0:HmrunA} and \cref{def:corr}, this follows from
  \cref{security0:ih:corr} and the following conditions:
  \begin{enumerate}
    \item \(\nextdir(\micro_{aux}) = \nextdir(\micro_{aux}')\), which we have
          just shown;
    \item
          \(\transient(\conf{\mem_{n-1}, \reg_{n-1}, \buf_{n-1}, \micro_{n-1}}) \iff \transient(\conf{\mem_{n-1}', \reg_{n-1}', \buf_{n-1}', \micro_{n-1}'})\),
          which directly follows from \(|\buf_{n-1}| = |\buf_{n-1}'|\) (by \cref{def:rsbufproj} and \cref{security0:ih:buf}) and
          \cref{security0:ih:trans};
    \item In the case \(\directive = \fetch\), we additionally have to show
          \[\neg \transient(\conf{\mem_{n-1}, \reg_{n-1}, \buf_{n-1}, \_}) \implies \lstpc(\conf{\mem_{n-1}, \reg_{n-1}, \buf_{n-1}, \micro_{n-1}}) = \lstpc(\conf{\mem_{n-1}', \reg_{n-1}', \buf_{n-1}', \_})\]
          In other words, assuming
          \(\neg \transient(\conf{\mem_{n-1}, \reg_{n-1}, \buf_{n-1}, \_})\),
          we need to show that \(\reg_{seq}(\pc) = \reg_{seq}'(\pc)\) where
          \(\conf{\cdot, \reg_{seq}} = \deepapl{\conf{\mem_{n-1}, \reg_{n-1}}}{\buf_{n-1}}\)
          and
          \(\conf{\cdot, \reg_{seq}'} = \deepapl{\conf{\mem_{n-1}', \reg_{n-1}'}}{\buf_{n-1}'}\).
          From \cref{lemma:deepapl_apl}, we can instead show that
          \(\apl(\buf_{n-1},\reg_{n-1})(\pc) = \apl(\buf_{n-1}',\reg_{n-1}')(\pc)\).
          From the application of \cref{lemma:apl_rloweq} with
          \cref{security0:ih:reg,security0:ih:buf}, we have
          \(\rsproj{\apl(\buf_{n-1}, \reg_{n-1})} = \rsproj{\apl(\buf_{n-1}',\reg_{n-1}')}\),
          meaning that
          \(\rsproj{\apl(\buf_{n-1},\reg_{n-1})(\pc)} = \rsproj{\apl(\buf_{n-1}',\reg_{n-1}')(\pc)}\)
          Finally, because the security level of \(\pc\) is \(\botsec\) (cf.
          \cref{lemma:pc_low}) we have from \cref{lemma:apl_low_var} that the
          security level of \(\apl(\buf_{n-1}, \reg_{n-1})(\pc)\) and
          \(\apl(\buf_{n-1}', \reg_{n-1}')(\pc)\) is also \(\botsec\), which by
          \(\rsproj{\apl(\buf_{n-1},\reg_{n-1})(\pc)} = \rsproj{\apl(\buf_{n-1}',\reg_{n-1}')(\pc)}\)
          means
          \(\apl(\buf_{n-1},\reg_{n-1})(\pc) = \apl(\buf_{n-1}',\reg_{n-1}')(\pc)\).
          This concludes the proof that \(\reg_{seq}(\pc) = \reg_{seq}'(\pc)\).
  \end{enumerate}
  This concludes the proof of \cref{security0:corr}.

  \medskip
  Knowing that the directive that is applied is the same in both configurations,
  we can proceed to show the remaining goals by case analysis on the directive
  \(\directive\). We consider the hardware evaluation rules for the directives
  \(\fetch\), \(\execute{i}\), and \(\retire\). For simplicity, we rename
  \(\micro_{aux}\) and \(\micro_{aux}'\) as \(\micro_{n-1}\) and
  \(\micro_{n-1}'\) respectively and, because \(\micro_{aux} = \micro_{aux}'\),
  we let \cref{security0:ih:micro} denote the equality of these renamed
  architectural states.

  \proofcase{Fetch directive.} All \textsc{fetch} rules (cf.\
  \cref{fig:fetch-rules}), start by evaluating the value of \(\pc\) to get the
  current location using respectively \(\apl(\buf_{n-1}, \reg_{n-1})\) and
  \(\apl(\buf_{n-1}', \reg_{n-1}')\). From the application of
  \cref{lemma:apl_rloweq} with \cref{security0:ih:reg,security0:ih:buf}, we have:
  \begin{equation}
    \tag{Hapl}\label{security0:fetch:apl_rloweq}
    \rsproj{\apl(\buf_{n-1}, \reg_{n-1})} = \rsproj{\apl(\buf_{n-1}',\reg_{n-1}')}
  \end{equation}

  The goal now is to show that both executions evaluate \(\pc\) to the same
  value. From \cref{security0:ih:step}, we know that the evaluation of the
  expressions do not get stuck in the first configuration. Hence from
  \cref{lemma:expr_rloweq}, and \cref{security0:fetch:apl_rloweq}, we know that it
  is not stuck in the second configuration either. Hence the values of \(\pc\)
  in both configurations is respectively given by:
  \begin{equation*}
    \slevel{l}{s} \mydef \semexpr{\pc}{\apl(\buf_{n-1}, \reg_{n-1})} \text{ and } %
    \slevel{l'}{s'} \mydef \semexpr{\pc}{\apl(\buf_{n-1}', \reg_{n-1}')}
  \end{equation*}
  From the expression evaluation rules (cf. \cref{fig:eval_expr}), we have:
  \begin{equation*}
    \slevel{l}{s} = \apl(\buf_{n-1}, \reg_{n-1})(\pc) \text{ and }
    \slevel{l'}{s'} = \apl(\buf_{n-1}', \reg_{n-1}')(\pc)
  \end{equation*}

  Additionally, from \cref{security0:fetch:apl_rloweq}, we have
  \(\rsproj{\slevel{l}{s}} = \rsproj{\slevel{l'}{s'}}\). %
  Finally, from \cref{lemma:pc_low} and \cref{lemma:apl_low_var}, we have
  \(\level{s} = \level{s'} = \botsec\), meaning that \(\val{l} = \val{l'}\).

  \newgoal{security0:step} %
  In all \textsc{fetch} rules the only hypothesis that can block the execution
  is if the evaluation of expression is stuck. However, we have shown that, from
  \cref{security0:fetch:apl_rloweq,lemma:expr_rloweq,security0:ih:step}, the
  expression evaluation in the second rule is not stuck, which entails
  \cref{security0:step}.

  \newgoal{security0:reg}, \cref{security0:mem}, and \cref{security0:micro}: %
  Because the register map, the memory, and the microarchitectural context are not
  updated in any of the \textsc{fetch} rules, \cref{{security0:reg}},
  \cref{security0:mem}, and \cref{security0:micro} directly follow from
  \cref{security0:ih:reg}, \cref{security0:ih:mem}, and
  \cref{security0:ih:micro}.

  \newgoal{security0:buf}, \cref{security0:trans}: %
  Because both configurations evaluate \(\pc\) to the same value \(\val{l}\) and
  fetch the same instruction (i.e.,\ the instruction at \(\prog{l}\)), then they
  either both evaluate \textsc{fetch-predict-branch-jmp} or they both evaluate
  \textsc{fetch-others}. %
  We show the remaining goals \cref{security0:buf} and \cref{security0:trans} separately for each \textsc{fetch} rule:

  \newrule{fetch-predict-branch-jmp}: An instruction that moves control to the
  next target is added to the buffer such that:
  \begin{equation*}
    \buf_{n} = \buf_{n-1} \cdot \mov{\pc}{\slevel{l_{n}}{\botsec}} @ \val{l} \text{ and }
    \buf_{n}' = \buf_{n-1}' \cdot \mov{\pc}{\slevel{l_{n}'}{\botsec}} @ \val{l'}
  \end{equation*}
  where %
  \(\val{l_{n}} = \predict(\micro_{n-1}) \text{ and } %
  \val{l_{n}'} = \predict(\micro_{n-1})\).

  To show \cref{security0:buf}, we have to show that \(\sbufproj{\buf_{n}} = \sbufproj{\buf_{n}'}\), which
  from \cref{def:sbufproj} and \cref{security0:ih:buf}, amounts to that
  \(\sbufproj{\mov{\pc}{\slevel{\val{l_{n}}}{\botsec}} @ \val{l}} = \sbufproj{\mov{\pc}{\slevel{\val{l_{n}'}}{\botsec}} @ \val{l'}}\).
  We have already shown that \(\val{l} = \val{l'}\). Hence it just remains to
  show \(\val{l_{n}} = \val{l_{n}'}\) which directly follows from
  \cref{security0:ih:micro} and \cref{hyp:deterministic}.

  To show \cref{security0:trans}, by \cref{security0:ih:trans}, the definition
  of \(\transient\) in \cref{def:trans}, and the definition of \(\buf_{n}\) and
  \(\buf_{n}'\), we simply need to show
  \(\transient(\conf{\mem_{n}, \reg_{n}, \buf_{n}, \micro_{n}}) \iff \transient(\conf{\mem_{n}', \reg_{n}', \buf_{n}', \micro_{n}'})\).
  We first consider the case where
  \(\transient(\conf{\mem_{n-1}, \reg_{n-1}, \buf_{n-1}, \_}) = true\) and then
  the case where
  \(\transient(\conf{\mem_{n-1}, \reg_{n-1}, \buf_{n-1}, \_}) = false\).
  \begin{itemize}
    \item In the case,
          \(\transient(\conf{\mem_{n-1}, \reg_{n-1}, \buf_{n-1}, \_}) = true\),
          by \cref{security0:ih:trans} we also have
          \(\transient(\conf{\mem_{n-1}', \reg_{n-1}', \buf_{n-1}', \_}) = true\).
          By definition of \(\transient\) (cf. \cref{def:corr}), \(\buf_{n}\)
          and \(\buf_{n}'\) we also have \(\transient(\conf{\mem_{n}, \reg_{n}, \buf_{n}, \_}) = \transient(\conf{\mem_{n}', \reg_{n}', \buf_{n}', \_}) = true\);
    \item In the case,
          \(\transient(\conf{\mem_{n-1}, \reg_{n-1}, \buf_{n-1}, \_}) = false\),
          by \cref{security0:ih:trans} we also have
          \(\transient(\conf{\mem_{n-1}', \reg_{n-1}', \buf_{n-1}', \_}) = false\).
          By definition of \(\transient\) (cf. \cref{def:trans}), \(\buf_{n}\)
          and \(\buf_{n}'\) and by \(\mem_{n-1} = \mem_{n}\), \(\mem_{n-1}' = \mem_{n}'\) and \(\reg_{n-1} = \reg_{n}\), \(\reg_{n-1}' = \reg_{n}'\) we have to show: %
          \[\goodpred(\mov{\pc}{\slevel{l_{n}}{\botsec}} @ \val{l}, %
          \deepapl{\conf{\mem_{n-1}, \reg_{n-1}}}{\buf_{n-1}}) \iff %
          \goodpred(\mov{\pc}{\slevel{l_{n}}{\botsec}} @ \val{l}, %
          \deepapl{\conf{\mem_{n-1}', \reg_{n-1}'}}{\buf_{n-1}'})\] %
          We focus on the case where the fetched instruction
          \(\prog{\val{l}} = \beqz{e}{\val{l_{br}}}\) (the case
          \(\prog{\val{l}} = \jmp{e}\) is similar). %
          From the definition of \(\goodpred\) (cf.\ \cref{def:trans}), this
          amounts to showing
          \(\semexpr{e}{r_{seq}} = 0 \iff \semexpr{e}{r_{seq}'} = 0\) where
          \(r_{seq} = \deepapl{\conf{\mem_{n-1}, \reg_{n-1}}}{\buf_{n-1}}\) and
          \(r_{seq}' = \deepapl{\conf{\mem_{n-1}', \reg_{n-1}'}}{\buf_{n-1}'}\).
          Notice that because there are no mispredicted instructions in
          \(\buf_{n-1}\) and \(\buf_{n-1}'\) and because reorder buffers are
          well-formed (\cref{lemma:wfbuf}), we have
          \(\buf_{n-1} \in \prefix(\buf_{n-1}')\) and
          \(\buf_{n-1}' \in \prefix(\buf_{n-1}')\). By \cref{security0:ih:arch},
          and because \(|\buf_{n-1}| = |\buf_{n-1}'|\) we get that
          \(\conf{\_, \reg_{seq}} = \aconf_{j} %
          \text{ and } \conf{\_, \reg_{seq}'} = \aconf_{j}'\)
          where \(j = \corr{\arun}{\mrun}(n-1)(|\buf_{n-1}|)\). %

          From \cref{lemma:deepapl_apl_expr}, and the definition of \(\val{l}\),
          we know that
          \(\semexpr{\pc}{\reg_{seq}} = \semexpr{\pc}{\reg_{seq}'} = \val{l}\).
          Hence, because \(\prog{l} = \beqz{e}{\val{l_{br}}}\), we know that we
          can apply the \textsc{branch} rule from the sequential semantics
          (\cref{fig:sequential-semantics}) to \(\aconf_{j}\) and
          \(\aconf_{j}'\). %
          Consequently, we have \(\sarch{}{}{\aconf_{j}}{\_}{\obs}\) and
          \(\sarch{}{}{\aconf_{j}'}{\_}{\obs'}\) where
          \(\obs = (\semexpr{e}{\reg_{seq}} = 0)\) and
          \(\obs' = (\semexpr{e}{\reg_{seq}} = 0)\). Moreover, because the
          program is constant-time \cref{hyp:ct-no-decl}, we have \(\obs = \obs'\),
          which concludes
          \(\semexpr{e}{r_{seq}} = 0 \iff \semexpr{e}{r_{seq}'} = 0\) and, in
          turn,
          \(\transient(\conf{\mem_{n}, \reg_{n}, \buf_{n}}) \iff \transient(\conf{\mem_{n}', \reg_{n}', \buf_{n}'})\).
  \end{itemize}

  \newrule{fetch-others}: In the rule \textsc{fetch-others}, two instructions
  are added to the buffer:
  \begin{gather*}
    \buf_{n} = \buf_{n-1} \cdot \prog{l} @ \varepsilon \cdot \mov{\pc}{\slevel{\val{l_{n}}}{\botsec}} @\varepsilon\\
    \buf_{n}' = \buf_{n-1}' \cdot \prog{l'} @ \varepsilon \cdot \mov{\pc}{\slevel{\val{l_{n}'}}{\botsec}} @\varepsilon
  \end{gather*}
  where \(\val{l_{n}} = \semexpr{\pc+1}{apl(\reg_{n-1}, buf_{n-1})}\) and
  \(\val{l_{n}'} = \semexpr{\pc+1}{apl(\reg_{n-1}', buf_{n-1}')}\). We have
  already shown that \(\val{l} = \val{l'}\). In the same way, we can show that
  \(\val{l_{n}} = \val{l_{n}'}\). Consequently, the same instructions are added
  to the reorder buffers in both executions. %
  Therefore, \cref{security0:buf} follows from \cref{def:rsbufproj} and
  \cref{security0:ih:buf}. %
  Additionally, because the register map and memory are not modified and because
  instructions added to the buffer have tag \(\varepsilon\),
  \cref{security0:trans}, follows directly from \cref{security0:ih:trans}.

  \medskip\noindent%
  This concludes the proof for the \(\fetch\) directive.

  \proofcase{Execute \(i\).} From \cref{security0:ih:step}, we can apply an
  \textsc{execute} directive with an index \(i\) in the first configuration.
  Consequently, we have that: %
  \begin{equation*}
    \buf_{n-1} = \buf_{a} \cdot inst @ T \cdot \buf_{b} \text{ such that }
    |\buf_{a}| = i-1
  \end{equation*}
  From \cref{security0:ih:buf,def:rsbufproj}, there is also a \(i^{th}\)
  instruction \(inst'\) in \(\buf_{n-1}\) such that:
  \(\buf_{n-1}' = \buf_{a}' \cdot inst' @ T' \cdot \buf_{b}'\). Additionally we have:
  \begin{gather*}
    \tag{Hbufa}\label{security0:execute:bufa}
    \rsbufproj{\buf_{a}} = \rsbufproj{\buf_{a}'}\\
    \tag{Hinst}\label{security0:execute:instr}
    \rsbufproj{inst @ T} = \rsbufproj{inst' @ T'}\\
    \tag{Hbufb}\label{security0:execute:bufb}
    \rsbufproj{\buf_{b}} = \rsbufproj{\buf_{b}'}
  \end{gather*}
  From the application of \cref{lemma:raplsan} with
  \cref{security0:execute:bufa}, \cref{security0:ih:reg} we also have:
  \begin{equation}
    \rsproj{\aplsan(\buf_{a}, \reg_{n-1})} = \rsproj{\aplsan(\buf_{a}', \reg_{n-1}')}\tag{Haplsan}\label{security0:execute:aplsan}
  \end{equation}
  and from the application of \cref{lemma:apl_rloweq} with
  \cref{security0:execute:bufa}, \cref{security0:ih:reg} we have:
  \begin{equation}
    \rsproj{\apl(\buf_{a}, \reg_{n-1})} = \rsproj{\apl(\buf_{a}', \reg_{n-1}')}\tag{Hapl}\label{security0:execute:apl}
  \end{equation}

  In all the \textsc{execute} rules, expressions are evaluated using
  \(\aplsan(\buf_{a}, \reg_{n-1})\) or \(\apl(\buf_{a}, \reg_{n-1})\) in the
  first execution or \(\aplsan(\buf_{a}', \reg_{n-1}')\) or
  \(\apl(\buf_{a}', \reg_{n-1}')\). From \cref{security0:ih:step}, we know that
  the evaluation of expressions is not stuck in the first execution. Therefore,
  from the application of \cref{lemma:expr_rloweq} with
  \cref{security0:execute:aplsan} and \cref{security0:execute:apl}, we know
  that:
  \begin{equation}
    \tag{Hexpr}\label{security0:execute:expr_stuck}
    \text{The evaluation of expressions does not get stuck in the second
      configuration.}
  \end{equation}
  In all the \textsc{execute} rules, the memory and the reorder buffer are not
  modified. Therefore, \cref{security0:reg,security0:mem} directly follow from
  induction hypotheses \cref{security0:ih:reg,security0:ih:mem}. %
  We show the remaining goals \cref{security0:step}, \cref{security0:buf},
  \cref{security0:micro}, and \cref{security0:trans} for each \textsc{execute}
  rule.

  \newrule{branch-commit} and \textsc{branch-rollback}: From the hypotheses of
  the rule, the instruction \(inst @ T\) that is executed in the first
  configuration is \(\mov{\pc}{\slevel{l}{\_}} @ \val{l_{0}}\). From
  \cref{security0:execute:instr} and \cref{def:rsbufproj}, we have that
  \(inst' @ T' = \mov{\pc}{\slevel{l'}{\_}} @ \val{l_{0}}\). %
  Additionally, from \cref{lemma:pc_low} and \cref{security0:execute:instr}, we
  have
  \begin{equation}
    \tag{Hloc}\label{security0:branch-commit:location}
    \val{l} = \val{l'}
  \end{equation}
  Moreover, considering that the first execution evaluates either
  \textsc{branch-commit} or \textsc{branch-rollback}, we have
  \(\prog{l_0} = \beqz{e}{\val{l_{br}}}\). Therefore, the second configuration
  can only evaluate the rules \textsc{branch-commit} or
  \textsc{branch-rollback}.

  The rule computes conditions \(\val{c}\) and \(\val{c'}\) such that:
  \begin{equation*}
    \slevel{c}{\_} \mydef \semexpr{e}{\aplsan(\buf_{a}, \reg_{n-1})} \text{ and } %
    \slevel{c'}{\_} \mydef \semexpr{e}{\aplsan(\buf_{a}', \reg_{n-1}')}
  \end{equation*}
  We now show that \((\val{c} = 0) \iff (\val{c'} = 0)\). We first consider the case in which both
  configurations are in sequential execution; and then the case in which both
  configurations are in speculative execution (note that from
  \cref{security0:execute:bufa} and \cref{lemma:speculations}, other cases are
  impossible):
  \begin{enumerate}
    \item Both configurations are in sequential execution. In this case, because
          there are no mispredicted instructions in \(\buf_{a}\) and
          \(\buf_{a}'\) and because reorder buffers are well-formed
          (\cref{lemma:wfbuf}), we have \(\buf_{a} \in \prefix(\buf_{n-1}')\)
          and \(\buf_{a}' \in \prefix(\buf_{n-1}')\). Let
          \(\conf{\mem_{a}, \reg_{a}} = \deepapl{\conf{\mem_{n-1},\reg_{n-1}}}{\buf_{a}}\)
          and
          \(\conf{\mem_{a}', \reg_{a}'} = \deepapl{\conf{\mem_{n-1}',\reg_{n-1}'}}{\buf_{a}'}\).
          By \cref{security0:ih:arch}, and because
          \(|\buf_{a}| = |\buf_{a}'| = i-1\) we get that
          \(\conf{\mem_{a}, \reg_{a}} = \aconf_{j} %
          \text{ and } \conf{\mem_{a}', \reg_{a}'} = \aconf_{j}' %
          \text{ where } j = \corr{\arun}{\mrun}(n-1)(i-1)\). %

          From \cref{def:wfbuf}, we know that
          \(\semexpr{\pc}{\reg_{a}} = \semexpr{\pc}{\reg_{a}'} = \val{l_0}\). Hence,
          because \(\prog{l_{0}} = \beqz{e}{\val{l_{br}}}\), we know that we can
          apply the \textsc{branch} rule from the sequential semantics
          (\cref{fig:sequential-semantics}) to \(\aconf_{j}\) and
          \(\aconf_{j}'\). %
          Consequently, we have \(\sarch{}{}{\aconf_{j}}{\_}{\obs}\) and
          \(\sarch{}{}{\aconf_{j}'}{\_}{\obs'}\) where
          \(\obs = (\semexpr{e}{\reg_{a}} = 0)\) and
          \(\obs' = (\semexpr{e}{\reg_{a}} = 0)\). Moreover, because the program
          is constant-time \cref{hyp:ct-no-decl}, we have \(\obs = \obs'\).
          By definition of \(\val{c}\) and \(\val{c'}\) and by
          \cref{lemma:deepapl_aplsan_expr}, we have that
          \(\slevel{c}{\_} = \semexpr{e}{\reg_{a}}\) and %
          \(\slevel{c'}{\_} = \semexpr{e}{\reg_{a}'}\). Hence, by
          \(\obs = \obs'\), we get \((\val{c} = 0) \iff (\val{c' = 0})\).
    \item Both configurations are in speculative execution. In this case, from
    \cref{def:aplsan}, we have:
    \begin{gather*} %
      \slevel{\val{c}}{\_} = \semexpr{e}{\sproj{\apl(\buf_{a}, \reg_{n-1})}} \text{ and } %
      \slevel{\val{c'}}{\_} = \semexpr{e}{\sproj{\apl(\buf_{a}', \reg_{n-1}')}}
    \end{gather*} %
          From the application of \cref{lemma:rsproj_sproj_reg} with
          \cref{security0:execute:apl}, we have
          \(\sproj{\apl(\buf_{a}, \reg_{n-1})} = \sproj{\apl(\buf_{a}', \reg_{n-1}')}\),
          which entails \(\val{c} = \val{c'}\) and concludes
          \((\val{c} = 0) \iff (\val{c'} = 0)\).
  \end{enumerate}

  Then, the rule computes the next target, which from
  \((\val{c} = 0) \iff (\val{c'} = 0)\), is the same in both executions:
  \begin{equation*}
    \val{l_{next}} \mydef \ite{\val{c} = 0}{\val{l_{br}}}{\val{l_0} + 1} \text{ and } %
  \end{equation*}
  From \cref{security0:branch-commit:location}, this also means that the
  condition \(\val{l_{next}} = \val{l}\), which selects whether
  \textsc{branch-commit} or \textsc{branch-rollback} is applied, has the same
  value in both executions. Hence either both executions evaluate the rule
  \textsc{branch-commit} or both executions evaluate the rule
  \textsc{branch-rollback}.

  \begin{itemize}
    \item[\cref{security0:micro}] The microarchitectural context is not modified
          by the rules so \cref{security0:micro} directly follows from
          \cref{security0:ih:micro}.

    \item[\cref{security0:step}] From \cref{security0:execute:expr_stuck}, we
          have that the evaluation of expressions is not stuck. The hypotheses
          \(\val{l_{next}} = \val{l}\) and \(\val{l_{next}} \neq \val{l}\) do
          not block the execution but just selects which rule is applied: if
          \(\val{l_{next}} = \val{l}\) the rule \textsc{branch-commit} is
          applied, otherwise the rule \textsc{branch-rollback} is applied. Hence
          in both cases a rule can be applied, which entails
          \cref{security0:step}.

    \item[\cref{security0:buf}] and \cref{security0:trans}. We first consider
          the case where both executions evaluate the rule
          \textsc{branch-commit} and then the case where both execution evaluate
          the rule \textsc{branch-rollback} (we have shown before that other
          cases are not possible).
          \begin{itemize}
            \item Case \textsc{branch-commit}: %
            By the hypothesis of the rule and
            \cref{security0:branch-commit:location}, the reorder buffers
            are updated such that:
            \begin{gather*}
              \buf_{n} = \buf_{a} \cdot \mov{\pc}{\slevel{\val{l}}{\botsec}} @ \varepsilon \cdot \buf_{b} \text{ and }
              \buf_{n}' = \buf_{a}' \cdot \mov{\pc}{\slevel{\val{l}}{\botsec}} @ \varepsilon \cdot \buf_{b}'
            \end{gather*}
            Therefore, \cref{security0:buf} follows from \cref{def:rsbufproj},
            \cref{security0:execute:bufa}, \cref{security0:execute:bufb} and
            from
            \(\rsbufproj{\mov{\pc}{\slevel{\val{l}}{\botsec}} @ \varepsilon} = \rsbufproj{\mov{\pc}{\slevel{\val{l}}{\botsec}} @ \varepsilon}\).

                  Because \(\mem_{n-1} = \mem_{n}\), \(\mem_{n-1}' = \mem_{n}'\)
                  and \(\reg_{n-1} = \reg_{n}\) \(\mem_{n-1} = \mem_{n-1}'\),
                  showing \cref{security0:trans} amounts to showing
                  \begin{equation}
                    \forall 0 \leq j \leq |\buf_{n-1}|.\ \transient(\conf{\mem_{n-1}, \reg_{n-1}, \buf_{n}\bufrange{0}{j}, \_}) = \transient(\conf{\mem_{n-1}', \reg_{n-1}', \buf_{n}\bufrange{0}{j}, \_})
                    \tag{G}\label{security0:branch-commit:trans}
                  \end{equation}
                  For \(0 \leq j < i\), \cref{security0:branch-commit:trans}
                  follows directly from \cref{security0:ih:trans} and the
                  definition of \(\buf_{n}\) and \(\buf_{n}'\). %
                  For \(j = i\), \cref{security0:branch-commit:trans} follows
                  from the fact that \cref{security0:branch-commit:trans} holds
                  for \(0 \leq j < i\) and by the fact that both
                  \(\buf_{n}\bufindex{i}\) and \(\buf_{n}\bufindex{i}\) have tag
                  \(\varepsilon\). %
                  Finally, for \(i < j \leq |\buf_{n}|\), first notice that by the definition of \(\buf_{n}\) and \(\buf_{n}'\), by
                  \cref{def:deepapl} and because the predictions for
                  \(\buf_{n-1}\bufindex{i}\) and \(\buf_{n-1}'\bufindex{i}\)
                  were correct, we have:
                  \begin{align*}
\deepapl{\conf{\mem_{n-1}, \reg_{n-1}}}{\buf_{n}\bufrange{0}{i}} =&
\deepapl{\conf{\mem_{n-1}, \reg_{n-1}}}{\buf_{n-1}\bufrange{0}{i}} \text{ and }\\
\deepapl{\conf{\mem_{n-1}', \reg_{n-1}'}}{\buf_{n}'\bufrange{0}{i}} =&
\deepapl{\conf{\mem_{n-1}', \reg_{n-1}'}}{\buf_{n-1}'\bufrange{0}{i}}
                  \end{align*}
                  Hence, \cref{security0:branch-commit:trans} follows from \cref{security0:ih:trans} and from the fact
                  that \cref{security0:branch-commit:trans} holds for
                  \(0 \leq j \leq i\). \lestodo{Be more precise if time allows.}

            \item Case \textsc{branch-rollback}: By the hypothesis of the rule,
                  the reorder buffers are updated such that:
            \begin{gather*}
              \buf_{n} = \buf_{a} \cdot \mov{\pc}{\slevel{\val{l_{next}}}{\botsec}} @ \varepsilon \text{ and }
              \buf_{n}' = \buf_{a}' \cdot \mov{\pc}{\slevel{\val{l_{next}}}{\botsec}} @ \varepsilon
            \end{gather*}
                  Therefore, \cref{security0:buf} follows from
                  \cref{def:rsbufproj}, \cref{security0:execute:bufa} and from
                  \(\rsbufproj{\mov{\pc}{\slevel{\val{l_{next}}}{\botsec}} @ \varepsilon} = \rsbufproj{\mov{\pc}{\slevel{\val{l_{next}}}{\botsec}} @ \varepsilon}\).

                  The proof for \cref{security0:trans} is similar to the case \textsc{branch-commit}.
          \end{itemize}
  \end{itemize}

  \newrule{jmp-commit} and \textsc{jmp-rollback}: The proof is similar to the case \textsc{branch-commit} and \textsc{branch-rollback}. %

  \newrule{execute-assign}: From the hypotheses of the rule, the instruction
  \(inst @ T\) that is executed in the first configuration is
  \(inst = \mov{r}{e} @ T\) where \(e \not\in \SVals\). %
  From \cref{security0:execute:instr} and \cref{def:rsbufproj}, we have that
  \(inst' @ T' = \mov{r}{e} @ T\). Therefore, the second configuration can only
  evaluate the rule \textsc{execute-assign}.

  The rule evaluates the expression \(e\) to values \(\val{v}\) (in the first
  configuration) and \(\val{v'}\) (in the second configuration) such that:
  \begin{equation*}
    \slevel{v}{s} \mydef \semexpr{e}{\apl(\buf_{a}, \reg_{n-1})} \text{ and } \slevel{v'}{s'} \mydef \semexpr{e}{\apl(\buf_{a}', \reg_{n-1}')}
  \end{equation*}
  From \cref{security0:execute:apl}, and \cref{lemma:expr_rloweq} we have that
  \(\rsproj{\slevel{v}{s}} = \rsproj{\slevel{v}{s'}}\).

  \begin{itemize}
    \item[\cref{security0:micro}] The microarchitectural context is not modified
          by the rule so \cref{security0:micro} directly follows from
          \cref{security0:ih:micro}.

    \item[\cref{security0:step}] The first hypothesis that can block the
          execution in the second configuration is \(e \not\in \SVals\). We have
          show that \(e\) is the same in both executions and from
          \cref{security0:ih:step} we have \(e \not\in \SVals\). %
          The second hypothesis that can block the execution in the second
          configuration is if the evaluation of the expression gets stuck. From
          \cref{security0:execute:expr_stuck}, we know that it does not get
          stuck. Hence the rule \textsc{execute-assign} can also be applied in
          the second configuration, which entails \cref{security0:step}.

    \item[\cref{security0:buf}] The reorder buffers are updated such that:
          \begin{equation*}
            \buf_{n} = \buf_{a} \cdot \mov{r}{\slevel{v}{s}} \cdot \buf_{b} \text{
              and
            } \buf_{n}' = \buf_{a}' \cdot \mov{r}{\slevel{v'}{s'}} \cdot \buf_{b}'
          \end{equation*}
          From \cref{security0:execute:bufa}, \cref{security0:execute:bufb}, and
          \cref{def:rsproj}, we only need to show
          \(\rsbufproj{\mov{r}{\slevel{v}{s}}} =
          \rsbufproj{\mov{r}{\slevel{v'}{s'}}}\), %
          which amounts to show
          \(\rsbufproj{\slevel{v}{s}} = \rsbufproj{\slevel{v'}{s'}}\). This
          directly follows from the application of
          \cref{lemma:val_rsbufproj_rsproj} with
          \(\rsproj{\slevel{v}{s}} = \rsproj{\slevel{v}{s'}}\).

    \item[\cref{security0:trans}] Because \(\mem_{n-1} = \mem_{n}\),
          \(\mem_{n-1}' = \mem_{n}'\) and \(\reg_{n-1} = \reg_{n}\)
          \(\mem_{n-1} = \mem_{n-1}'\), showing \cref{security0:trans} amounts
          to showing
          \begin{equation}
            \forall 0 \leq j \leq |\buf_{n-1}|.\ \transient(\conf{\mem_{n-1}, \reg_{n-1}, \buf_{n}\bufrange{0}{j}, \_}) = \transient(\conf{\mem_{n-1}', \reg_{n-1}', \buf_{n}\bufrange{0}{j}, \_})
            \tag{G}\label{security0:assign:trans}
          \end{equation}
          Notice that from by the definition of \(\buf_{n}\) and \(\buf_{n}'\),
          and by \cref{def:deepapl}, we have:
\begin{align*}
\deepapl{\conf{\mem_{n-1}, \reg_{n-1}}}{\buf_{n}\bufrange{0}{i}} =&
\deepapl{\conf{\mem_{n-1}, \reg_{n-1}}}{\buf_{n-1}\bufrange{0}{i}} \text{ and }\\
\deepapl{\conf{\mem_{n-1}', \reg_{n-1}'}}{\buf_{n}'\bufrange{0}{i}} =&
\deepapl{\conf{\mem_{n-1}', \reg_{n-1}'}}{\buf_{n-1}'\bufrange{0}{i}}
\end{align*}
          Hence, because no other instructions are modified in the ROB,
          \cref{security0:assign:trans} follows from
          \cref{security0:ih:trans}. \lestodo{be more precise if time allows}.
\end{itemize}

\newrule{execute-load-predict}: From the hypotheses of the rule,
the instruction \(inst @ T\) that is executed in the first configuration is
\(inst @ T = \load{x}{e} @ T\). %
From \cref{security0:execute:instr} and \cref{def:rsbufproj}, we have that
\(inst' @ T' = \load{x}{e} @ T\). Therefore, the second configuration can only
evaluate the rule \textsc{execute-load-predict}.

  \begin{itemize}
    \item[\cref{security0:micro}] The microarchitectural context is not modified
          by the rule so \cref{security0:micro} directly follows from
          \cref{security0:ih:micro}.

    \item[\cref{security0:step}] From \cref{security0:execute:expr_stuck}, we
          have that the evaluation of expressions is not stuck. Because no other
          hypothesis can block the evaluation of the rule, it can also be
          applied in the second configuration, which entails
          \cref{security0:step}.

    \item[\cref{security0:buf}] The reorder buffers are updated such that:
          \begin{equation*}
            \buf_{n} = \buf_{a} \cdot \mov{x}{\slevel{v}{\botsec}} @ \val{l} \cdot \buf_{b} \text{
              and
            } \buf_{n}' = \buf_{a}' \cdot \mov{x}{\slevel{v'}{\botsec}} @ \val{l'}  \cdot \buf_{b}'
          \end{equation*}
          where
          \begin{gather*}
            \val{v}  = \predict(\micro_{n-1}) \text{ and } \val{v'} = \predict(\micro_{n-1}') \\
            \slevel{\val{l}}{s} = \semexpr{\pc}{\apl(\buf_{a}, \reg_{n-1})} \text{ and
            } \slevel{\val{l'}}{s'} = \semexpr{\pc}{\apl(\buf_{a}', \reg_{n-1}')}
          \end{gather*}
          From \cref{security0:execute:bufa}, \cref{security0:execute:bufb}, and
          \cref{def:rsbufproj}, we only need to show
          \(\rsbufproj{\mov{x}{\slevel{v}{\botsec}} @ \val{l}} = \rsbufproj{\mov{x}{\slevel{v'}{\botsec}} @ \val{l'}}\)
          which amounts to showing\(\val{v} = \val{v'}\) and
          \(\val{l} = \val{l'}\).

          \begin{itemize}
            \item \(\val{v} = \val{v'}\) follows from the fact that \(\predict\)
                  is deterministic (cf. \cref{hyp:deterministic}) and
                  \cref{security0:ih:micro}.
            \item From the expression evaluation rules (cf.
                  \cref{fig:eval_expr}), we have:
                  \begin{equation*}
                    \slevel{l}{s} = \apl(\buf_{a}, \reg_{n-1})(\pc) \text{ and }
                    \slevel{l'}{s'} = \apl(\buf_{a}', \reg_{n-1}')(\pc)
                  \end{equation*}
                  Additionally, from \cref{security0:execute:apl} we have
                  \(\rsproj{\slevel{l}{s}} = \rsproj{\slevel{l'}{s'}}\).
                  Finally, from \cref{lemma:pc_low} and
                  \cref{lemma:apl_low_var}, we have
                  \(\level{s} = \level{s'} = \botsec\), meaning that
                  \(\val{l} = \val{l'}\).
            \end{itemize}
    \item[\cref{security0:trans}] Because \(\mem_{n-1} = \mem_{n}\),
          \(\mem_{n-1}' = \mem_{n}'\) and \(\reg_{n-1} = \reg_{n}\)
          \(\mem_{n-1} = \mem_{n-1}'\), and by definition of \(\transient\)
          (cf.\ \cref{def:trans}) showing \cref{security0:trans} amounts to
          showing for all \(j\) such that \(0 \leq j \leq |\buf_{n-1}|\),
          \begin{equation}
            \forall 0 \leq j \leq |\buf_{n-1}|.\ \transient(\conf{\mem_{n-1}, \reg_{n-1}, \buf_{n}\bufrange{0}{j}, \_}) = \transient(\conf{\mem_{n-1}', \reg_{n-1}', \buf_{n}\bufrange{0}{j}, \_})
            \tag{G}\label{security0:load-predict:trans}
          \end{equation}
          For \(0 \leq j < i\), \cref{security0:load-predict:trans} follows
          directly from
          \(\buf_{n}\bufrange{0}{i-1} = \buf_{n-1}\bufrange{0}{i-1}\),
          \(\buf_{n}'\bufrange{0}{i-1} = \buf_{n-1}'\bufrange{0}{i-1}\) and
          \cref{security0:ih:trans}. %
          For \(j \geq i\), we consider two cases separately. First the case
          where
          \(\transient(\mem_{n-1}, \reg_{n-1}, \buf_{n}\bufrange{0}{i-1})\)
          and second, the case where
          \(\neg\transient(\mem_{n-1}, \reg_{n-1}, \buf_{n}\bufrange{0}{i-1})\).
          \begin{itemize}
            \item In case
                  \(\transient(\mem_{n-1}, \reg_{n-1}, \buf_{n}\bufrange{0}{i-1})\),
                  we also have
                  \(\transient(\mem_{n-1}', \reg_{n-1}', \buf_{n}'\bufrange{0}{i-1})\)
                  from
                  \(\buf_{n}\bufrange{0}{i-1} = \buf_{n-1}\bufrange{0}{i-1}\),
                  \(\buf_{n}'\bufrange{0}{i-1} = \buf_{n-1}'\bufrange{0}{i-1}\)
                  and \cref{security0:ih:trans}. This in turns implies
                  \(\transient(\mem_{n-1}, \reg_{n-1}, \buf_{n}\bufrange{0}{j})\)
                  and
                  \(\transient(\mem_{n-1}, \reg_{n-1}, \buf_{n}\bufrange{0}{j})\),
                  which concludes \cref{security0:load-predict:trans};
            \item In case
                  \(\neg \transient(\mem_{n-1}, \reg_{n-1}, \buf_{n}\bufrange{0}{i-1})\),
                  we also have
                  \(\neg \transient(\mem_{n-1}', \reg_{n-1}', \buf_{n}'\bufrange{0}{i-1})\)
                  from
                  \(\buf_{n}\bufrange{0}{i-1} = \buf_{n-1}\bufrange{0}{i-1}\),
                  \(\buf_{n}'\bufrange{0}{i-1} = \buf_{n-1}'\bufrange{0}{i-1}\)
                  and \cref{security0:ih:trans}.

                  Let us first focus on \(j = i\). By definition of
                  \(\transient\) (cf.\ \cref{def:trans}), and because
                  \cref{security0:load-predict:trans} holds for \(j < i\), we
                  simply have to show:
\[\goodpred(\mov{\pc}{\slevel{l_{n}}{\botsec}} @ \val{l}, %
  \deepapl{\conf{\mem_{n-1}, \reg_{n-1}}}{\buf_{n-1}}) \iff %
  \goodpred(\mov{\pc}{\slevel{l_{n}}{\botsec}} @ \val{l}, %
  \deepapl{\conf{\mem_{n-1}', \reg_{n-1}'}}{\buf_{n-1}'})\]
          From the definition of \(\goodpred\) (cf.\ \cref{def:trans}), this
          amounts to showing
\[\semexpr{e}{\reg_{j-1}} = \val{a} %
  \wedge \memsec(a) = \botsec %
  \wedge \mem(\val{a}) = \val{v} \iff
  \semexpr{e}{\reg_{j-1}} = \val{a'} %
  \wedge \memsec(a') = \botsec %
  \wedge \mem'(\val{a'}) = \val{v}\]
                  with
                  \(\reg_{j-1} = \deepapl{\conf{\mem_{n-1}, \reg_{n-1}}}{\buf_{n}\bufrange{0}{j-1}}\)
                  and
                  \(\reg_{j-1}' = \deepapl{\conf{\mem_{n-1}', \reg_{n-1}'}}{\buf_{n}'\bufrange{0}{j-1}}\).
                  Notice that because there are no mispredicted instructions in
                  \(\buf_{n}\bufrange{0}{j-1}\) and
                  \(\buf_{n}'\bufrange{0}{j-1}\) and because reorder buffers are
                  well-formed (\cref{lemma:wfbuf}), we have
                  \(\buf_{n}\bufrange{0}{j-1} \in \prefix(\buf_{n-1}')\) and
                  \(\buf_{n}'\bufrange{0}{j-1} \in \prefix(\buf_{n-1}')\). %
                  By \cref{security0:ih:arch}, and because
                  \(|\buf_{n}\bufrange{0}{j-1}| = |\buf_{n}'\bufrange{0}{j-1}| = j - 1\)
                  we get that \(\conf{\mem_{n-1}, \reg_{n-1}} = \aconf_{j} %
                  \text{ and } \conf{\mem_{n-1}', \reg_{n-1}'} = \aconf_{j}' %
                  \text{ where } j = \corr{\arun}{\mrun}(n-1)(j-1)\). %

                  From \cref{lemma:deepapl_apl_expr}, and the definition of
                  \(\val{l}\), we know that
                  \(\semexpr{\pc}{\reg_{j-1}} = \semexpr{\pc}{\reg_{j-1}'} = \val{l}\).
                  Hence, because \(\prog{l} = \load{x}{e}\), we know that we can
                  apply the \textsc{load} rule from the sequential semantics
                  (\cref{fig:sequential-semantics}) to \(\aconf_{j}\) and
                  \(\aconf_{j}'\). %
                  Consequently, we have \(\sarch{}{}{\aconf_{j}}{\_}{\obs}\) and
                  \(\sarch{}{}{\aconf_{j}'}{\_}{\obs'}\) where \(\obs = \val{a}\)
                  and \(\obs' = \val{a'}\). Moreover, because the program is
                  constant-time \cref{hyp:ct-no-decl}, we have \(\val{a} = \val{a'}\).
                  This, in turns, implies
                  \(\memsec(a) = \botsec \iff \memsec(a') = \botsec\) and if
                  \(\memsec(a) = \botsec\), we get
                  \(\mem(\val{a}) = \mem'(\val{a'})\) from
                  \cref{security0:ih:mem} which concludes
                  \(\mem(\val{a}) = \val{v} \iff \mem'(\val{a'}) = \val{v}\).
                  This concludes \cref{security0:load-predict:trans} for \(j = i\).

                  For \(i < j \leq |\buf_{n}|\), we get by the definition of
                  \(\buf_{n}\) and \(\buf_{n}'\), and by \cref{def:deepapl} that
                  \begin{align*}
\deepapl{\conf{\mem_{n-1}, \reg_{n-1}}}{\buf_{n}\bufrange{0}{i}} =&
\deepapl{\conf{\mem_{n-1}, \reg_{n-1}}}{\buf_{n-1}\bufrange{0}{i}} \text{ and }\\
\deepapl{\conf{\mem_{n-1}', \reg_{n-1}'}}{\buf_{n}'\bufrange{0}{i}} =&
\deepapl{\conf{\mem_{n-1}', \reg_{n-1}'}}{\buf_{n-1}'\bufrange{0}{i}}
                  \end{align*}
                  Hence, \cref{security0:load-predict:trans} follows from \cref{security0:ih:trans} and from the fact
                  that \cref{security0:load-predict:trans} holds for
                  \(0 \leq j \leq i\).
          \end{itemize}
  \end{itemize}

  \newrule{execute-load-commit} and \textsc{execute-load-rollback}: From the
  hypotheses of the rule, the instruction that is executed in the first
  configuration is \(inst = \mov{x}{\slevel{\val{v}}{\_}} @ \val{l_{0}}\). From
  \cref{security0:ih:buf} and \cref{def:rsbufproj}, we have that
  \(inst = \mov{x}{\slevel{\val{v'}}{\_}} @ \val{l_{0}}\). Therefore, the second
  configuration can only evaluate one of the rules \textsc{execute-load-commit}
  or \textsc{execute-load-rollback}.

  From the hypothesis \(\prog{\val{l_{0}}} = \load{x}{e}\), both rules evaluate
  the expression \(e\) to values \(\val{a}\) (in the first configuration) and
  \(\val{a'}\) (in the second configuration) such that:
  \begin{equation*}
    \slevel{a}{s} \mydef \semexpr{e}{\aplsan(\buf_{a}, \reg_{n-1})} \text{ and } %
    \slevel{a'}{s'} \mydef \semexpr{e}{\aplsan(\buf_{a}', \reg_{n-1}')}
  \end{equation*}

  We now show that the index of the load is the same in both configurations:
  \begin{equation}
    \val{a} = \val{a'}\tag{Ha}\label{security0:execute-load-commit:a}
  \end{equation}
  To show this, we first consider the case in which both
  configurations are in sequential execution; and then the case in which both
  configurations are in speculative execution (note that from
  \cref{security0:execute:bufa} and \cref{lemma:speculations}, other cases are
  impossible):
  \begin{enumerate}
    \item Both configurations are in sequential execution. In this case, because
          there are no mispredicted instructions in \(\buf_{a}\) and
          \(\buf_{a}'\) and because reorder buffers are well-formed
          (\cref{lemma:wfbuf}), we have \(\buf_{a} \in \prefix(\buf_{n-1}')\)
          and \(\buf_{a}' \in \prefix(\buf_{n-1}')\). Let
          \(\conf{\mem_{a}, \reg_{a}} = \deepapl{\conf{\mem_{n-1},\reg_{n-1}}}{\buf_{a}}\)
          and
          \(\conf{\mem_{a}', \reg_{a}'} = \deepapl{\conf{\mem_{n-1}',\reg_{n-1}'}}{\buf_{a}'}\).
          By \cref{security0:ih:arch}, and because
          \(|\buf_{a}| = |\buf_{a}'| = i-1\) we get that
          \(\conf{\mem_{a}, \reg_{a}} = \aconf_{j} %
          \text{ and } \conf{\mem_{a}', \reg_{a}'} = \aconf_{j}' %
          \text{ where } j = \corr{\arun}{\mrun}(n-1)(i-1)\). %

          From \cref{def:wfbuf}, we know that
          \(\semexpr{\pc}{\reg_{a}} = \semexpr{\pc}{\reg_{a}'} = l_0\). Hence,
          because \(\prog{l_{0}} = \load{\register{x}}{e}\), we know that we can
          apply the \textsc{load} rule from the sequential semantics
          (\cref{fig:sequential-semantics}) to \(\aconf_{j}\) and
          \(\aconf_{j}'\). %
          Consequently, we have \(\sarch{}{}{\aconf_{j}}{\_}{\obs}\) and
          \(\sarch{}{}{\aconf_{j}'}{\_}{\obs'}\) where
          \(\obs = \semexpr{e}{\reg_{a}}\) and
          \(\obs' = \semexpr{e}{\reg_{a}}\). Moreover, because the program
          is constant-time \cref{hyp:ct-no-decl}, we have \(\obs = \obs'\).
          By definition of \(\val{a}\) and \(\val{a'}\) and by
          \cref{lemma:deepapl_aplsan_expr}, we have that
          \(\slevel{a}{\_} = \semexpr{e}{\reg_{a}}\) and %
          \(\slevel{a'}{\_} = \semexpr{e}{\reg_{a}'}\). Hence, by
          \(\obs = \obs'\), we get \(\val{a} = \val{a'}\).
    \item Both configurations are in speculative execution: from
          \cref{def:aplsan}, we have
          \(\slevel{a}{s} = \semexpr{e}{\sproj{\apl(\buf_{a}, \reg_{n-1})}} \text{
          and
          } \slevel{a}{s'} = \semexpr{e}{\sproj{\apl(\buf_{a}', \reg_{n-1}')}}\).
          From the application of \cref{lemma:rsproj_sproj_reg} with
          \cref{security0:execute:apl}, we have
          \(\sproj{\apl(\buf_{a}, \reg_{n-1})} = \sproj{\apl(\buf_{a}', \reg_{n-1}')}\),
          which concludes \(\val{a} = \val{a'}\).
  \end{enumerate}
  \smallskip
  \begin{itemize}
    \item[\cref{security0:step}] From \cref{security0:execute:expr_stuck}, we
          have that the evaluation of expressions is not stuck. %
          Additionally, from \cref{security0:ih:step}, we have
          \(\store{\_}{\_} \not\in \buf_{a}\), and from
          \cref{security0:execute:bufa} and \cref{def:rsbufproj}, we have
          \(\store{\_}{\_} \not\in \buf_{a}'\). %
          Hypotheses
          \(\val{v} = \mem(\val{a}) \wedge \memsec(\val{a}) = \botsec\) (in rule
          \textsc{execute-load-commit}) and
          \(\val{v} \neq \mem(\val{a}) \vee \memsec(\val{a}) = \topsec\) (in
          rule \textsc{execute-load-rollback}) do not block the execution but
          just select which rule is applied. Finally, in case the rule
          \textsc{execute-load-rollback} is applied, we have from
          \cref{security0:execute:bufb} and \cref{lemma:pc_low} that
          \(\buf_{b}\bufindex{1} = \buf_{b}'\bufindex{1} = \incrpc{\slevel{l}{\botsec}} @\varepsilon\).
          These conditions are sufficient to make a step in the second
          execution, which entails \cref{security0:step}.
    \item[\cref{security0:buf}] In the rule \textsc{execute-load-commit}, the
          reorder buffers are updated such that:
          \begin{equation*}
            \buf_{n} = \buf_{a} \cdot \mov{x}{\slevel{\mem(a)}{\memsec(a)}} \cdot \buf_{b} \text{
              and
            } \buf_{n}' = \buf_{a}' \cdot \mov{x}{\slevel{\mem'(a)}{\memsec(a)}} \cdot \buf_{b}'
          \end{equation*}
          whereas in the rule \textsc{execute-load-rollback}, younger parts of
          the reorder buffer \(\buf_{b}\) and \(\buf_{b}'\) are forgotten giving
          reorder buffers:
          \begin{equation*}
            \buf_{n} = \buf_{a} \cdot \mov{x}{\slevel{\mem(a)}{\memsec(a)}} \cdot \buf_{b}\bufindex{1} \text{
              and
            } \buf_{n}' = \buf_{a}' \cdot \mov{x}{\slevel{\mem'(a)}{\memsec(a)}} \cdot \buf_{b}'\bufindex{1}
          \end{equation*}
          From \cref{security0:execute:bufa,security0:execute:bufb} and
          \cref{def:rsbufproj}, it is sufficient to show that:
          \begin{enumerate}
            \item\label{security0:load-commit:item:rule} Both configurations
                  evaluate the same rule (either \textsc{execute-load-commit} or
                  \textsc{execute-load-rollback}). By the hypotheses of the rule
                  and \cref{security0:execute-load-commit:a}, this amounts to
                  show \(\val{v} = \val{v'}\). Because \(buf_{n}\) and
                  \(\buf_{n}'\) are well-formed (cf. \cref{lemma:wfbuf}) and
                  that the security level of predicted values in well-formed
                  buffers is \(\botsec\), we have
                  \(\level{s} = \level{s'} = \botsec\). By
                  \cref{security0:execute:bufa} and \cref{def:rsbufproj}, it
                  gives us \(\val{v} = \val{v'}\);
            \item\label{security0:load-commit:item:low}
                  \(\sbufproj{\mov{x}{\slevel{\mem(a)}{\memsec(a)}}} = \sbufproj{\slevel{\mem'(a)}{\memsec(a)}}\).
                  This follows from \cref{security0:ih:mem} and
                  \cref{def:rsbufproj}: if \(\memsec(\val{a}) = \botsec\) then
                  \(\mem(\val{a}) = \mem(\val{a})'\) which concludes our goal.
                  Otherwise
                  \(\sbufproj{\mov{x}{\slevel{\mem(\val{a})}{\memsec(\val{a})}}} = \sbufproj{\slevel{\mem'(\val{a})}{\memsec(\val{a})}} = \dummy\).
          \end{enumerate}

    \item[\cref{security0:micro}] In both rules, the microarchitectural contexts
          are updated with the index \(\val{a}\) (resp. \(\val{a'}\)).
          Therefore, \cref{security0:micro} follows from
          \cref{hyp:deterministic}, \cref{security0:ih:micro}, and
          \cref{security0:execute-load-commit:a}.

    \item[\cref{security0:trans}] The proof is similar to the proof for the case
          \textsc{execute-branch-predict}, \textsc{execute-branch-rollback}.
\end{itemize}

\newrule{execute-store}: From the hypotheses of the rule, the instruction that
is executed in the first configuration is \(inst = \store{e_{a}}{e_{v}} @ T\)
where \(e_{a}, e_{v} \not\in \SVals\). %
From \cref{security0:execute:instr} and \cref{def:rsbufproj}, we have that
\(\rsbufproj{inst} = \store{e_{a}}{e_{v}} @ T\). %
Therefore, the second configuration can only evaluate the rule
\textsc{execute-store}.

The rule \textsc{execute-store} evaluates the expressions \(e_{a}\) and
\(e_{v}\) to values \(\val{a}, \val{v}\) (in the first configuration) and
\(\val{a'}, \val{v'}\) (in the second configuration):
\begin{gather*}
  \slevel{a}{\_} \mydef \semexpr{e_{a}}{\aplsan(\buf_{a}, \reg_{n-1})} \text{
    and
  } \slevel{a'}{\_} \mydef \semexpr{e_{a}}{\aplsan(\buf_{a}', \reg_{n-1}')}\\ %
  \slevel{v}{s} \mydef \semexpr{e_{a}}{\apl(\buf_{a}, \reg_{n-1})} \text{ and
  } \slevel{v'}{s} \mydef \semexpr{e_{v}}{\apl(\buf_{a}', \reg_{n-1}')}
\end{gather*}

\begin{itemize}
  \item[\cref{security0:micro}] The microarchitectural context is not modified by
        the rule so \cref{security0:micro} directly follows from
        \cref{security0:ih:micro}.

  \item[\cref{security0:step}] From \cref{security0:execute:expr_stuck}, we have
        that the evaluation of expressions is not stuck. Because no other
        hypothesis can block the evaluation of the rule, it can also be applied
        in the second configuration, which entails \cref{security0:step}.

  \item[\cref{security0:buf}] The reorder buffers are updated such that:
  \begin{gather*}
    \buf_{n} = \buf_{a} \cdot \store{\slevel{a}{\botsec}}{\slevel{v}{s}} \cdot \buf_{b} \text{
      and
    }\\ \buf_{n}' = \buf_{a}' \cdot \store{\slevel{a'}{\botsec}}{\slevel{v'}{s'}} \cdot \buf_{b}'
  \end{gather*}
  From \cref{security0:execute:bufa}, \cref{security0:execute:bufb} and
  \cref{def:rsbufproj}, we only need to show:
  \begin{equation*}
    \rsbufproj{\store{\slevel{a}{\botsec}}{\slevel{v}{s}}} = \rsbufproj{\store{\slevel{a'}{\botsec}}{\slevel{v'}{s'}}}
  \end{equation*}
  which amounts to showing:
  \begin{equation*}
    \rsbufproj{\slevel{a}{\botsec}} = \rsbufproj{\slevel{a'}{\botsec}} \text{ and } \rsbufproj{\slevel{v}{s}} = \rsbufproj{\slevel{v'}{s'}}
  \end{equation*}
  With the same reasoning used in case \textsc{execute-load-commit} to
  prove \cref{security0:execute-load-commit:a}, we can show that
  \(\val{a} = \val{a'}\). Finally, from \cref{security0:execute:apl}, we
  have \(\rsproj{\slevel{v}{s}} = \rsproj{\slevel{v'}{s'}}\), which by
  \cref{lemma:val_rsbufproj_rsproj} entails %
  \(\sbufproj{\slevel{v}{s}} = \sbufproj{\slevel{v'}{s'}}\).

  \item[\cref{security0:trans}] The proof is similar to the proof for the case
        \textsc{execute-assign}.
\end{itemize}

\medskip\noindent%
This concludes the proof for the \(\execute{i}\) directive.

\proofcase{Retire.} %
  For the retire directive, we have to consider the rules \textsc{retire-assign},
  \textsc{retire-store-low}, and \textsc{retire-store-high}.

  In all \textsc{retire} rules, the first instruction \(inst\) is removed from
  the reorder buffer:
  \begin{equation*}
    \buf_{n-1} = inst @ \varepsilon \cdot buf_{n} \text{ and  } %
    \buf_{n-1}' = inst' @ \varepsilon \cdot buf_{n}'
  \end{equation*}
  Therefore, \cref{security0:buf} follows from \cref{security0:ih:buf} and
  \cref{def:rsbufproj}. Additionally, we have:
  \begin{equation}
    \tag{Hinst}\label{security0:retire:instr}
    \rsbufproj{inst} = \rsbufproj{inst'}
  \end{equation}

  From \cref{security0:ih:step}, the first configuration can make a step. We
  first consider the case where the rule \textsc{retire-assign} is applied; and
  then the case where the rules \textsc{retire-store-low} or
  \textsc{retire-store-high} are applied.

  \newrule{retire-assign}: %
  From the hypotheses of the rules, we have \(inst = \mov{r}{\slevel{v}{s}}\),
  meaning that \(\rsbufproj{inst} = \mov{r}{\rsbufproj{\slevel{v}{s}}}\). %
  Therefore, from \cref{security0:retire:instr}, we have
  \(\rsbufproj{inst'} = \mov{r}{\rsbufproj{\slevel{v}{s}}}\) meaning that
  \(inst' = \mov{r}{\slevel{v'}{s'}}\). Consequently, in the second
  configuration, the only rule that can be applied is \textsc{retire-assign}.

  Additionally, we have
  \(\rsbufproj{\slevel{v}{s}} = \rsbufproj{\slevel{v'}{s}}\), which by
  \cref{lemma:val_rsbufproj_rsproj} gives us
  \(\rsproj{\slevel{v}{s}} = \rsproj{\slevel{v'}{s}}\).

  In \textsc{retire-assign}, the memory and the microarchitectural contexts are
  not modified. Therefore, \cref{security0:mem} (resp. \cref{security0:micro})
  directly follows from \cref{security0:ih:mem} (resp.
  \cref{security0:ih:micro}). As we have already shown \cref{security0:buf}, it
  remains to show \cref{security0:step} and \cref{security0:reg}. %

  \begin{itemize}
    \item[\cref{security0:step}] We have shown that %
          \(\buf_{n-1}' = \mov{r}{\rsbufproj{\slevel{v}{s}}} \cdot \buf_{n}'\),
          hence the rule \textsc{retire-assign} can also be applied in the
          second configuration, which entails \cref{security0:step}.
    \item[\cref{security0:reg}] The register maps are updated such that
          \(\reg_{n} \mydef \reg_{n-1}\change{x}{\slevel{v}{s}}\) and
          \(\reg_{n}' \mydef \reg_{n-1}'\change{x}{\slevel{v'}{s'}}\).
          Therefore, \cref{security0:reg}, follows from
          \(\sproj{\slevel{v}{s}} = \sproj{\slevel{v'}{s'}}\) and \cref{security0:ih:reg}.
    \item[\cref{security0:trans}] Notice that
          \(\buf_{n} = \buf_{n-1}\bufrange{2}{|\buf_{n-1}|}\) and
          \(\buf_{n}' = \buf_{n-1}'\bufrange{2}{|\buf_{n-1}'|}\). Additionally, we
          have \(\reg_{n} \mydef \reg_{n-1}\change{\register{r}}{\slevel{v}{s}}\) and
          \(\reg_{n}' \mydef \reg_{n-1}'\change{\register{r}}{\slevel{v'}{s'}}\), as well
          as \(\mem_{n-1} = \mem_{n}\) and \(\mem_{n-1}' = \mem_{n}'\).
          Hence, to show \cref{security0:trans}, we need to show:
\[
  \transient(\conf{\mem_{n}, \reg_{n-1}\change{\register{r}}{\slevel{v}{s}}, \buf_{n-1}\bufrange{2}{|\buf_{n-1}|}}) \iff \transient(\conf{\mem_{n}', \reg_{n-1}\change{\register{r}}{\slevel{v'}{s'}}, \buf_{n-1}'\bufrange{2}{|\buf_{n-1}'|}})
\]
  This simply follows from \cref{security0:ih:trans}, \cref{def:deepapl} and \cref{def:trans}.
  \end{itemize}

  \newrule{retire-store-low} or \textsc{retire-store-high}: %
  From the hypotheses of the rules, we have
  \(inst = \store{\slevel{a}{s_{a}}}{\slevel{v}{s_{v}}}\), meaning that
  \(\rsbufproj{inst} = \store{\rsbufproj{\slevel{a}{s_{a}}}}{\rsbufproj{\slevel{v}{s_{v}}}}\).
  Therefore, from \cref{security0:retire:instr}, we have
  \(\rsbufproj{inst'} = \store{\rsbufproj{\slevel{a}{s_{a}}}}{\rsbufproj{\slevel{v}{s_{v}}}}\)
  meaning that \(inst' = \store{\slevel{a'}{s_{a}'}}{\slevel{v'}{s_{v}'}}\).
  Consequently the only rules that can be applied in the second configuration
  are \textsc{retire-store-low} and \textsc{retire-store-high}.

  Additionally, we have:
  \(\rsbufproj{\slevel{a}{s_{a}}} = \rsbufproj{\slevel{a'}{s_{a}'}} \text{ and
  } \rsbufproj{\slevel{v}{s_{v}}} = \rsbufproj{\slevel{v'}{s_{v}'}}\), which
  from \cref{lemma:val_rsbufproj_rsproj} entails
  \(\rsproj{\slevel{a}{s_{a}}} = \rsproj{\slevel{a'}{s_{a}'}}\) and %
  \(\rsproj{\slevel{v}{s_{v}}} = \rsproj{\slevel{v'}{s_{v}'}}\). %
  Finally, because \(\buf_{n}\) and \(\buf_{n-1}\) are well-formed (cf.\
  \cref{lemma:wfbuf}) and because store addresses have security level \(\botsec\)
  in well-formed buffers (by \cref{def:wfbuf}), we have
  \(\level{s_{a}} = \level{s_{a}'} = \botsec\). From %
  \(\rsproj{\slevel{a}{s_{a}}} = \rsproj{\slevel{a'}{s_{a}'}}\) this entails
  \begin{gather}
    \val{a} = \val{a'}\tag{Ha}\label{security0:retire:store:a}
  \end{gather}

  In both rules, the register map is not modified. Therefore,
  \cref{security0:reg} directly follows from \cref{security0:ih:reg}. As we have
  already shown \cref{security0:buf}, it remains to show \cref{security0:step},
  \cref{security0:mem}, \cref{security0:micro} and \cref{security0:trans}.

  \begin{itemize}
    \item[\cref{security0:step}] We have shown that %
          \(\buf_{n-1}' = \store{\slevel{a'}{s_{a}'}}{\slevel{v'}{s_{v}'}} \cdot \buf_{n}'\),
          hence the rule \textsc{retire-store-low} or the rule
          \textsc{retire-store-high} can also be applied in the second
          configuration, which entails \cref{security0:step}.

    \item[\cref{security0:micro}] In both rules, the microarchitectural context is
          modified such that \(\micro_{n} = \update(\micro_{n-1}, \val{a})\) and
          \(\micro_{n}' = \update(\micro_{n-1}', \val{a'})\). Because
          \(\update\) is deterministic (cf. \cref{hyp:deterministic}),
          \cref{security0:micro} directly follows from \cref{security0:ih:micro}
          and \cref{security0:retire:store:a}.

    \item[\cref{security0:mem}] In both rules, the memories are updated such
          that \(\mem_{n} \mydef \mem_{n-1}\change{\val{a}}{\val{v}}\) and
          \(\mem_{n}' \mydef \mem_{n-1}'\change{\val{a'}}{\val{v'}}\). Notice
          that from
          \(\rsproj{\slevel{v}{s_{v}}} = \rsproj{\slevel{v'}{s_{v}'}}\) and
          \cref{def:rsproj}, we have \(\level{s_{v}} = \level{s_{v}'}\). Hence,
          because we also have \(\memsec(\val{a}) = \memsec(\val{a'})\) from
          \cref{security0:retire:store:a}, both executions evaluate the same
          rule. We first consider the case in which both executions evaluate the
          rule \textsc{retire-store-low} and then, the case where both
          executions evaluate the rule \textsc{retire-store-high}:
            \begin{itemize}
              \item In case \textsc{retire-store-low}, we need to show
                    \(\val{v} = \val{v'}\). By \cref{security0:ih:arch}, we get
                    that \(\conf{\mem_{n-1}, \reg_{n-1}} = \aconf_{j} %
                    \text{ and } \conf{\mem_{n-1}', \reg_{n-1}'} = \aconf_{j}' %
                    \text{ where } j = \corr{\arun}{\mrun}(n-1)(0)\). %
                    From \cref{def:wfbuf}, we know that
                    \(\prog{\semexpr{\pc}{\reg_{1-n}}} = \prog{\semexpr{\pc}{\reg_{n-1}'}} = \store{e_{a}}{e_{v}}\).
                    Additionally, we have \(\val{v} = \semexpr{e_{v}}{\reg}\)
                    and \(\val{v'} = \semexpr{e_{v}}{\reg'}\) and
                    \(\val{a} = \semexpr{e_{a}}{\reg} = \semexpr{e_{a}}{\reg'}\).
                    Consequently, we can apply the \textsc{store} rule in the
                    sequential semantics (\cref{fig:sequential-semantics}) to
                    \(\aconf_{j}\) and \(\aconf_{j}'\). Additionally, from
                    \(\val{a} = \semexpr{e_{a}}{\reg} = \semexpr{e_{a}}{\reg'}\)
                    and \(\memsec(\val{a}) = \botsec\), we know that the
                    evaluation of the rule in both configurations produce
                    non-empty declassification traces \(\declassify{\val{v}}\)
                    and \(\declassify{\val{v'}}\). Moreover, from
                    \cref{hyp:ct-no-decl}, we have
                    \(\declassify{\val{v}} = \declassify{\val{v'}}\).
                    Finally, by the rule \textsc{retire-store-low} we have
                    \(\mem_{n} \mydef \mem_{n-1}\change{\val{a}}{\val{v}}\) and
                    \(\mem_{n}' \mydef \mem_{n-1}'\change{\val{a}}{\val{v}}\).
                    Therefore \cref{security0:mem} follows from
                    \cref{security0:ih:mem}.
              \item In case \textsc{retire-store-high}, we have
                    \(\memsec(\val{a}) = \topsec\) so the public memory is not
                    modified, meaning that \cref{security0:mem} follows from
                    \cref{security0:ih:mem} and \cref{def:sproj}.
            \end{itemize}
     \item[\cref{security0:trans}] The proof is similar to the case \textsc{retire-assign}.
  \end{itemize}

  \medskip\noindent%
  This concludes the proof for the \(\retire\) directive, and in turn, for
  \cref{lemma:security0}.
\end{proof}

\securityzero*

\begin{proof}
  Proof follows from the direct application of \cref{lemma:security0}.
\end{proof}

\subsection{Correspondence between patched architectural and patched hardware semantics}\label{sec:proof-corr-patch}
\paragraph{Notations} %
Through this section, we call a (architectural or hardware) configuration \(c\)
with a declassification trace \(\declassified\), denoted
\(\dconf{c}{\declassified}\), a \emph{patched configuration}. We let
\(\darun{\declassified}\) (resp.\ \(\dmrun{\declassified}\)) denote a sequence
of patched architectural configuration or (resp.\ patched hardware
configuration) resulting from a execution in the architectural (resp. hardware)
semantics patched with \(\declassified\). %
Given a sequence of patched architectural configurations
\(\darun{\declassified}\), we let \(\dconf{\aconf_{j}}{\declassified_{j}}\) with
\(0 \leq j \leq |\arun|\) denote the \(j^{th}\) configuration in \(\arun\) (the
same holds for \(\mrun\)).

The deep update of a patched architectural configuration
\(\dconf{\conf{\mem,\reg}}{\declassified}\) and reorder buffer \(\buf\), denoted
\(\deepapl{\dconf{\conf{\mem,\reg}}{\declassified}}{\varepsilon}\), applies all pending
instructions in \(\buf\) to the configuration
\(\dconf{\conf{\mem,\reg}}{\declassified}\). That the deep update for patched
configurations is similar to the deep update defined in \cref{def:deepapl}
except that it replaces values stored to low-memory with values from the
declassification trace.
\begin{definition}[Patched deep update]\label{def:deepapl-patch} %
  For any reorder buffer \(\buf\) and patched architectural configuration
  \(\dconf{\conf{\mem,\reg}}{\declassified}\):
  \begin{align*}
    \deepapl{\dconf{\conf{\mem,\reg}}{\declassified}}{\varepsilon} & \mydef \dconf{\conf{\mem,\reg}}{\declassified} \\
    \deepapl{\dconf{\conf{\mem,\reg}}{\declassified}}{\register{x} \gets e @ T} & \mydef
    \begin{cases}
     \dconf{\conf{\mem,\reg\change{\register{x}}{\semexpr{e}{\reg}}}}{\declassified} &
     \text{if } T = \varepsilon\\
     \dconf{\conf{\mem,\reg\change{\register{x}}{\mem(\semexpr{e_{a}}{\reg})}}}{\declassified} &
     \text{if } T = \val{l} \wedge \prog{\val{l}} = \load{\register{x}}{e_{a}}\\
    \end{cases} \\
    \deepapl{\dconf{\conf{\mem,\reg}}{\declassified}}{\pc \gets e @ T} & \mydef
    \begin{cases}
      \dconf{\conf{\mem,\reg\change{\pc}{\semexpr{e}{\reg}}}}{\declassified} & %
      \text{if } T = \varepsilon\\
      \dconf{\conf{\mem,\reg\change{\pc}{\semexpr{e_{l}}{\reg}}}}{\declassified} & %
      \text{if } T = \val{l} \wedge \prog{\val{l}} = \jmp{e_{l}}\\
      \dconf{\conf{\mem,\reg\change{\pc}{\val{l'}}}}{\declassified} & %
      \text{if } T = \val{l} \wedge \prog{\val{l}} = \beqz{e_{c}}{\val{l'}} %
      \wedge \semexpr{e_{c}}{\reg} = 0\\
      \dconf{\conf{\mem,\reg\change{\pc}{\val{l + 1}}}}{\declassified} & %
      \text{if } T = \val{l} \wedge \prog{\val{l}} = \beqz{e_{c}}{\val{l'}} %
      \wedge \semexpr{e_{c}}{\reg} \neq 0\\
    \end{cases} \\
    \deepapl{\dconf{\conf{\mem,\reg}}{\declassified}}{\load{x}{e_{a}} @ T} & \mydef \dconf{\conf{\mem,\reg\change{\register{x}}{\mem(\semexpr{e_{a}}{\reg})}}}{\declassified}\\
    \deepapl{\dconf{\conf{\mem,\reg}}{\declassified}}{\store{e_{a}}{e_{v}} @ T} & \mydef %
    \begin{cases}
      \dconf{\conf{\mem\change{\val{a}}{\declassify{\val{v}}},\reg}}{\declassify{\declassified'}} & %
      \text{if } \val{a} \mydef \semexpr{e_{a}}{\reg} \wedge \memsec(\val{a}) = \botsec \wedge \declassified = \declassify{\val{v}} \cdot \declassify{\declassified'}\\
      \dconf{\conf{\mem\change{\val{a}}{\semexpr{e_{v}}{\reg}},\reg}}{\declassified} & %
      \text{if } \val{a} \mydef \semexpr{e_{a}}{\reg} \wedge \memsec(\val{a}) = \topsec\\
    \end{cases} \\
    \deepapl{\dconf{\conf{\mem,\reg}}{\declassified}}{(inst @ T \cdot \buf)} & \mydef
    \deepapl{(\deepapl{\dconf{\conf{\mem,\reg}}{\declassified}}{inst @ T})}{\buf}
  \end{align*}
\end{definition}

The well-formed buffer predicate for patched configurations,
\(\wf(\buf, \dconf{\conf{\mem, \reg}}{\declassified})\), can be obtained from
\cref{def:wfbuf} by replacing the deep projection with the patched deep
projection; \(\transient(\dconf{\mconf}{\declassified})\),
\(\goodpred(inst @ \varepsilon, \dconf{\conf{\mem, \reg}}{\declassified})\)
\(\prefix(\buf, \dconf{\conf{\mem, \reg}}{\declassified})\), and
\(\corr{\darun{\declassified}}{\dmrun{\declassified}}\)
can be obtained similarly.

The following lemma states that for all non-transient buffer \(buf\) with
respect to a configuration \(\conf{\mem, \reg}\), the register map obtained by
the function \(\apl\) is equivalent (when defined) to the register map obtained
by the deep update of \(\buf\) with \(\conf{\mem, \reg}\).
\begin{lemma}\label{lemma:deepapl-patch_apl}
  Let \(\dconf{\conf{\mem,\reg}}{\declassified}\) be an architectural
  configuration and \(\buf\) a reorder buffer such that
  \(\neg\transient(\conf{\mem,\reg,\buf,\_})\). Let
  \(\dconf{\conf{\_,\reg'}}{\_} = \deepapl{\dconf{\conf{\mem,\reg}}{\declassified}}{\buf}\).
  We have that for all \(\register{r}\) if
  \(\apl(\buf, \reg)(\register{r}) = \slevel{\val{v}}{s}\) then
  \(\reg'(\register{r}) = \slevel{\val{v}}{s}\).
\end{lemma}
\begin{proof}
  The proof is similar to the proof for \cref{lemma:deepapl_apl}. In particular,
  it goes by induction on the size of the ROB, \cref{def:apl} and
  \cref{def:deepapl-patch}. The only case that differ is the case
  \(inst = \store{e_{a}}{e_{v}} @ T\). Notice that the instruction is ignored by
  \(\apl\) but the memory is modified (and possibly patched using the
  declassification trace) in \(\deepapl{}{}\). However, it may only impacts the
  result of subsequent \(inst = \load{x}{e} @ T\) instructions but these
  instructions result in undefined \(\register{x}\) with \(apl\), which
  immediately concludes our goal.
\end{proof}

It follows that \cref{lemma:deepapl_apl_expr,lemma:deepapl_aplsan_expr} can also
be applied to patched configurations.

The following lemma states that for each patched hardware state in a patched
hardware run, for all the prefixes of its reorder buffer, there exists a
corresponding patched architectural state in the corresponding patched architectural run.
\lestodo{Proofread!}
\begin{lemma}[Correctness of the \(\corr{\darun{\declassified}}{\dmrun{\declassified}}\)
  relation]\label{lemma:corr-patch}
  Let \(\aconf{}_{0} = \conf{\mem, \reg}\) be an initial architectural
  configuration, \(\micro\) be a microarchitectural context and \(\declassified\)
  a declassification trace. Additionally, let \(\darun{\declassified}\) be the
  longest contract run, starting from \(\dconf{\aconf_{0}}{{\declassified}}\),
  and \(\dmrun{\declassified}\) be the longest \name{} run starting from
  \(\dconf{\conf{\mem, \reg, \varepsilon, \micro}}{\declassified}\).
  Additionally, for all \(0 \leq n < |\mrun|\), we let
  \(\dconf{\mconf_{n}}{\declassified_{n}}\) denote the \(n^{th}\) configuration
  in \(\mrun\), and for all \(0 \leq j < |\arun|\), we let \(\arun[j]\) be the
  \(j^{th}\) architectural configuration in \(\arun\). \lestodo{What about
    termination?}

  For all \(0 \leq n < |\mrun|\), we have:
  \begin{equation}
    \text{For all } \buf \in \prefix(\buf_{n}, \dconf{\conf{\mem_{n},\reg_{n}}}{\declassified_{n}}),
    \deepapl{\dconf{\conf{\mem_{n},\reg_{n}}}{\declassified_{n}}}{\buf} =
    \arun[j] \text{ where } j = \corr{\darun{\declassified}}{\dmrun{\declassified}}(n)(|\buf|). \tag{G}\label{lemma:corr-patch:goal}
  \end{equation}
\end{lemma}
\begin{proof}
  The proof goes by induction on \(n\).

  \proofcase{Base case \((n = 0)\).} We have
  \(\mconf_{0} = \dconf{\conf{\mem, \reg, \varepsilon, \micro}}{\declassified}\), hence
  \(\buf_{0} = \varepsilon\). %
  From \cref{def:prefixes}, we have that
  \(\prefix(\varepsilon) = \{\varepsilon\}\). Additionally, from
  \cref{def:corr}, we have that %
  \(j = \corr{\darun{\declassified}}{\dmrun{\declassified}}(0)(|\varepsilon|) = 0\).
  From \cref{def:deepapl}, we have that
  \(\deepapl{\dconf{\conf{\mem,\reg}}{\declassified}}{\varepsilon} = \dconf{\conf{\mem,\reg}}{\declassified} = \arun[0]\),
  which concludes \cref{lemma:corr-patch:goal}.

  \proofcase{Inductive case.} %
  Assume that the hypothesis holds for hardware runs of length \(n-1\), namely:
  \begin{equation}
    \text{For all } \buf \in \prefix(\buf_{n-1}, \dconf{\conf{\mem_{n-1},\reg_{n-1}}}{\declassified_{n-1}}),
    \deepapl{\dconf{\conf{\mem_{n-1},\reg_{n-1}}}{\declassified_{n-1}}}{\buf} =
    \arun[j] \text{ where } j = \corr{\darun{\declassified}}{\dmrun{\declassified}}(n-1)(|\buf|). \tag{IH}\label{lemma:corr-patch:IH}
  \end{equation}

  We show that \cref{lemma:corr-patch:goal} holds at step \(n\). The proof is
  similar to the proof for \cref{lemma:corr} except for the rules
  \textsc{fetch-others}, \textsc{execute-store} and \textsc{retire-store}.

  \newrule{fetch-others}: from the evaluation rules (cf.\
  \cref{fig:fetch-rules}), we have
  \(\buf_{n} = \buf_{n-1} \cdot inst @ \varepsilon \cdot \pc \gets \slevel{\val{l+1}}{\botsec} @ \varepsilon\)
  where
  \(\slevel{\val{l}}{\_} \mydef \semexpr{\pc}{\apl(\buf_{n-1}, \reg_{n-1})}\),
  \(inst = \prog{\val{l}}\) and \(inst \not\in \{\beqz{\_}{\_}, \jmp{\_}\}\).
  Additionally, we have \(\mem_{n} = \mem_{n-1}\), \(\reg_{n} = \reg_{n-1}\) and
  \(\declassified_{n} = \declassified_{n-1}\). As in the proof for
  \cref{lemma:corr}, there are several cases to consider. We focus here on the
  case \(\neg\transient(\mconf_{n-1})\), which means
  \(\prefix(\buf_{n}, \dconf{\conf{\mem_{n}, \reg_{n}}}{\declassified_{n}}) = \prefix(\buf_{n-1}, \dconf{\conf{\mem_{n-1}, \reg_{n-1}}}{\declassified_{n-1}}) \cup \{\buf_{n}\}\);
  and in particular the case
  \(\buf \in \prefix(\buf_{n}, \dconf{\conf{\mem_{n}, \reg_{n}}}{\declassified_{n}}) = \buf_{n} = \buf_{n-1} \cdot inst @ \varepsilon \cdot \pc \gets \slevel{\val{l+1}}{\botsec} @ \varepsilon\)
  with \(inst = \store{e_{a}}{e_{v}}\). The other cases are similar to the proof
  for \cref{lemma:corr}. In particular, when we need to show that the sequential
  semantics and the patched deep update would produce the same changes on a
  configuration, we can observe that the declassification trace is simply
  propagated by the sequential semantics and by the patched deep update
  \cref{def:deepapl-patch} and not modified.

  In that case, because \(\neg\transient(\mconf_{n-1})\) and because reorder
  buffers are well-formed (\cref{lemma:wfbuf}), we have from \cref{def:prefixes}
  that \(\buf_{n-1} \in \prefix(\buf_{n-1}, \dconf{\conf{\mem_{n-1}, \reg_{n-1}}}{\declassified_{n-1}})\). %
  Hence, from \cref{lemma:corr-patch:IH}, we have
  \[\deepapl{\dconf{\conf{\mem_{n-1},\reg_{n-1}}}{\declassified_{n-1}}}{\buf_{n-1}}%
    = \arun[j] \text{ where } j =
    \corr{\darun{\declassified}}{\dmrun{\declassified}}(n-1)(|\buf_{n-1}|)\] %
  In the following, we let
  \(\arun[j] = \dconf{\conf{\reg_{j}, \mem_{j}}}{\declassified_{j}}\). From
  \cref{lemma:deepapl_apl_expr} and
  \(\slevel{\val{l}}{\_} = \semexpr{\pc}{\apl(\buf_{n-1}, \reg_{n-1})}\), we
  have \(\reg_{j}(\pc) = \slevel{\val{l}}{\_}\) and hence
  \(\lstpc(\dconf{\mconf_{n-1}}{\declassified_{n-1}}) = \val{l}\), meaning that
  \(\prognott{\lstpc(\mconf_{n-1})} = \store{e_{a}}{e_{v}}\). From
  this, \cref{def:corr}, and \(|\buf| = |\buf_{n-1}| + 2\), we get
  \(\corr{\darun{\declassified}}{\dmrun{\declassified}}(n)(|\buf|) = %
  \corr{\darun{\declassified}}{\dmrun{\declassified}}(n-1)(|\buf_{n-1}|) + 1 = %
  j + 1\). %

  Additionally, from \cref{def:deepapl-patch}, we have that
  \[\deepapl{\dconf{\conf{\mem_{n},\reg_{n}}}{\declassified_{n}}}{\buf} = \deepapl{(\deepapl{(\deepapl{\dconf{\conf{\mem_{n},\reg_{n}}}{\declassified_{n}}}{\buf_{n-1}})}{\store{e_{a}}{e_{v}} @ \varepsilon})}{\pc \gets \slevel{\val{l+1}}{\botsec} @ \varepsilon}\]

  From \(\dconf{\conf{\mem_{n-1},\reg_{n-1}}}{\declassified_{n-1}} = \dconf{\conf{\mem_{n},\reg_{n}}}{\declassified_{n}}\), and the
  definition of \(\arun[j]\), this gives us
  \(\deepapl{\dconf{\conf{\mem_{n},\reg_{n}}}{\declassified_{n}}}{\buf} = %
  \deepapl{(\deepapl{\arun[j]}{\store{e_{a}}{e_{v}} @ \varepsilon})}{\pc \gets \slevel{\val{l+1}}{\botsec} @ \varepsilon}\).
  Hence, in order to show
  \(\deepapl{\dconf{\conf{\mem_{n},\reg_{n}}}{\declassified_{n}}}{\buf} = \arun[{\corr{\darun{\declassified}}{\dmrun{\declassified}}(n)(|\buf|)}]\)
  (which concludes our goal \cref{lemma:corr-patch:goal} for this case), it remains to
  show
  \[\deepapl{(\deepapl{\arun[j]}{\store{e_{a}}{e_{v}} @ \varepsilon})}{\pc \gets \slevel{\val{l+1}}{\botsec} @ \varepsilon} = \arun[j+1]\]

  Let \(\val{a} = \semexpr{e_{a}}{r_{j}}\). Next, we consider the case where
  \(\val{a}\) corresponds to public memory and the case where \(\val{a}\)
  corresponds to secret memory.

  \begin{itemize}
    \item In case \(\memsec(\val{a}) = \topsec\), from \cref{def:deepapl-patch},
          we have %
  \(\deepapl{(\deepapl{\arun[j]}{\store{e_{a}}{e_{v}} @ \varepsilon})}{\pc \gets \slevel{\val{l+1}}{\botsec} @ \varepsilon} = %
    \dconf{\conf{\mem_{j}\change{\val{a}}{\semexpr{e_{v}}{r_{j}}}, \reg_{j}\change{\pc}{\val{l+1}}}}{\declassified_{j}}\)
  Moreover, from the evaluation rules of the sequential semantics
  (\cref{fig:sequential-patched-semantics}, rule \textsc{store-high}), we also have
  \(\arun[j+1] = \dconf{\conf{\mem_{j}\change{\val{a}}{\semexpr{e_{v}}{r_{j}}}, \reg_{j}\change{\pc}{\val{l+1}}}}{\declassified_{j}}\),
  which concludes
  \(\deepapl{(\deepapl{\arun[j]}{\store{e_{a}}{e_{v}} @ \varepsilon})}{\pc \gets \slevel{\val{l+1}}{\botsec} @ \varepsilon} = \arun[j+1]\).
    \item In case \(\memsec(\val{a}) = \botsec\), from \cref{def:deepapl-patch},
          we have %
  \(\deepapl{(\deepapl{\arun[j]}{\store{e_{a}}{e_{v}} @ \varepsilon})}{\pc \gets \slevel{\val{l+1}}{\botsec} @ \varepsilon} = %
  \dconf{\conf{\mem_{j}\change{\val{a}}{\declassify{\val{v}}}, \reg_{j}\change{\pc}{\val{l+1}}}}{\declassify{\declassified'}}\) where %
  \(\declassified_{j} = \declassify{\val{v}} \cdot \declassify{\declassified'}\).
  Moreover, from the evaluation rules of the sequential semantics
  (\cref{fig:sequential-patched-semantics}, rule \textsc{store-patched}), we also have
  \(\arun[j+1] = \dconf{\conf{\mem_{j}\change{\val{a}}{\declassify{\val{v}}}, \reg_{j}\change{\pc}{\val{l+1}}}}{\declassify{\declassified'}}\),
  which concludes
  \(\deepapl{(\deepapl{\arun[j]}{\store{e_{a}}{e_{v}} @ \varepsilon})}{\pc \gets \slevel{\val{l+1}}{\botsec} @ \varepsilon} = \arun[j+1]\).
  \end{itemize}

  \newrule{execute-store}: In this case, the directive that is applied is
  \(\directive = \execute{j}\) and from \cref{def:corr}, we have
  \(\corr{\darun{\declassified}}{\dmrun{\declassified}}(n) = \corr{\darun{\declassified}}{\dmrun{\declassified}}(n-1)\).
  Additionally, from the hardware evaluation rules (cf.\
  \cref{fig:execute-mem-rules}), we have \(\dconf{\conf{\mem_{n}, \reg_{n}}}{\declassified_{n}} = \dconf{\conf{\mem_{n-1}, \reg_{n-1}}}{\declassified_{n-1}}\), and:
\begin{align*}
  \buf_{n-1} =~& \buf_{n-1}\bufrange{1}{j-1} \cdot
                 \store{e_{a}}{e_{v}} @ T \cdot
                 \buf_{n-1}\bufrange{j+1}{|buf_{n-1}|}\\
  \buf_{n}   =~& \buf_{n-1}\bufrange{1}{j-1} \cdot
                 \store{\slevel{\val{a}}{\botsec}}{\slevel{\val{v}}{\_}} @ T \cdot
                 \buf_{n-1}\bufrange{j+1}{|buf_{n-1}|}
\end{align*}
with \(e_{a} \not\in \SVals\), \(e_{v} \not\in \SVals\),
\(\slevel{\val{a}}{\_} = \semexpr{e_{a}}{\apl(\buf_{n-1}\bufrange{1}{j-1}, \reg_{n-1})}\), and
\(\slevel{\val{v}}{\_} = \semexpr{e_{v}}{\apl(\buf_{n-1}\bufrange{1}{j-1}, \reg_{n-1})}\).

Let \(\buf\) be an arbitrary prefix in
\(\prefix(\buf_{n}, \dconf{\conf{\mem_{n}, \reg_{n}}}{\declassified_{n}})\). We can consider two cases:
\begin{itemize}
  \item \(|buf| < j\). The proof for this case is similar to the proof for the
        corresponding case in \cref{lemma:corr}.
  \item \(|buf| \geq j\). Notice that, by \cref{def:prefixes} and
        \(\buf \in \prefix(\buf_{n}, \dconf{\conf{\mem_{n}, \reg_{n}}}{\declassified_{n}})\), there is no
        misspredicted instruction in \(\buf\) (except possibly the last one).
        Therefore, it follows from
        \(\dconf{\conf{\mem_{n}, \reg_{n}}}{\declassified_{n}} = \dconf{\conf{\mem_{n-1}, \reg_{n-1}}}{\declassified_{n-1}}\), that
        \(\buf' = \buf\bufrange{1}{j-1} \cdot \store{e_{a}}{e_{v}} @ T \cdot \buf\bufrange{j+1}{|buf|}\)
        belongs to \(\prefix(\buf_{n-1}, \dconf{\conf{\mem_{n-1}, \reg_{n-1}}}{\declassified_{n-1}})\). %
        Hence, from \cref{lemma:corr-patch:IH}, we have that
        \[\deepapl{\dconf{\conf{\mem_{n-1},\reg_{n-1}}}{\declassified_{n-1}}}{\buf'} = %
 \arun[\corr{\darun{\declassified}}{\dmrun{\declassified}}(n-1)(|\buf'|)]\]
        From this, \(|\buf'| = |\buf|\),
        \(\dconf{\conf{\mem_{n}, \reg_{n}}}{\declassified_{n}} = \dconf{\conf{\mem_{n-1}, \reg_{n-1}}}{\declassified_{n-1}}\), and
        \(\corr{\darun{\declassified}}{\dmrun{\declassified}}(n) = \corr{\darun{\declassified}}{\dmrun{\declassified}}(n-1)\), we get that
\[\deepapl{\dconf{\conf{\mem_{n}, \reg_{n}}}{\declassified_{n}}}{\buf'}
 = \arun[\corr{\darun{\declassified}}{\dmrun{\declassified}}(n)(|\buf|)]\]
        Let
        \(\dconf{\conf{\mem', \reg'}}{\declassified'} = \deepapl{\dconf{\conf{\mem_{n}, \reg_{n}}}{\declassified_{n}}}{\buf_{n-1}\bufrange{1}{j-1}}\).
        Notice that from \cref{lemma:deepapl-patch_apl}, \(\reg_{n} = \reg_{n-1}\),
\(\slevel{\val{a}}{\_} = \semexpr{e_{a}}{\apl(\buf_{n-1}\bufrange{1}{j-1}, \reg_{n-1})}\), and
\(\slevel{\val{v}}{\_} = \semexpr{e_{v}}{\apl(\buf_{n-1}\bufrange{1}{j-1}, \reg_{n-1})}\),
        we get \(\slevel{\val{a}}{\_} = \semexpr{e_{a}}{\reg'}\) and \(\slevel{\val{v'}}{\_} = \semexpr{e_{v}}{\reg'}\).
        From there, we consider two cases:
        \begin{itemize}
          \item \(\memsec(\val{a}) = \botsec\). In this case, by \cref{def:deepapl-patch} and rewriting we get
        \begin{align*}
        \deepapl{\dconf{\conf{\mem_{n}, \reg_{n}}}{\declassified_{n}}}{\buf'} =~ %
        &\deepapl{\dconf{\conf{\mem', \reg'}}{\declassified'}}{\store{e_{a}}{e_{v}} @ T \cdot \buf\bufrange{j+1}{|buf|}} \\=~ %
        &\deepapl{\dconf{\conf{\mem'\change{\val{a}}{\declassify{\val{v'}}}, \reg'}}{\declassified''}}{\buf\bufrange{j+1}{|buf|}} \text{ where }  %
          \declassified' = \declassify{\val{v}} \cdot \declassified''\\
          =~& \deepapl{\dconf{\conf{\mem', \reg'}}{\declassified'}}{\store{\slevel{\val{a}}{\_}}{\slevel{\val{v}}{\_}} @ T \cdot \buf\bufrange{j+1}{|buf|}} \\
          =~& \deepapl{\dconf{\conf{\mem_{n},\reg_{n}}}{\declassified_{n}}}{\buf}
        \end{align*}
        which means \(\deepapl{\dconf{\conf{\mem_{n},\reg_{n}}}{\declassified_{n}}}{\buf} = \arun[\corr{\darun{\declassified}}{\dmrun{\declassified}}(n)(|\buf|)]\)
and concludes \cref{lemma:corr-patch:goal}.
          \item \(\memsec(\val{a}) = \topsec\). In this case, by \cref{def:deepapl-patch} and rewriting we get
        \begin{align*}
        \deepapl{\dconf{\conf{\mem_{n}, \reg_{n}}}{\declassified_{n}}}{\buf'} =~ %
        &\deepapl{\dconf{\conf{\mem', \reg'}}{\declassified'}}{\store{e_{a}}{e_{v}} @ T \cdot \buf\bufrange{j+1}{|buf|}} \\=~ %
        &\deepapl{\dconf{\conf{\mem'\change{\val{a}}{\val{v}}, \reg'}}{\declassified'}}{\buf\bufrange{j+1}{|buf|}} \\
          =~& \deepapl{\dconf{\conf{\mem', \reg'}}{\declassified'}}{\store{\slevel{\val{a}}{\_}}{\slevel{\val{v}}{\_}} @ T \cdot \buf\bufrange{j+1}{|buf|}} \\
          =~& \deepapl{\dconf{\conf{\mem_{n},\reg_{n}}}{\declassified_{n}}}{\buf}
        \end{align*}
        which concludes \cref{lemma:corr-patch:goal}.
        \end{itemize}
      \end{itemize}

  \newrule{retire-store-low} and \textsc{retire-store-high}. %
  In this case, from \cref{def:corr},
  we have \(\corr{\darun{\declassified}}{\dmrun{\declassified}}(n) = \shift(\corr{\darun{\declassified}}{\dmrun{\declassified}}(n-1))\).
  Moreover, from the hardware evaluation rules in \cref{fig:retire-rules}, we
  have \(\buf_{n-1} = \store{\slevel{\val{a}}{\_}}{\slevel{\val{v}}{\_}} @ \varepsilon \cdot \buf_{n}\).
  From here, we consider the cases \(\memsec(a) = \botsec\) and
  \(\memsec(a) = \topsec\) separately. We only detail the case
  \(\memsec(a) = \botsec\) as the proof for the other case is similar to the
  corresponding proof case in the proof of \cref{lemma:corr}. %

  In the case \(\memsec(a) = \botsec\), the rule \textsc{retire-store-low} is
  applied and we have
  \(\declassified_{n-1} = \declassify{\val{v'}} \cdot \declassified_{n}\), \(\mem_{n} = \mem_{n-1}\change{\val{a}}{\declassify{\val{v'}}}\),
  and
  \(\reg_{n} = \reg_{n-1}\). Additionally, because \(\buf_{n-1}\) is well-formed
  (cf. \cref{lemma:wfbuf}) and from the definition of well-formed buffers
  (\cref{def:wfbuf}), we have that
  \(\buf_{n} = \incrpc{\slevel{\val{l}}{\botsec}} \cdot \buf_{n}'\). Hence, from
  \cref{def:prefixes}, we have
    \begin{align*}
      \prefix(\buf_{n-1}, \dconf{\conf{\mem_{n-1}, \reg_{n-1}}}{\declassified_{n-1}\}}) =~&
      \{\varepsilon \} \cup %
      \{\store{\slevel{\val{a}}{\_}}{\slevel{\val{v}}{\_}} @ \varepsilon \cdot \pc \gets \slevel{\val{l}}{\botsec} @ \varepsilon \cdot \buf'~|~ \\
      & %
      \buf' \in \prefix(\buf_{n}', \deepapl{\dconf{\conf{\mem_{n-1}, \reg_{n-1}}}{\declassified_{n-1}}}{\store{\slevel{\val{a}}{\_}}{\slevel{\val{v}}{\_}} @ \varepsilon \cdot \pc \gets \slevel{\val{l}}{\botsec} @ \varepsilon})\}\\
      \prefix(\buf_{n}, \dconf{\conf{\mem_{n}, \reg_{n}}}{\declassified_{n}\}}) =~& \{\varepsilon \} \cup %
      \{\pc \gets \slevel{\val{l}}{\botsec} @ \varepsilon \cdot \buf' ~|~ %
      \buf' \in \prefix(\buf_{n}', \deepapl{\conf{\mem_{n}, \reg_{n}}}{\pc \gets \slevel{\val{l}}{\botsec} @ \varepsilon})\}\\
    \end{align*}
    Let \(\buf\) be an arbitrary prefix in \(\buf_{n}\). By \cref{def:prefixes}
    and the definition of \(\buf_{n}\) and \(\buf_{n-1}\), we have that
    \(\store{\slevel{\val{a}}{\_}}{\slevel{\val{v}}{\_}} @ \varepsilon \cdot \buf\) is
    a prefix in \(\prefix(\buf_{n-1}, \dconf{\conf{\mem_{n-1}, \reg_{n-1}}}{\declassified_{n-1}\}})\). Hence, from \cref{lemma:corr:IH}, we have
    \[\deepapl{\dconf{\conf{\mem_{n-1},\reg_{n-1}}}{\declassified_{n-1}}}{\store{\slevel{\val{a}}{\_}}{\slevel{\val{v}}{\_}} @ \varepsilon \cdot \buf} = \arun[j] %
      \text{ where
      } j = \corr{\darun{\declassified}}{\dmrun{\declassified}}(n-1)(|\buf| + 1)\]
    Hence, by \(\declassified_{n-1} = \declassify{\val{v'}} \cdot \declassified_{n}\) and \cref{def:deepapl-patch}, we get %
    \(\deepapl{\dconf{\conf{\mem_{n-1}\change{\val{a}}{\val{\declassify{\val{v'}}}},\reg_{n-1}}}{\declassified_{n}}}{\buf} = \arun[j]\).
    By \(\mem_{n} = \mem_{n-1}\change{\val{a}}{\val{\declassify{\val{v'}}}}\)
    and \(\reg_{n} = \reg_{n-1}\), this gives us
    \(\deepapl{\dconf{\conf{\mem_{n},\reg_{n}}}{\declassified_{n}}}{\buf} = \arun[j]\).
    Finally, %
    by definition of \(\shift\), we get %
    \[\deepapl{\dconf{\conf{\mem_{n},\reg_{n}}}{\declassified_{n}}}{\buf} = \arun[\shift(\corr{\darun{\declassified}}{\dmrun{\declassified}}(n-1)(|\buf|))]\] %
    which from
    \(\corr{\darun{\declassified}}{\dmrun{\declassified}}(n) = \shift(\corr{\darun{\declassified}}{\dmrun{\declassified}}(n-1))\),
    entails
    \(\deepapl{\dconf{\conf{\mem_{n},\reg_{n}}}{\declassified_{n}}}{\buf} = \arun[\corr{\darun{\declassified}}{\dmrun{\declassified}}(n)(|\buf|)]\) and concludes \cref{lemma:corr-patch:goal}.

\medskip%
\noindent%
We have shown for any evaluation rule in the hardware semantics that if
\cref{lemma:corr-patch:goal} holds at step \(n-1\), then it also holds at step \(i\).
This concludes the proof of \cref{lemma:corr-patch}.
\end{proof}

\subsection{Proof of security for constant-time programs with
  declassification}\label{sec:proof-decl}
The following hypotheses are used to prove security for constant-time programs
with declassification:

\hypdeterministic*

\begin{restatable}{hypothesis}{hypct2}\label{hyp:ct2}%
  The program is constant-time up to declassification (according to \cref{def:ct-decl}). In particular, it means that for all configurations
  \(\conf{\reg_{0},\mem_{0}}\) and \(\conf{\reg_{0}',\mem_{0}'}\) such that
  \(\sproj{\reg_{0}} = \sproj{\reg_{0}'}\) and
  \(\sproj{\mem_{0}} = \sproj{\mem_{0}'}\), and for all number of step \(n\),
    \begin{equation*}
      \sarch{n}{\declassified}{\conf{\mem_{0}, \reg_{0}}}%
      {\conf{\mem_{n}, \reg_{n}}}{\leaked{\obs}} \implies%
      \sarchpatched{n}{\conf{\mem_{0}',\reg_{0}'}}%
      {\conf{\mem_{n}', \reg_{n}'}}{\leaked{\obs'}}{\declassified}{\varepsilon} ~\wedge~ \leaked{\obs} = \leaked{\obs'}
  \end{equation*}
\end{restatable}

\noindent%
Additionally, the following contract emphasizes that software-developer should
make sure to not unintentionally declassify secrets by writing them to public
locations.
\hypdeclassification*

The following lemma expresses that given a n-step patched execution, which from
an initial declassification trace \(\declassify{\declassified_{i}}\) produces a
declassification trace \(\declassify{\declassified_{f}}\), adding a suffix
\(\declassify{\declassify{\declassified}}\) to the declassification trace does
not change the execution and the final declassification trace becomes
\(\declassify{\declassified_{f}} \cdot \declassified\):
\begin{lemma}\label{lemma:patch_length_declassified}
  For all configuration \(\conf{\mem_{0}', \reg_{0}', \buf_{0}, \micro_{0}}\),
  number of steps \(n\) and declassification traces
  \(\declassify{\declassified_{i}}\), \(\declassify{\declassified_{f}}\), and
  \(\declassify{\declassified}\):
\begin{gather*}
  \smicropatched{}{n}{\conf{\mem_{0}, \reg_{0}, \buf_{0}, \micro_{0}}}{\conf{\mem_{n}, \reg_{n}, \buf_{n}, \micro_{n}}}{\declassified_{i}}{\declassified_{f}} \implies \\
  \smicropatched{}{n}{\conf{\mem_{0}, \reg_{0}, \buf_{0}, \micro_{0}}}{\conf{\mem_{n}, \reg_{n}, \buf_{n}, \micro_{n}}}{\declassified_{i} \cdot \declassified}{\declassified_{f}\cdot \declassified}
\end{gather*}
\end{lemma}
\begin{proof}
  The proof goes by induction on the number of steps. The base case directly
  follows from \cref{def:n-step-patched}: for a 0-step execution, the
  declassification trace is simply forwarded:
  \(\smicropatched{}{0}{c}{c}{\declassified_{i}}{\declassified_{i}}\).
  Therefore, adding a suffix to the declassification trace does not change the
  execution and declassification trace is forwarded with its suffix:
  \(\smicropatched{}{0}{c}{c}{\declassified_{i} \cdot \declassified_{d}}{\declassified \cdot \declassified_{d}}\).

  We assume that the hypothesis holds for \(n-1\) steps and show that it still
  holds after \(n\) steps. From \cref{def:n-step-patched}, this amounts to showing:
  \begin{gather}
    \smicropatched{}{}{\conf{\mem_{n-1}, \reg_{n-1}, \buf_{n-1}, \micro_{n-1}}}{\conf{\mem_{n}, \reg_{n}, \buf_{n}, \micro_{n}}}{\declassified_{i}}{\declassified_{f}} \implies\notag \\
    \smicropatched{}{}{\conf{\mem_{n-1}, \reg_{n-1}, \buf_{n-1}, \micro_{n-1}}}{\conf{\mem_{n}, \reg_{n}, \buf_{n}, \micro_{n}}}{\declassified_{i} \cdot \declassified}{\declassified_{f} \cdot \declassified} \tag{G}\label{patch_length_declassified:goal}
  \end{gather}

  The proof proceeds by case analysis on the hardware evaluation. %
  Notice that all the rules except \textsc{retire-store-patched} do not use the
  declassification trace and simply forward it to the next configuration (i.e., \(\declassify{\declassified_{i}} = \declassify{\declassified_{f}}\)).
  Therefore, for these rules, adding a suffix \(\declassified\) to the
  declassification trace \(\declassify{\declassified_{i}}\) does not change the
  execution: the same rule is applied and the same configuration is returned
  because the semantics is deterministic (cf.\ \cref{hyp:deterministic}).
  Moreover, the declassification trace with its suffix
  \(\declassify{\declassified_{i}}\cdot \declassified\) is forwarded to the next
  configuration, which concludes \cref{patch_length_declassified:goal}.

  For the rule \textsc{retire-store-patched}, we have
  \(\declassify{\declassified_{i}} = \declassify{\val{v}} \cdot \declassify{\declassified_{f}}\).
  In this case, the application of the rule with
  \(\declassify{\declassified_{i}} \cdot \declassified\) does not change the
  execution (the same microarchitectural configuration is returned) and produces
  a final declassification trace
  \(\declassify{\declassified_{f}} \cdot \declassified\), which conclude
  \cref{patch_length_declassified:goal}.
\end{proof}

\newpar%
The following lemma expresses that the hardware semantics and the architectural semantics produce the same declassification traces:
\begin{lemma}\label{lemma:decl-eq}
  For all initial architectural configuration \(\conf{\reg,\mem}\) and
  microarchitectural context \(\micro\), let \(\arun\) be the longest
  architectural run starting from \(\conf{\reg,\mem}\) and producing a
  declassification trace \(\declassified\); and \(\mrun\) be the longest \name{}
  run starting from \(\mconf_{0} = \conf{\mem, \reg, \varepsilon, \micro}\) and
  producing a declassification trace \(\declassify{\declassified'}\). Then we
  have \(\declassified = \declassify{\declassified'}\).
\end{lemma}
\begin{proof}
  The proof goes by induction on the hardware evaluation.

  First, note that the sequence of retired instructions (excluding
  \(\incrpc{\_}\)) corresponds to the sequence of architectural instructions (by
  \cref{def:corr}). Hence, for the \(i^{th}\) instruction that is retired in the
  hardware semantics (excluding \(\incrpc{\_}\)) and corresponding to
  configuration \(n\), we have that \(\corr{\arun}{\mrun}(n)(0) = i\). %
  Second, note that the only rule in the hardware semantics that produces a
  declassification trace is \textsc{retire-store-low}.

  Hence it suffices to show that the sequence of retired instructions produces
  the same declassification events than the architectural semantics.

  Let us consider the \(i^{th}\) instruction that is retired in the hardware
  semantics (excluding \(\incrpc{\_}\)) and corresponding to configuration
  \(n\). We show that the corresponding architectural configuration
  \(\aconf_{i}\) produces the same declassification trace.

  If the retired instruction is retired using \textsc{retire-store-low} (cf.
  \cref{fig:retire-rules}), then the retired instruction is a \lstinline{store},
  it writes a value \(\val{v}\) at an address \(\val{a}\) such that
  \(\memsec(\val{a}) = \botsec\) and it produces a declassification trace
  \(\declassify{val{v}}\) that is the value written to memory by the store. From
  \cref{lemma:corr} (and because \(\varepsilon\) is a prefix of \(\buf_{n}\)),
  we have that the corresponding state \(\aconf_{i}\) (cf.
  \cref{fig:sequential-semantics}) also evaluates the rule \textsc{store} with
  index \(\val{a}\) and value \(\val{v}\). Hence it also produces a
  declassification trace \(\declassify{val{v}}\).

  If the retired instruction is retired using \textsc{retire-store-high} of
  \textsc{retire-assign}, we can show in a similar way that the declassification
  trace will be empty in the hardware and in the architectural semantics.

  We have show that declassification events only happen when retiring
  instructions, that the sequence of retired instructions corresponds to
  \(\arun\), and that retired instructions produce the same declassification
  events as their corresponding architectural step. Hence the declassification
  traces produced by \(\arun\) and \(\mrun\) are equal.
\end{proof}

\newpar%
This is the main lemma:
\begin{lemma}\label{lemma:security_patched}
  Assuming \cref{hyp:deterministic}, for all program \(\prog{}\) satisfying
  \cref{hyp:ct2} (and assuming \cref{hyp:declassification}), for all number of
  steps \(n\), memories \(\mem_{0}\) and \(\mem_{0}'\), register maps
  \(\reg_{0}\) and \(\reg_{0}'\), and microarchitectural context \(\micro_{0}\),
  \begin{gather}
    \smemproj{\mem_{0}} = \smemproj{\mem_{0}'} ~\wedge~ \sproj{\reg_{0}'} = \sproj{\reg_{0}'} ~\wedge~ %
    \smicro{}{n}{\conf{\mem_{0}, \reg_{0}, \varepsilon, \micro_{0}}}{\conf{\mem_{n}, \reg_{n}, \buf_{n},  \micro_{n}}}{\declassified_{0:n}} \implies \notag \\
    \smicropatched{}{n}{\conf{\mem_{0}', \reg_{0}', \varepsilon, \micro_{0}}}{\conf{\mem_{n}', \reg_{n}', \buf_{n}', \micro_{n}'}}{\declassified_{0:n}}{\varepsilon} ~\wedge \tag{Gstepn}\label{security_patched:stepfull} \\
    \rsbufproj{\buf_{n}} = \rsbufproj{\buf_{n}'} ~\wedge \tag{Gbuf}\label{security_patched:buf} \\
    \rsproj{\reg_{n}} = \rsproj{\reg_{n}'} ~\wedge \tag{Greg}\label{security_patched:reg} \\
    \smemproj{\mem_{n}} = \smemproj{\mem_{n}'} ~\wedge \tag{Gmem}\label{security_patched:mem} \\
    \micro_{n} = \micro_{n}' \tag{Gmicro}\label{security_patched:micro} ~\wedge \\
    \corr{\arun}{\mrun}(n) = \corr{\darun{\declassified}'}{\dmrun{\declassified}'}(n)\tag{Gcorr}\label{security_patched:corr} ~\wedge \\
    \forall 0 \leq i \leq |\buf_{n}|.\
    \transient(\conf{\mem_{n}, \reg_{n}, \buf_{n}\bufrange{0}{i}, \_}) = \transient(\dconf{\conf{\mem_{n}', \reg_{n}', \buf_{n}'\bufrange{0}{i}, \_}}{\declassify{\declassified_{n}}})\tag{Gtrans}\label{security_patched:trans}
  \end{gather}
  where \(\arun\) is the longest architectural run starting from
  \(\conf{\mem_{0},\reg_{0}}\), \(\mrun\) is the longest \name{} run starting
  from \(\conf{\mem_{0}, \reg_{0}, \varepsilon, \micro}\) and producing a
  declassification trace \(\declassified\), \(\darun{\declassified}'\) is the
  longest patched architectural run starting from
  \(\dconf{\conf{\mem_{0}',\reg_{0}'}}{\declassified}\),
  \(\dmrun{\declassified}'\) is the longest patched \name{} run starting from
  \(\dconf{\conf{\mem_{0}', \reg_{0}', \varepsilon, \micro}}{\declassified}\),
  and \(\declassify{\declassified_{n}}\) is the declassification trace of
  \(\dmrun{\declassified}[n]\).
\end{lemma}
\begin{proof}
  Assume a program \(\prog{}\) satisfying \cref{hyp:ct2}, memories \(\mem_{0}\)
  and \(\mem_{0}'\), register maps \(\reg_{0}\) and \(\reg_{0}'\), and a
  microarchitectural context \(\micro_{0}\), %
  such that:
  \begin{gather}
    \tag{Hloweq} \smemproj{\mem_{0}} = \smemproj{\mem_{0}} \text{ and } \sproj{\reg_{0}} = \sproj{\reg_{0}'} \label{security_patched:h:loweq} \\
    \tag{Hstepn} \smicro{}{n}{\conf{\mem_{0}, \reg_{0}, \varepsilon, \micro_{0}}}{\conf{\mem_{n}, \reg_{n}, \buf_{n}, \micro_{n}}}{\declassified_{0:n}} \label{security_patched:h:stepfull}
  \end{gather}
  Under these hypothesis, we show that \cref{security0:stepfull},
  \cref{security0:buf}, \cref{security0:mem}, \cref{security0:micro}, \cref{security0:corr}, and \cref{security0:trans} The proof goes by induction on the number of steps \(n\).

  \proofcase{Base case \((n = 0)\):}
  \begin{itemize}
    \item[\eqref{security_patched:stepfull}] From \cref{def:n-step}, a 0-step
          execution from the first configuration produces a declassification
          trace \(\declassify{\varepsilon}\):
          \(\smicro{}{0}{\conf{\mem_{0}, \reg_{0}, \varepsilon, \micro_{0}}}{\conf{\mem_{0}, \reg_{0}, \buf_{0}, \micro_{0}}}{\varepsilon}\).
          Hence, from \cref{def:n-step-patched}, in the second configuration, we have:
          \(\smicropatched{}{0}{\conf{\mem_{0}', \reg_{0}', \buf_{0}', \micro_{0}'}}{\conf{\mem_{0}', \reg_{0}', \buf_{0}', \micro_{0}'}}{\varepsilon}{\varepsilon}\)
    \item[\eqref{security0:buf}] \(\sbufproj{\buf_{0}} = \sbufproj{\buf_{0}'} = \varepsilon\)
    \item[\eqref{security0:reg}]
          \(\rsproj{\reg_{0}} = \rsproj{\reg_{0}'}\) follows from the
          application of \cref{lemma:sproj_rsproj_reg} with
          \cref{security_patched:h:loweq} and \cref{lemma:reg_sval},
    \item[\eqref{security0:mem}] \(\smemproj{\mem_{0}} = \smemproj{\mem_{0}'}\)
          directly follows from \cref{security0:h:loweq},
    \item[\eqref{security0:micro}] \(\micro_{0} = \micro_{0}'\) because the
          initial microarchitectural context is the same in both initial
          configurations,
    \item[\eqref{security0:corr}]
          \(\corr{\arun}{\mrun}(0) = \corr{\darun{\declassified}'}{\dmrun{\declassified}'}(0) = \{0 \mapsto 0\}\) by \cref{def:corr},
    \item[\eqref{security0:trans}]
          \(\forall 0 \leq i \leq |\buf_{0}|.\ \transient(\conf{\mem_{0}, \reg_{0}, \buf_{0}\bufrange{0}{i}, \micro_{0}}) = \transient(\dconf{\conf{\mem_{0}', \reg_{0}', \buf_{0}'\bufrange{0}{i}, \micro_{0}'}}{\declassify{\declassified_{0}}})\)
          follows directly by definition of \(\transient\) (cf.
          \cref{def:trans}) and \(\buf_{0} = \buf_{0}' = \varepsilon\).
  \end{itemize}

  \proofcase{Inductive case:}
  We assume that \cref{lemma:security_patched} holds at step \(n - 1\), meaning that:
  \begin{gather}
    \smemproj{\mem_{0}} = \smemproj{\mem_{0}'} ~\wedge~ \sproj{\reg_{0}'} = \sproj{\reg_{0}'} ~\wedge~ %
    \smicro{}{n-1}{\conf{\mem_{0}, \reg_{0}, \varepsilon, \micro_{0}}}{\conf{\mem_{n-1}, \reg_{n-1}, \buf_{n-1},  \micro_{n-1}}}{\declassified_{0:n-1}} \implies \notag \\
    \tag{IHstepn} \smicropatched{}{n-1}{\conf{\mem_{0}', \reg_{0}', \varepsilon, \micro_{0}}}{\conf{\mem_{n-1}', \reg_{n-1}', \buf_{n-1}', \micro_{n-1}'}}{\declassified_{0:n-1}}{\varepsilon} \label{security_patched:ih:stepfull} \\
    \tag{IHbuf} \rsbufproj{\buf_{n-1}} = \rsbufproj{\buf_{n-1}'} \label{security_patched:ih:buf} \\
    \tag{IHreg} \rsproj{\reg_{n-1}} = \rsproj{\reg_{n-1}'} \label{security_patched:ih:reg} \\
    \tag{IHmem} \smemproj{\mem_{n-1}} = \smemproj{\mem_{n-1}'} \label{security_patched:ih:mem} \\
    \tag{IHmicro} \micro_{n-1} = \micro_{n-1}' \label{security_patched:ih:micro} \\
    \tag{IHcorr} \corr{\arun}{\mrun}(n-1) = \corr{\darun{\declassified}'}{\dmrun{\declassified}'}(n-1) \label{security_patched:ih:corr} \\
    \forall 0 \leq i \leq |\buf_{n-1}|.\
    \transient(\conf{\mem_{n-1}, \reg_{n-1}, \buf_{n-1}\bufrange{0}{i}, \_}) = \transient(\dconf{\conf{\mem_{n-1}', \reg_{n-1}', \buf_{n-1}'\bufrange{0}{i}, \_}}{\declassify{\declassified_{n-1}}})\tag{IHtrans}\label{security_patched:ih:trans}
  \end{gather}
  Under these hypotheses, we show that \cref{lemma:security0} still holds at
  step \(n\).

  First, note that because the semantics is deterministic, we have
  \begin{equation*}
    \declassified = \declassify{\declassified_{0:n-1}} \cdot \declassify{\declassified_{n-1}} \text{ and can let } %
    \declassify{\declassified_{n-1}} = \declassify{e} \cdot \declassify{\declassified_{n}}
    \text{ where } \declassify{e} \text{ is either a value or } \varepsilon
  \end{equation*}

  \proofcase{Simplify goal \cref{security_patched:stepfull}:}
  By \cref{def:n-step} and \cref{security_patched:h:stepfull} we
  have:
  \begin{equation*}
    \smicro{}{n-1}{\conf{\mem_{0}, \reg_{0}, \varepsilon, \micro_{0}}}{\conf{\mem_{n-1}, \reg_{n-1}, \buf_{n-1},  \micro_{n-1}}}{\declassify{\declassified_{0:n-1}}}
  \end{equation*}
  which together with \cref{security_patched:h:loweq}, gives us that
  \cref{security_patched:ih:stepfull}, \cref{security_patched:ih:buf},
  \cref{security_patched:ih:reg} \cref{security_patched:ih:mem} and
  \cref{security_patched:ih:micro} hold at step \(n-1\).
  Additionally, by \cref{def:n-step} and \cref{security_patched:h:stepfull} we
  have:
  \begin{gather*}
    \smicro{}{}{\conf{\mem_{n-1}, \reg_{n-1}, \buf_{n-1}, \micro_{n-1}}}{\conf{\mem_{n}, \reg_{n}, \buf_{n}, \micro_{n}}}{\declassify{e}} \tag{IHstep}\label{security_patched:ih:step}
  \end{gather*}
  such that
  \(\declassify{\declassified_{0:n}} = \declassify{\declassified_{0:n-1}} \cdot \declassify{e}\).
  Therefore, from \cref{lemma:patch_length_declassified} and
  \cref{security_patched:ih:stepfull}, we have that:
  \begin{equation*}
    \smicropatched{}{n-1}{\conf{\mem_{0}', \reg_{0}', \varepsilon, \micro_{0}}}{\conf{\mem_{n-1}', \reg_{n-1}', \buf_{n-1}', \micro_{n-1}'}}{\declassified_{0:n}}{\declassify{e}}
  \end{equation*}
  Hence proving \cref{security_patched:stepfull}
  amounts to showing:
  \begin{equation}
    \smicropatched{}{}{\conf{\mem_{n-1}', \reg_{n-1}', \buf_{n-1}', \micro_{n-1}'}}{\conf{\mem_{n}', \reg_{n}', \buf_{n}', \micro_{n}'}}{\declassify{e}}{\varepsilon} \tag{Gstep}\label{security_patched:step}
  \end{equation}
  Consequently, we have to show that \cref{security_patched:step},
  \cref{security_patched:buf}, \cref{security_patched:reg} \cref{security_patched:mem},
  \cref{security_patched:micro}, \cref{security_patched:corr} and
  \cref{security_patched:trans} hold at step \(n\).

  \proofcase{Transform \cref{security_patched:ih:corr}:}
  First, observe that because the semantics is deterministic (cf.\
  \cref{hyp:deterministic}), we have that:
  \begin{gather}
    \mrun_{n-1} = \conf{\mem_{n-1}, \reg_{n-1}, \buf_{n-1}, \micro_{n-1}} \text{ and } %
    \dmrun{\declassified}'[n-1] = \dconf{\conf{\mem_{n-1}', \reg_{n-1}', \buf_{n-1}', \micro_{n-1}'}}{\declassify{\declassified_{n-1}}} %
    \tag{H$\mrun_{n-1}$}\label{security_patched:HmrunA}
  \end{gather}
  From this, \cref{lemma:corr-patch} and \cref{security_patched:ih:corr}, we have that for all
  \(\buf \in \prefix(\buf_{n-1}, \conf{\mem_{n-1},\reg_{n-1}})\) and
  \(\buf' \in \prefix(\buf_{n-1}', \conf{\mem_{n-1}',\reg_{n-1}'})\) %
  such that %
  \(|\buf| = |\buf'|\),
  \begin{equation}
    \deepapl{\conf{\mem_{n-1},\reg_{n-1}}}{\buf} = \aconf_{j} %
    \text{ and } \deepapl{\dconf{\conf{\mem_{n-1}',\reg_{n-1}'}}{\declassified}}{\buf} = \darun{\declassified}'[j] %
    \text{ where } j = \corr{\arun}{\mrun}(n-1)(|\buf|) \tag{IHarch}\label{security_patch:ih:arch} %
  \end{equation}

  Notice that from \cref{lemma:decl-eq}, we have that the declassification trace
  produced by \(\arun{}\) is \(\declassified\). From \cref{hyp:ct2}, it means
  that the observations produced by \(\arun{}\) and \(\darun{\declassified}\)
  are the same. In particular, for all \(j\), we have that
  \begin{equation*}
  \sarch{}{}{\aconf_{j}}{\_}{\obs} \text{ then } \sarch{}{}{\darun{\declassified}[j]'}{\_}{\obs'}
  \text{ and } \leaked{\obs} = \leaked{\obs'}
  \tag{Hct}\label{security_patch:ct} %
  \end{equation*}

  \proofcase{Proof overview.}
  The proof is similar to the proof of \cref{lemma:security0}. It proceeds by
  case analysis on the hardware evaluation, knowing that both configuration
  evaluate the same directive.
  There are two main points that differ from the proof for
  \cref{lemma:security0}:
  \begin{itemize}
    \item First, we have to show that low-memories are equal after the
          evaluation of the rule \textsc{retire-store-low} (which becomes
          \textsc{retire-store-patched} in the patched semantics);

    \item Second, \cref{security_patched:step} requires to show that the second
          configuration can make a step in the \emph{patched} execution and that
          the \emph{final declassification trace is empty}. %
          For all rules except \textsc{retire-store-patched}, the declassification
          trace that is produced is empty:
          \begin{equation*}
          \smicro{}{}{\conf{\mem_{n-1}, \reg_{n-1}, \buf_{n-1}, \micro_{n-1}}}{\conf{\mem_{n}, \reg_{n}, \buf_{n}, \micro_{n}}}{\varepsilon}
          \end{equation*}

          In the patched execution, the declassification trace is simply
          propagated to the next configuration. Hence the final declassification
          trace in the patched execution is also \(\declassify{\varepsilon}\):
          \begin{equation*}
            \smicropatched{}{}{\conf{\mem_{n-1}', \reg_{n-1}', \buf_{n-1}', \micro_{n-1}'}}{\conf{\mem_{n}', \reg_{n}', \buf_{n}', \micro_{n}'}}{\varepsilon}{\varepsilon}
          \end{equation*}
          Notice that the rules are the same in the standard and the patched
          executions. Moreover, whenever we use the hypothesis
          \cref{hyp:ct-no-decl} in the proof for \cref{lemma:security0}, we can
          apply the same kind of reasoning here with
          \cref{security_patch:ih:arch} and \cref{security_patch:ct} \lestodo{Detail this
          case}. Hence the proof that the patch execution can actually make a
          step and the proofs for goals \cref{security_patched:buf},
          \cref{security_patched:mem}, \cref{security_patched:reg},
          \cref{security_patched:micro}, \cref{security_patched:corr}, and
          \cref{security_patched:trans} are similar to the proof for
          \cref{lemma:security0}.
  \end{itemize}
  Consequently, we only need to adapt the proof of \cref{lemma:security0} for
  the rule \textsc{retire-store-patched}.

  \proofcase{Case \textsc{retire-store-patched}.} %
  In all retire rules for store, the first instruction \(inst\) is removed from
  the reorder buffer:
  \begin{equation*}
    \buf_{n-1} = inst @ \varepsilon \cdot buf_{n} \text{ and  } \buf_{n-1}' = inst' @ \varepsilon \cdot buf_{n}'
  \end{equation*}
  Therefore, \cref{security_patched:buf} follows from
  \cref{security_patched:ih:buf}. Additionally, from
  \cref{security_patched:ih:buf} and \cref{def:rsbufproj}, we have
  \begin{equation}
    \tag{Hinst}\label{security_patched:retire:instr}
    \rsbufproj{inst} = \rsbufproj{inst'}
  \end{equation}
  From the hypotheses of the rules, we have
  \(inst = \store{\slevel{a}{s_{a}}}{\slevel{v}{s_{v}}}\), meaning that
  \(\rsbufproj{inst} = \store{\rsbufproj{\slevel{a}{s_{a}}}}{\rsbufproj{\slevel{v}{s_{v}}}}\).
  Therefore, from \cref{security_patched:retire:instr}, we have
  \(\rsbufproj{inst'} = \store{\rsbufproj{\slevel{a}{s_{a}}}}{\rsbufproj{\slevel{v}{s_{v}}}}\)
  meaning that \(inst' = \store{\slevel{a'}{s_{a}'}}{\slevel{v'}{s_{v}'}}\).
  Consequently, from the hardware evaluation rules (cf. \cref{fig:retire-rules}), the only rules that can be applied in the second configuration
  are the rules \textsc{retire-store-patched} and
  \textsc{retire-store-high}.

  Additionally, we have:
  \(\rsbufproj{\slevel{a}{s_{a}}} = \rsbufproj{\slevel{a'}{s_{a}'}} \text{ and
  } \rsbufproj{\slevel{v}{s_{v}}} = \rsbufproj{\slevel{v'}{s_{v}'}}\), which from
  \cref{lemma:val_rsbufproj_rsproj} entails
  \(\rsproj{\slevel{a}{s_{a}}} = \rsproj{\slevel{a'}{s_{a}'}}\) and
  \(\rsproj{\slevel{v}{s_{v}}} = \rsproj{\slevel{v'}{s_{v}'}}\).
  Finally, because \(\buf_{n}\) and \(\buf_{n-1}\) are well-formed (cf.\
  \cref{lemma:wfbuf}) and because store addresses have security level \(\botsec\)
  in well-formed buffers (by \cref{def:wfbuf}), we have
  \(\level{s_{a}} = \level{s_{a}'} = \botsec\). From
  \(\rsproj{\slevel{a}{s_{a}}} = \rsproj{\slevel{a'}{s_{a}'}}\) this entails
  \(\val{a} = \val{a'}\).
  In particular, it means that \(\memsec(\val{a}) = \memsec(\val{a})\). The
  proof for \textsc{retire-store-high} is similar to the proof for
  \cref{lemma:security0}.

  In both \textsc{retire-store-low} and \textsc{retire-store-patched}, the
  register map is not modified. Therefore, \cref{security_patched:reg} directly
  follows from \cref{security_patched:ih:reg}. As we have already shown
  \cref{security_patched:buf}, it remains to show \cref{security_patched:step},
  \cref{security_patched:mem}, \cref{security_patched:micro} and \cref{security_patched:trans}.

  \begin{itemize}
    \item[\cref{security_patched:micro}] See \cref{lemma:security0}.

    \item[\cref{security_patched:step}] The first execution evaluates the rule
          \textsc{retire-store-low}, producing the declassification trace
          \(\declassify{\val{v}}\). Therefore we have
          \(\declassify{e} = \declassify{\val{v}}\) and, from
          \textsc{retire-store-patched}, %
          \(\smicropatched{}{}{\conf{\mem_{n-1}', \reg_{n-1}', \buf_{n-1}', \micro_{n-1}'}}{\conf{\mem_{n}', \reg_{n}', \buf_{n}', \micro_{n}'}}{\declassify{\val{v}}}{\varepsilon}\),
          which concludes \cref{security_patched:step}.

    \item[\cref{security_patched:mem}] %
          The first execution (cf.\ rule \textsc{retire-store-low}) updates its
          memory such that
          \(\mem_{n} \mydef \mem_{n-1}\change{\val{a}}{\val{v}}\) and produces
          the declassification trace \(\declassify{\val{v}}\). Notice that from
          \cref{hyp:declassification} the declassification of \(\val{v}\) is
          intentional and it is secure to consider that \(\mem_{n}(\val{a})\) is
          public.
          In the second execution (cf.\ rule \textsc{retire-store-patched}),
          we have \(\declassify{e} = \declassify{\val{v}}\)
          and the memory is updated with the declassified value
          \(\mem_{n}' \mydef \mem_{n-1}'\change{\val{a}}{\declassify{\val{v}}}\). Therefore,
          \cref{security_patched:mem} follows from
          \cref{security_patched:ih:mem}.

    \item[\cref{security_patched:trans}]  Notice that
          \(\buf_{n} = \buf_{n-1}\bufrange{2}{|\buf_{n-1}|}\) and
          \(\buf_{n}' = \buf_{n-1}'\bufrange{2}{|\buf_{n-1}'|}\). Additionally, we
          have \(\mem_{n} \mydef \mem_{n-1}\change{\val{a}}{\declassify{\val{v}}}\) and
          \(\mem_{n}' \mydef \mem_{n-1}'\change{\val{a}}{\declassify{\val{v}}}\), as well
          as \(\reg_{n-1} = \reg_{n}\) and \(\reg_{n-1}' = \reg_{n}'\).
          Hence, to show \cref{security_patched:trans}, we need to show:
\[
  \transient(\conf{\mem_{n-1}\change{\val{a}}{\declassify{\val{v}}}, \reg_{n-1}, \buf_{n-1}\bufrange{2}{|\buf_{n-1}|}}) \iff \transient(\dconf{\conf{\mem_{n-1}'\change{\val{a}}{\declassify{\val{v}}}, \reg_{n-1}}}{\declassify{\declassified_{n}}}, \buf_{n-1}'\bufrange{2}{|\buf_{n-1}'|})
\]
  This simply follows from \cref{security_patched:ih:trans}, \cref{def:deepapl-patch} and \cref{def:trans}.%

  \end{itemize}
\end{proof}

\thmpatcheddeclassification*
\begin{proof}
  Proof follows from the direct application of \cref{lemma:security_patched}.
\end{proof}

\fi{}

\end{document}